\documentclass[twocolumn, twocolappendix]{aastex63}
\usepackage{natbib}
\usepackage{float}
\usepackage{tabularx}
\usepackage{multirow}
\usepackage{rotating}
\usepackage{footnote}
\usepackage{subfigure}
\usepackage{ae,aecompl}
\usepackage{graphicx}	
\usepackage{amsmath}	
\usepackage{amssymb}	
\usepackage[switch]{lineno} 
\received{---}
\revised{---}
\accepted{}
\submitjournal{AJ}

\shorttitle{Spectroscopic study of Ba and CEMP-s stars}
\shortauthors{Goswami \& Goswami}
\graphicspath{{./}{figures/}}

\begin{document}
\title{Spectroscopic study of Ba and CEMP-s stars\footnote{Based [in part] on data collected at Subaru Telescope, which is operated by the National Astronomical Observatory of Japan.}\footnote{[Part of] the data are retrieved from the JVO portal (\url{http://jvo.nao.ac.jp/portal}) operated by the NAOJ}: Mass distribution of AGB progenitors}

\correspondingauthor{Partha Pratim Goswami}
\email{partha.pg@iiap.res.in, ppg0024@gmail.com}

\author{Partha Pratim Goswami}
\affiliation{Indian Institute of Astrophysics, Koramangala, Bangalore
    560034, India; partha.pg@iiap.res.in; aruna@iiap.res.in
}
\affiliation{Pondicherry University, R.V. Nagar, Kalapet, 
605014 Puducherry, India
}

\author{Aruna Goswami}
\affiliation{Indian Institute of Astrophysics, Koramangala, Bangalore
    560034, India; partha.pg@iiap.res.in; aruna@iiap.res.in
}



\begin{abstract}

We have performed detailed high-resolution spectroscopic analysis on seven metal-poor stars (BD+75~348, BD+09~3019, HD238020, HE0319--0215, HE0507--1653, HE0930--0018, HE1023--1504) and derived their atmospheric parameters T$_{eff}$, log$g$, [Fe/H], and mictroturbulent velocity ($\xi$).  The metallicity range is found to be --2.57$<$[Fe/H]$<$--0.42. The elemental abundances of 17 light elements and 12 heavy elements are estimated.  We have classified  BD+75~348 and  BD+09~3019 as strong Ba stars, HD238020 as a mild Ba star, and the remaining four objects as CEMP-s stars.  We have estimated the masses of the stars from Hertzsprung-Russel (HR) diagram, and, compiling the data of 205 Ba stars from literature, estimated the mass distribution of Ba stars. We have also estimated the initial masses of the companion AGBs of the program stars  as well as  the masses of the companion AGBs of 159 Ba and 36 CEMP-s stars from literature, with the help of a parametric-model-based analysis using FRUITY models. While the primary mass distribution of mild Ba stars peaks at 3.7~M$_{\odot}$, for the strong Ba stars the peak appears at 2.5~M$_{\odot}$. We, therefore, propose  that the initial masses of the progenitor AGBs  dominantly control the formation of mild and strong Ba stars. However, a clear overlap, in the range 1.3--4.0 M$_{\odot}$, noticed between the progenitor masses of both the subclasses of Ba stars, may indicate that other  factors, such as metallicities and the orbital periods, may also have significant contributions. The progenitor AGBs' mass distribution of CEMP-s stars is found to peak at 2.03~M$_{\odot}$.

\end{abstract}


\section{Introduction}
\label{sec:introduction}

Barium (Ba) stars and a subclass of carbon enhanced metal-poor (CEMP) stars, known as CEMP-s stars, exhibit enhanced abundances of elements produced by slow (s-) neutron (n)-capture process. Ba stars, first recognized by \citet{Bidelman_Keenan_1951}, are identified by the presence of strong absorption lines of Ba~II at 4554 {\rm \AA} \& Sr~II at 4077 {\rm \AA} in their spectra \citep{Warner_1965}. C$_{2}$, CN, and CH molecular bands are also prominent in Ba stars. Depending on the types of neutron-capture elements present, carbon enhanced metal-poor stars are divided into four subclasses \citep{beers2005discovery, Frebel_review_2018, Goswami_et_al_1_2021}, namely CEMP-s stars (enriched with s-process elements), CEMP-r stars (enriched with elements produced in rapid n-capture process or r-process), CEMP-r/s stars (enriched with elements produced in intermediate n-capture process or i-process \citep{cowan1977, Herwig_et_al_2011, Doherty_et_al_2015, Hampel_et_al_2016, Bannerjee_et_al_2018, Denissenkov_et_al_2019, Goswami_et_al_1_2021}) and CEMP-no stars (not enriched with n-capture elements). 

The formation scenario of CEMP-s \citep{Spite_et_al_2013} and Ba \citep{Cseh_et_al_2018, Cseh_et_al_2022} stars is similar. Both the types of stars (secondary) are extrinsic in nature and they get the s-process rich material through mass-transfer from a more massive companion (primary), which evolves faster and produces s-process elements in the AGB phase. CEMP-s and Ba stars are known to be in binary systems with invisible white-dwarf companions. The binary nature of these stars is proved by long-term radial velocity monitoring programs \citep{McClure1983, McClure1984, McClure1990, lucatello2005, Starkenburg_et_al.2014, jorissen2016rv, Hansen2016binaries}. Recent studies \citep{Roederer_2016, Hampel_et_al_2016, Hampel_et_al_2019, Koch_et_al_2019, Goswami_Goswami_2020, Goswami_et_al_1_2021} have revealed that one of the formation scenarios for CEMP-r/s stars might also be the binary evolution with contamination by AGB yields similar to the CEMP-s stars. The different physical conditions between the production of s- and i-process elements in AGB stars are still under debate. In our recent study \citep{Goswami_Goswami_2021}, we have reported an object HE~1005--1439, abundance pattern of which shows contamination by both s- and i-processes. We proposed that in AGB stars, s- and i-process can take place in succession without masking the signatures of each other. Although the formation scenarios of Ba, CEMP-s and CEMP-r/s stars are 
similar,  in this paper, we will discuss only CEMP-s and Ba stars.\\

\citet{Warner_1965} classified the Ba stars with Ba index 1 -- 5 based on the strength of Ba~II 4554 {\rm \AA} line. Ba1 and Ba5 signify stars showing the weakest and the strongest Ba lines respectively.  Later, several authors \citep{Lu_1991, Jorissen_et_al_1998, Yang_et_al_2016, Escorza_et_al_2017} used this Ba index to classify the Ba stars into two groups: mild Ba stars and strong Ba stars. The overabundance of n-capture elements is higher in strong Ba stars than that of mild Ba stars. In the Literature, lower enhancement of heavy elements in mild Ba stars are explained by two plausible formation scenarios- i) a longer orbital period of the binary system and ii) relatively weaker neutron-exposure in the AGB companion that pollutes the star \citep{Yang_et_al_2016}.

In this paper, we have reported the elemental abundances of seven stars BD+75~348, BD+09~3019, HD~238020, HE~0319--0215, HE~0507--1653, HE~0930--0018 and HE~1023--1504 based on detailed high-resolution spectroscopic analysis. We have estimated the masses of the program stars and their AGB companions. We have estimated the masses of several Ba stars found in the literature to find the mass distribution of these stars. We have also derived the initial mass distribution of the AGB companions of these Ba stars and a sample of literature CEMP-s stars. Here, we refer the stars (Ba and CEMP-s) that we are observing now as secondary stars and their companion AGBs as primary stars. 

We have organized the paper as follows. In Section~\ref{sec:previous_studies} we briefly discuss the summary of the earlier studies available in literature on our program stars. The details of the sources of the spectra are presented in Section~\ref{sec:source_of_spectra}. In Section~\ref{sec:photometric_temperature}, we have discussed the method of determination of photometric  temperatures. Section~\ref{sec:radial_velocity_stellar_atm_parameters} discusses the method of estimation of radial velocity and derivation of stellar atmospheric parameters of the program stars. In Section~\ref{sec:results} we have presented the Abundance analysis results (Section~\ref{sec:abundance_analysis}) and the kinematic analysis of the program stars (Section~\ref{sec:kinematics}). In Section~\ref{sec:discussion} we have presented i) the classification schemes of Ba and CEMP-s stars (Section~\ref{sec:classification1}), ii) classification of the program stars (Section~\ref{sec:classification2}), iii) determination of masses of the program stars (Section~\ref{sec:secondary_mass}), iv) determination of masses of the primary companions (AGB progenitors) of the program stars (Section~\ref{sec:primary_mass}), v) a comprehensive discussion on the mass distributions of a literature sample of Ba stars and their AGB progenitors and the mass distribution of the AGB progenitors of a literature sample of CEMP-s stars (Section~\ref{sec:mass_distribution}) and vi) formation scenarios of mild and strong Ba stars (Section~\ref{sec:formation}). Section~\ref{sec:conclusion} draws the conclusions.

\section{Previous Studies of the program stars}
\label{sec:previous_studies}
\subsection{BD+09~3019, HD~238020, HE~0930--0018 \& HE~1023--1504}

BD+09~3019 is included in the carbon star catalog of \citet{Stephenson1989}, CH star catalog of \citet{bartkevicius1996new} and the carbon star catalogue of \citet{Ji_lamost_dr2_carbon_2016} from LAMOST DR2 data. HD~238020 is included in the CH star catalogue of \citet{bartkevicius1996new}. The list of faint high-latitude carbon stars of \citet{christlieb2001} includes HE~0930--0018 and HE~1023--1504. Atmospheric parameters of BD+09~3019 and HE~0930--0018 are not reported previously in the literature.  \citet{McDonald2012} derived the effective temperatures of HD~238020 using the spectral energy distribution (SED) method of temperature calibration. Our estimate of temperature is more by 115 K for HD~238020 than that of \citet{McDonald2012}. \citet{kennedy2011} estimated the atmospheric parameters and the abundance of oxygen for HE~1023--1504. Our estimates of effective temperature and log~$g$ closely match the estimates of \citet{kennedy2011}. However, the metallicity ([Fe/H]) estimated by \citet{kennedy2011} is $\sim$ 0.8 dex lower than that of our estimate. In this work, first-time abundance estimates of several light elements from C through Zn and neutron-capture elements from Sr through Hf are presented for these four objects based on high-resolution spectroscopic analysis. \\

\subsection{BD+75~348, HE~0319--0215 \& HE~0507--1653}
BD+75~348 is listed in the carbon star catalog of \citet{Stephenson1989} and CH star catalog of \citet{bartkevicius1996new}. \citet{bergeat2001effective} and \citet{McDonald2012} estimated the effective temperature of BD+75~348 using the SED method of temperature calibration. \citet{zacs_et_al_2000} derived the atmospheric parameters as well as abundances of ten light and nine heavy elements for BD+75~348. The literature values of T$_{eff}$ for BD+75~348 range from 4700--4900 K. Our estimate of T$_{eff}$ ($\sim$ 4840 K) falls within the range. HE~0319--0215 \& HE~0507--1653 are included in the list of the faint high-latitude carbon stars of \citet{christlieb2001}. \citet{goswami2005ch} estimated the $^{12}$C/$^{13}$C ratio for HE~0319--0215 ($^{12}$C/$^{13}$C $\sim$ 4.7) and HE~0507--1653 ($^{12}$C/$^{13}$C $\sim$ 6.7) based on medium-resolution spectra. From the studies of \citet{Hansen_2016_III} on HE~0319--0215 and \citet{Hansen_2016_III} and \citet{jorissen2016rv} on HE~0507--1653 it is found that both the objects exhibit radial velocity variations with periods of 3078 days for HE~0319--0215 and 405 days for HE~0507--1653 confirming their binary nature. \citet{kennedy2011} derived atmospheric parameters and abundance of oxygen for HE~0319--0215. \citet{Hansen_2016_III} reported [Fe/H] = --2.30, [C/Fe] = 2.0, [Ba/Fe] = 0.52 for HE~0319--0215. \citet{karinkuzhi_et_al_i_process_2021} also derived the atmospheric parameters and elemental abundances of HE~0319--0215 and reported the object as a CEMP-r/s star. However, the estimates of [Fe/H] and [Ba/Fe] by the studies of \citet{Hansen_2016_III} and \citet{karinkuzhi_et_al_i_process_2021} differ by $\sim$ 0.6 dex and $\sim$ 1.23 dex respectively. These discrepancies in the literature values compelled us to re-investigate this object in detail using high-resolution spectroscopy. \citet{Schuler_et_al_2008} reported [Fe/H] = --1.42, [C/Fe] = 1.33 and [N/Fe] = 1.20 for HE~0507--1653. \citet{aoki2007carbon}  and \citet{Yong_et_al_2013}  derived atmospheric parameters and abundances of C, N, Na, Mg, Ca, Sc, Ti, Cr, Ni, Zn and Ba for HE~0507--1653. \citet{karinkuzhi_et_al_i_process_2021}  reported this object  as a CEMP-s star. However, this object also shows discrepancies in the atmospheric parameters and the abundance of Ba obtained by different groups. For instance, the ranges of T$_{eff}$, log~$g$ and [Fe/H] are 4880 to 5035 K, 1.50 to 2.40 dex and --1.81 to --1.32 dex, respectively, and the [Ba/Fe] ranges from 1.56 -- 1.89 dex. These differences prompted us to study this object in detail.

\section{Source of spectra}
\label{sec:source_of_spectra}

\begin{figure*}
     \begin{center}
\centering
        \subfigure[]{%
            \label{fig:sample_first}
            \includegraphics[height=8.0cm,width=8.5cm]{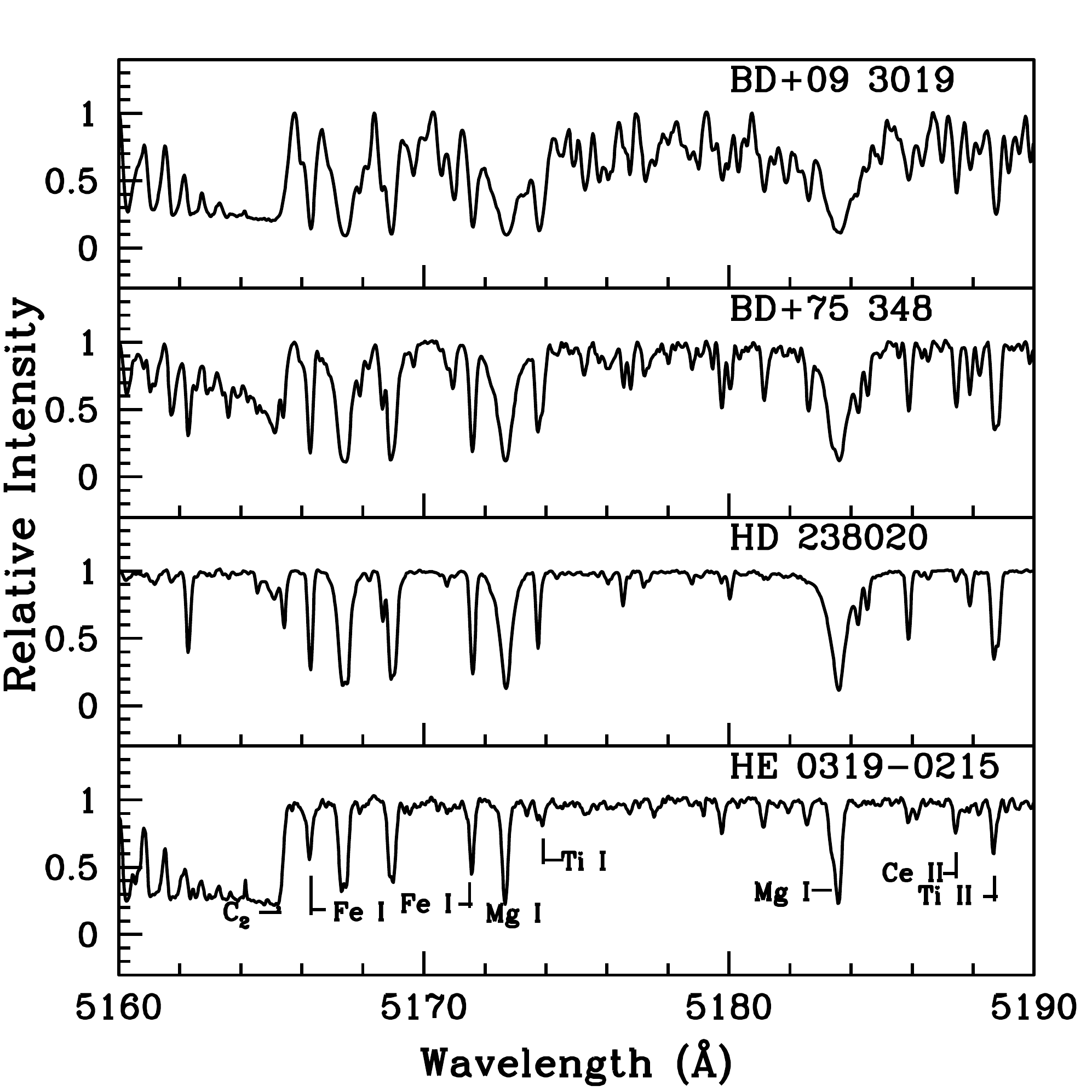}
        }%
        \subfigure[]{%
            \label{fig:sample_second}
            \includegraphics[height=8.0cm,width=8.5cm]{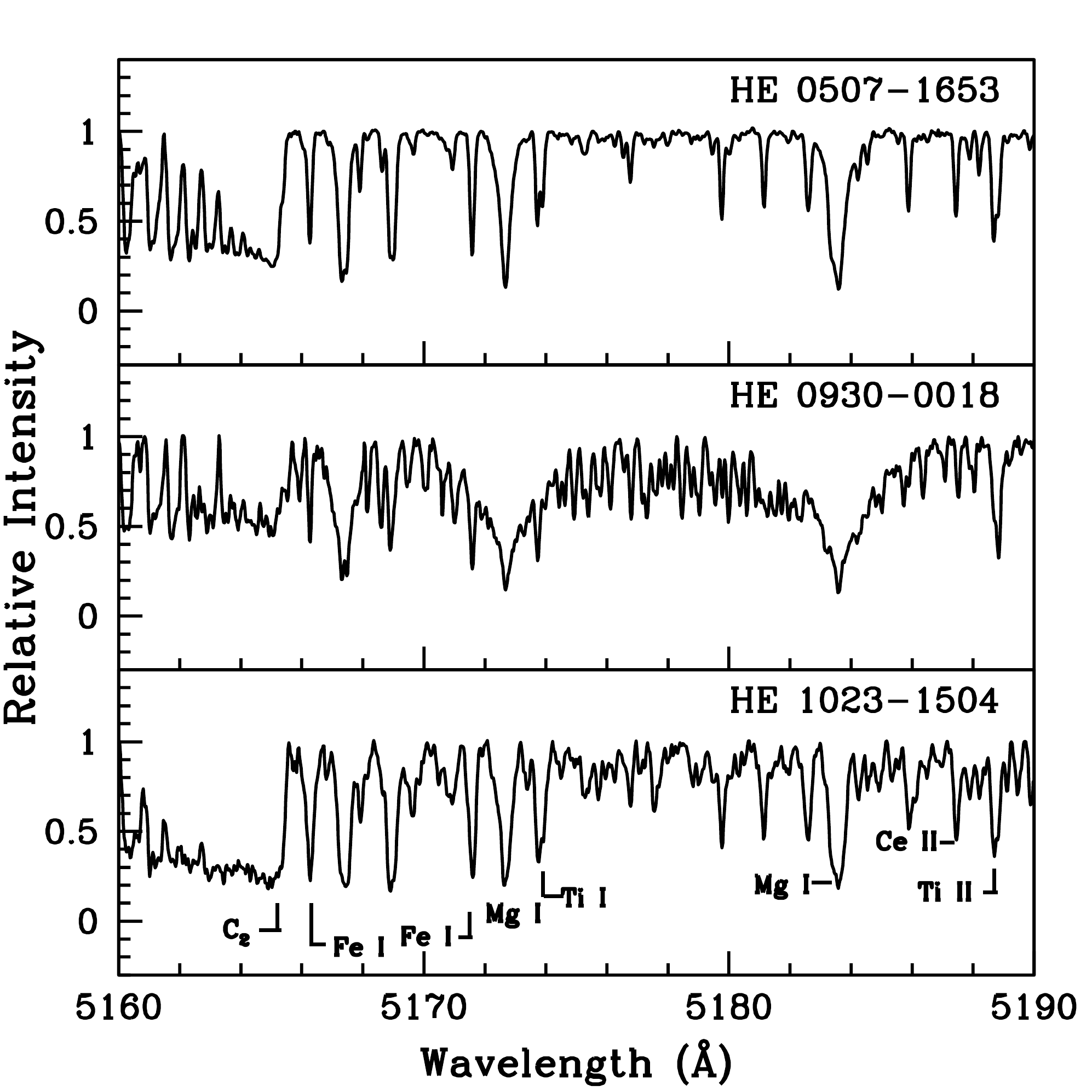}

        }\\ 

    \caption{ Sample spectra of the program stars in the wavelength region 5160 -- 5190 {\rm \AA}}%
   \label{fig:sample_spectra}
       \end{center}

\end{figure*}

We acquired high-resolution spectra of BD+09~3019, BD+75~348, and HD~238020 using the Hanle Echelle SPectrograph (HESP) attached to the 2m Himalayan Chandra Telescope (HCT) at the Indian Astronomical Observatory (IAO), Hanle. A 4K $\times$ 4K CCD detector with a pixel size of 15 $\micron$ is used. The wavelength of the spectra covers 3500--10~000 {\rm \AA} at a spectral resolution ($\lambda$/$\delta\lambda$) of 60~000. For data reduction, we have used Image Reduction and Analysis Facility (IRAF) software packages following a standard procedure. We have applied spectroscopic data reduction procedures such as trimming, bias subtraction, flat normalization, and extraction to the raw data. A high-resolution Th-Ar arc spectrum is used for wavelength calibration. High-resolution spectra (R $\sim$ 50~000) of HE~0319--0215, HE~0507--1653, HE~0930--0018 \& HE~1023--1504 are taken from the SUBARU archive\footnote{\url{http://jvo.nao.ac.jp/portal}} acquired using the High Dispersion Spectrograph (HDS) \citep{Noguchi_et_al_2002} attached to the 8.2 m Subaru Telescope. For HE~0319--0215, HE~0930--0018 \& HE~1023--1504, the wavelength coverage of the spectra spans from about 4020 to 6775 {\rm \AA}, with a gap of about 100 {\rm \AA} (from 5340 to 5440 {\rm \AA}). The wavelength coverage of the spectra of HE~0507--1653 spans from 4090 to 6870 {\rm \AA}, with a gap of 80 {\rm \AA} (from 5430 to 5510 {\rm \AA}). The gap in the wavelength coverage appears due to the physical spacing of the CCD detectors. We have continuum fitted the spectra using the task continuum in IRAF. The sample spectra of the program stars in the wavelength region 5160--5190 {\rm \AA} are shown in Figure~\ref{fig:sample_spectra}. The basic data for the program stars are presented in Table~\ref{tab:basicdata}.

{\footnotesize
\begin{table*}
\caption{ \bf{Basic data for the program stars}}
\label{tab:basicdata}
\scalebox{0.70}{
\begin{tabular}{lcccccccccccc}
\hline
Star Name         & RA$(2000)$   & Dec.$(2000)$   & B      & V      & J       & H      & K     & Exposure  & Date of Obs.  & Source      & S/N     & S/N    \\
                  &              &                &        &        &         &        &       & (seconds) &               & of spectrum & (4000 {\rm \AA}) & (6000 {\rm \AA}) \\
\hline
BD+75~348         & 08 46 11.64  &  +74 32 31.24  & 10.63  &  9.54  &  7.84   &  7.33  &  7.24 &   2400    & 08-03-2018    & HESP/HCT    & 50 & 150  \\
BD+09~3019        & 15 18 32.06  &   09 06 09.74  & 12.04  & 10.82  &  7.99   &  7.24  &  6.99 &   2700    & 24-05-2018    & HESP/HCT    & 20 &  40  \\
HD~238020         & 11 36 10.99  &  +56 50 17.66  &  9.34  &  8.49  &  6.90   &  6.51  &  6.38 &   1800    & 06-05-2017    & HESP/HCT    & 70 & 200  \\
HE~0319--0215     & 03 21 46.26  & --02 04 33.95  & 15.03  & 13.60  & 11.79   & 11.22  & 11.06 &   1800    & 08-12-2003    & HDS/SUBARU  & 40 &  70  \\
HE~0507--1653     & 05 09 16.56  & --16 50 04.69  & 13.63  & 12.51  & 10.88   & 10.43  & 10.32 &   1200    & 26-10-2002    & HDS/SUBARU  & 40 &  75  \\
HE~0930--0018     & 09 33 24.63  & --00 31 44.60  & 16.13  & 14.70  & 12.19   & 11.55  & 11.34 &   1800    & 09-12-2003    & HDS/SUBARU  & 10 &  30  \\
HE~1023--1504     & 10 25 55.55  & --15 19 17.08  & 16.26  & 14.40  & 12.32   & 11.61  & 11.42 &   1200    & 27-02-2005    & HDS/SUBARU  &  7 &  25  \\
\hline

\end{tabular}}
\end{table*}
}

\section{Photometric Temperatures}
\label{sec:photometric_temperature}

We have used broad-band colors, optical, and IR to determine the photometric temperatures of the program stars with color-temperature calibrations, based on the infrared flux method (IRFM), available for main-sequence \citep{Alonso1996} and giant stars \citep{alonso1999effective}. We have taken the 2MASS photometric magnitudes for J, H, and K from \citet{2MASS_2003}. We have followed the same procedure described in \citet{Goswami_et_al_2006, goswami_et_al_2016} and  \citet{Goswami_et_al_1_2021} as briefly outlined below. The photometric temperatures of the program stars corresponding to  J--H and V--K colors are estimated at different assumed metallicity values. The estimated photometric temperatures are listed in Table~\ref{tab:photometric_temp}. In order to select the model atmospheres to estimate the spectroscopic temperatures, we have used T$_{eff}$(J--K) as an initial guess as this temperature is independent of metallicity \citep{Alonso1996, alonso1999effective}.

{\footnotesize
\begin{table*}
\centering
\caption{\bf{Temperatures from  photometry}}
	\label{tab:photometric_temp}
\scalebox{0.85}{
\begin{tabular}{lllllllllllll}
\hline
Star Name      &$T_{eff}$&$T_{eff}$&$T_{eff}$&$T_{eff}$&$T_{eff}$&$T_{eff}$&T$_{eff}$&$T_{eff}$&T$_{eff}$&T$_{eff}$&T$_{eff}$& Spectroscopic \\
               &         & $(-0.5)$& $(-1.0)$& $(-1.5)$&$(-2.0)$ &$(-2.5)$ &$(-0.5)$ & $(-1.0)$& $(-1.5)$& $(-2.0)$& $(-2.5)$& estimates     \\
               & (J--K)  & (J--H)  & (J--H)  & (J--H)  & (J--H)  & (J--H)  & (V--K)  & (V--K)  & (V--K)  & (V--K)  & (V--K)  &               \\
\hline
BD+75~348      &  4856   &  4867   &  4891   &  4891   &   4867  &  4820   &  4749   &  4736   &  4729   &  4728   &  4733   &  4840         \\
BD+09~3019     &  3879   &  4038   &  4051   &  4047   &   4027  &   -     &  3810   &  3795   &   -     &   -     &    -    &  4220         \\ 
HD~238020      &  5085   &  5350   &  5382   &  5384   &   5358  &  5304   &  4941   &  4931   &  4928   &  4931   &  4941   &  5150         \\
HE~0319--0215  &  4479   &  4616   &  4636   &  4635   &   4612  &  4568   &  4530   &  4517   &  4510   &  4510   &  4515   &  4650         \\
HE~0507--1653  &  4935   &  5078   &  5105   &  5106   &   5081  &  5031   &  4855   &  4844   &  4839   &  4840   &  4848   &  4970         \\
HE~0930$-$0018 &  4164   &  4354   &  4370   &  4368   &   4346  &  4306   &  4002   &  3988   &  3979   &  3974   &    -    &  4190         \\
HE~1023--1504  &  4062   &  4144   &  4157   &  4154   &   4133  &   -     &  4211   &  4198   &  4190   &  4187   &    -    &  4440         \\
\hline
\end{tabular}}

The numbers in the parenthesis below $T_{eff}$ indicate the metallicity values at which the temperatures are calculated. \\
Temperatures are given in Kelvin.\\
\end{table*}
}

\section{Radial velocities and stellar atmospheric parameters}
\label{sec:radial_velocity_stellar_atm_parameters}

We have measured the radial velocities of the program stars using Doppler's formula from the shifts of the elemental absorption lines from the lab wavelengths. For this purpose, we have chosen a large number  of clean and unblended spectral lines of different elements in each star and compared them with the rest frame laboratory wavelength taking the spectra of Arcturus \citep{hinkle_et_al_2000} as a template. As Arcturus belongs to the giant class and has a comparable temperature as the program stars, we have chosen Arcturus for the comparison to have homogeneity in the analysis. We have also used the FXCOR package in IRAF and cross-checked our measurements from line-to-line analysis to the measurements provided by FXCOR. The measured radial velocities, after correcting for heliocentric motion, and the standard deviations from the mean value are presented in Table~\ref{tab:rad_vel}, along with the radial velocity information provided by \citet{gaia2018}. For the stars HE~0930--0018 and HE~1023--1504, radial velocities are not reported in \citet{gaia2018}. However, the other five stars show significant differences in radial velocities estimated by us with that of \citet{gaia2018}, implying the possibility of the stars being part of a binary system with a now invisible companion.

Stellar atmospheric parameters, T$_{eff}$, log~$g$, micro-turbulent velocity ($\xi$), and [Fe/H] are derived using the method described in \citet{Goswami_et_al_1_2021} as briefly discussed below. For this analysis, we have chosen a set of clean and unblended Fe~I and Fe~II lines (Table~\ref{tab:Fe_linelist}) from each star. The excitation potentials of the lines range from 0.0 to 6.0 eV. The atomic line information is taken from the kurucz atomic line database\footnote{\url{https://lweb.cfa.harvard.edu/amp/ampdata/kurucz23/sekur.html}}. After strong filtration of the clean lines, we have considered (85, 12), (36, 3), (136, 17), (30, 5), (36, 4), (19, 3), and (20, 2) numbers of (Fe~I, Fe~II) lines in the spectra of BD+75~348, BD+09~3019, HD~238020, HE~0319--0215, HE~0507--1653, HE~0930--0018 and HE~1023--1504 respectively. For our analysis, we have used the code MOOG \citep{Sneden_1973_MOOG} in its updated 2013 version. MOOG assumes conditions of local thermodynamic equilibrium (LTE). We have selected the model atmospheres from the Kurucz grid of model atmospheres\footnote{\url{http://kurucz.harvard.edu/grids.html}} with no convective overshooting, and the solar abundances are taken from \citet{asplund2009}. For determining the effective temperatures and micro-turbulent velocities of the program stars, we have used the conventional methods of excitation potential balance and equivalent width balance, respectively. The surface gravities of the program stars are determined by using the method of ionization equilibrium balance, in which the abundance of Fe measured from Fe~I and Fe~II lines are equated to fix the value of log~$g$. The abundance of Fe derived from Fe~I and Fe~II lines gives the metallicities of the stars. The atmospheric parameters of the program stars, along with the available literature values, are presented in Table~\ref{tab:atm_paracomp}.

{\footnotesize
\begin{table}
\centering
\caption{\bf{Radial velocities of the program stars. }}
\label{tab:rad_vel} 
\begin{tabular}{lcc}
\hline     
Star Name       & V$_{r}$            &  V$_{r}$                  \\
                &(Km s$^{-1}$)       &(Km s$^{-1}$)              \\
                &(Our estimates)     & $^{\it{(a)}}$                 \\      
\hline
BD+75~348       &   57.70 $\pm$ 1.72 &   56.91 $\pm$ 2.21     \\
BD+09~3019      &  --9.50 $\pm$ 1.77 &    8.21 $\pm$ 0.65     \\
HD~238020       & --24.62 $\pm$ 3.07 & --16.04 $\pm$ 0.15     \\
HE~0319$-$0215  &--257.43 $\pm$ 0.61 &--232.33 $\pm$ 4.38     \\
HE~0507--1653   &  349.81 $\pm$ 0.72 &  353.53 $\pm$ 2.55     \\
HE~0930$-$0018  &   45.73 $\pm$ 0.68 &        -               \\
HE~1023--1504   &--226.15 $\pm$ 0.39 &        -               \\
\hline
\end{tabular}

References: $^{\it{(a)}}$ \citet{gaia2018}\\

\end{table}
}

{\footnotesize
\begin{table*}
\centering
\caption{\bf{Equivalent widths (in m\r{A}) of Fe lines used for deriving atmospheric parameters.}}
\label{tab:Fe_linelist} 
\scalebox{0.73}{
\begin{tabular}{ccccccccccc}
\hline
Wavelength&Element&E$_{low}$ &   log gf  &  BD+75~348 & BD+09~3019 & HD~238020  &HE~0319--0215&HE~0507--1653&HE~0930--0018&HE~1023--1504  \\
(\r{A})   &       & (eV)     &           &            &            &            &             &             &             &               \\
\hline 
4187.04   & Fe I  & 2.45     & --0.548   & -          & -          & -          & -          &124.8 (6.12)& -          & -               \\
4202.03   &       & 1.48     & --0.708   & -          & -          & -          & -          &179.4 (6.02)& -          & -               \\
4203.57   &       & 1.01     & --3.869   & -          & -          & 73.1 (6.72)& -          & -          & -          & -               \\
\hline
\end{tabular}}

The numbers in the parenthesis in columns 5 to 11 give the derived abundances from the respective line.\\
\textbf{Note:} This table is available in its entirety online only.
A portion is shown here for guidance regarding its form and content.\\

\end{table*}}

{\footnotesize
\begin{table*}
\centering
\caption{\bf{The derived atmospheric parameters of our program stars and literature values. }}
\label{tab:atm_paracomp} 
\begin{tabular}{lccccccccc}
\hline     
Star Name       & T$_{eff}$ &log g  & $\xi    $   & [Fe I/H]            &  [Fe II/H]        & [Fe/H]   & Ref \\
                &    (K)    & (cgs) & (km s$^{-1}$) &                     &                   &          &     \\
\hline
BD+75~348       & 4840      & 2.00  & 1.43          & $-$0.41$\pm$ 0.20   & $-$0.42$\pm$ 0.18 & $-$0.42  & 1   \\
                & 4700      & 1.80  & 2.00          & $-$0.86$\pm$ 0.26   & $-$0.88$\pm$ 0.02 & $-$0.87  & 9   \\
                & 4900      &  -    &  -            &  -                  & -                 & -        & 8   \\
                & 4760      &  -    &  -            &  -                  & -                 & -        & 10  \\
\hline
BD+09~3019      & 4220      & 2.10  & 2.37          & $-$0.55$\pm$ 0.18   & $-$0.54$\pm$ 0.11 & $-$0.55  & 1   \\
\hline
HD~238020       & 5150      & 2.10  & 1.36          & $-$0.67$\pm$ 0.15   & $-$0.68$\pm$ 0.16 & $-$0.68  & 1   \\
                & 5035      & -     & -             & -                   &  -                & -        & 8   \\
\hline
HE~0319$-$0215  & 4650      & 0.50  & 1.33          & $-$2.58$\pm 0.10$   & $-$2.56$\pm 0.05$ & $-$2.57  & 1   \\
                & 4416      & 0.64  &  -            &     -               &    -              & $-$2.42  & 2   \\
                &  -        &  -    &  -            &     -               &    -              & $-$2.30  & 3   \\
                & 4738      & 0.66  & 2.16          &     -               &    -              & $-$2.90  & 4   \\
\hline
HE~0507--1653   & 4970      & 2.20  & 1.48          & $-$1.46$\pm 0.11$   & $-$1.42$\pm 0.08$ & $-$1.44  & 1   \\
                & 5000      & 2.40  & 2.00          & $-$1.38$\pm 0.19$   & $-$1.39$\pm 0.17$ & $-$1.38  & 5   \\
                &  -        &  -    &  -            &     -               &    -              & $-$1.42  & 6   \\
                & 4880      & 1.50  &  -            &     -               &    -              & $-$1.81  & 2   \\
                & 4935      & 1.88  &  -            &     -               &    -              & $-$1.32  & 7   \\
                & 5035      & 2.39  & 1.53          &     -               &    -              & $-$1.35  & 4   \\
\hline
HE~0930$-$0018  & 4190      & 2.65  & 1.86          & $-$1.39$\pm 0.11$   & $-$1.39$\pm 0.09$ & $-$1.39  & 1   \\
\hline
HE~1023--1504   & 4440      & 0.50  & 1.67          & $-$1.66$\pm 0.14$   & $-$1.66$\pm 0.07$ & $-$1.66  & 1   \\
                & 4421      & 0.66  &  -            &     -               &    -              & $-$2.50  & 2   \\
\hline
\end{tabular}

References: 1. Our work, 2. \citet{kennedy2011}. 3. \citet{Hansen_2016_III}, 4. \citet{karinkuzhi_et_al_i_process_2021}, 5. \citet{aoki2007carbon}, 6. \citet{Schuler_et_al_2008}, 7. \citet{Yong_et_al_2013}, 8. \citet{McDonald2012}, 9. \citet{zacs_et_al_2000}, 10. \citet{bergeat2001effective} \\          
\end{table*}
}

\section{Results}
\label{sec:results}
\subsection{Abundance analysis}
\label{sec:abundance_analysis}
We have determined the elemental abundances of the program stars by using an updated version of the code MOOG that assumes LTE conditions. We have interpolated the model atmospheres to our derived stellar parameters from the Kurucz grid of model atmospheres, with no convecting overshooting, for the program stars. We have determined the abundances of light elements C, N, O, Na, Mg, Si, Ca, Sc, Ti, V, Cr, Mn, Co, Ni, Cu, Zn, and neutron-capture elements Sr, Y, Zr, Ba, La, Ce, Pr, Nd, Sm, Eu, Dy and Hf for the program stars, depending upon the availability of the useful atomic lines or molecular bands of the elements. Abundances of C, N \& O and the elements showing hyperfine splitting such as Sc, V, Mn, Ba, La, and Eu are determined using spectrum synthesis calculations. The abundances of the other elements are derived with the help of the equivalent width method by measuring the equivalent widths of clean and unblended spectral lines (Table~\ref{tab:Elem_linelist1}) of the elements using the SPLOT task in IRAF. The atomic line information is taken from the Kurucz database of atomic line list. We have presented the abundance results in Tables~\ref{tab:abundances1}~\&~\ref{tab:abundances2}.

{\footnotesize
\begin{table*}
\centering
\caption{\bf{Equivalent widths (in m\r{A}) of lines used for calculation of elemental abundances.}}
\label{tab:Elem_linelist1} 
\scalebox{0.73}{
\begin{tabular}{ccccccccccc}
\hline
Wavelength&Element&E$_{low}$ &   log gf  &  BD+75~348 & BD+09~3019 & HD~238020  &HE~0319--0215&HE~0507--1653&HE~0930--0018&HE~1023--1504  \\
(\r{A})   &       & (eV)     &           &            &            &            &             &             &             &               \\
\hline 
5682.633  & Na I  & 2.10     & --0.700  &115.8 (6.25)&172.5 (6.08)& 64.4 (5.62)& 22.4 (4.71)& 45.2 (5.19)&112.1 (5.39)& 51.5 (5.06)\\
5688.205  &       & 2.10     & --0.450  &140.3 (6.42)&170.6 (5.81)& 87.1 (5.77)& 25.3 (4.53)& 58.7 (5.17)&136.8 (5.43)& -          \\
6154.23   &       & 2.10     & --1.560  & 59.1 (6.14)& -          & 16.1 (5.50)& -          & -          & -          & -          \\
6160.75   &       & 2.10     & --1.260  & 88.1 (6.30)& -          & 30.1 (5.55)& -          & 19.7 (5.21)& -          & -          \\
\hline
\end{tabular}}

The numbers in the parenthesis in columns 5 to 11 give the derived abundances from the respective line.\\
\textbf{Note:} This table is available in its entirety online only.
A portion is shown here for guidance regarding its form and content.\\

\end{table*}}

{\footnotesize
\begin{table*}
\centering
\caption{\bf{Absolute abundances (log~$\epsilon$) derived from different lines using spectral synthesis calculations. Hyperfine splitting contributions are taken from the sources given in Column 10.}}
\label{tab:Elem_linelist2} 
\scalebox{0.75}{
\begin{tabular}{cccccccccc}
\hline
Wavelength& Element &  BD+75~348 & BD+09~3019 & HD~238020  & HE~0319--0215 & HE~0507--1653 & HE~0930--0018 & HE~1023--1504 & Reference \\
(\r{A})   &         &            &            &            &               &               &               &               &           \\
\hline 
4415.557  &  Sc II  &  -         &  -         & 2.65       & 0.55          &  -            &  -            &  -            &   1$^{a}$ \\
5031.021  &         &  -         &  -         & 2.50       &  -            &  -            &  -            &  -            &   1$^{a}$ \\
5526.790  &         & 2.85       &  -         & 2.47       &  -            &  -            &  -            &  -            &   1$^{a}$ \\
5641.001  &         & 2.72       & 3.15       & 2.30       &  -            & 1.75          &  -            & 1.55          &   1$^{a}$ \\
5657.896  &         & 2.62       & 3.15       & 2.60       &  -            & 1.71          & 1.65          &  -            &   1$^{a}$ \\
5667.149  &         & 2.65       &  -         & 2.30       &  -            &  -            & 1.50          &  -            &   1$^{a}$ \\
5727.652  &  V I    & 3.12       & 3.88       & 3.23       & $<$ 1.60      & 2.80          & 2.66          & 2.53          &   2$^{a}$ \\
4754.042  &  Mn I   & 4.55       & 3.55       & 4.62       & 2.15          & 3.75          &  -            & 3.28          &   3$^{a}$ \\
4762.367  &         & 4.60       &  -         & 4.56       &  -            &  -            & 3.15          &  -            &   3$^{a}$ \\
4823.524  &         & 4.60       &  -         & 4.62       &  -            &  -            &  -            &  -            &   3$^{a}$ \\
5853.668  &  Ba II  & 3.47       &  -         &  -         & 1.37          & 2.70          & 1.55          & 2.60          &   4       \\
6141.713  &         & 3.29       &  -         & 1.75       & 1.60          & 2.90          & 2.20          &  -            &   4       \\
4921.776  &  La II  & 2.60       & 2.80       & 0.90       & 0.25          & 1.60          & 0.80          & 1.30          &   5       \\
6437.640  &  Eu II  &  -         &  -         &  -         &  -            & 0.10          &  -            &  -            &   6$^{a}$ \\
6645.064  &         & 0.80       & 1.02       &  -         & --1.20        & 0.12          &  -            & --0.25        &   7       \\
\hline
\end{tabular}}

{\textit{a}} - {\textit{linemake}}\\
{\bf{References:}} 1. \citet{Lawler_et_al_2019}, 2. \citet{Lawler_et_al_2014}, 3. \citet{Den_et_al_2011}, 4. \citet{McWilliam_1998}, 5. \citet{Jonsell_et_al_2006}, 6. \citet{Lawler_et_al_2001}, 7. \citet{Worley_et_al_2013}.\\

\end{table*}}

\subsubsection{C, N, O}
\begin{figure*}
     \begin{center}
\centering
        \subfigure[]{%
            \label{fig:C2_first}
            \includegraphics[height=8.0cm,width=8.5cm]{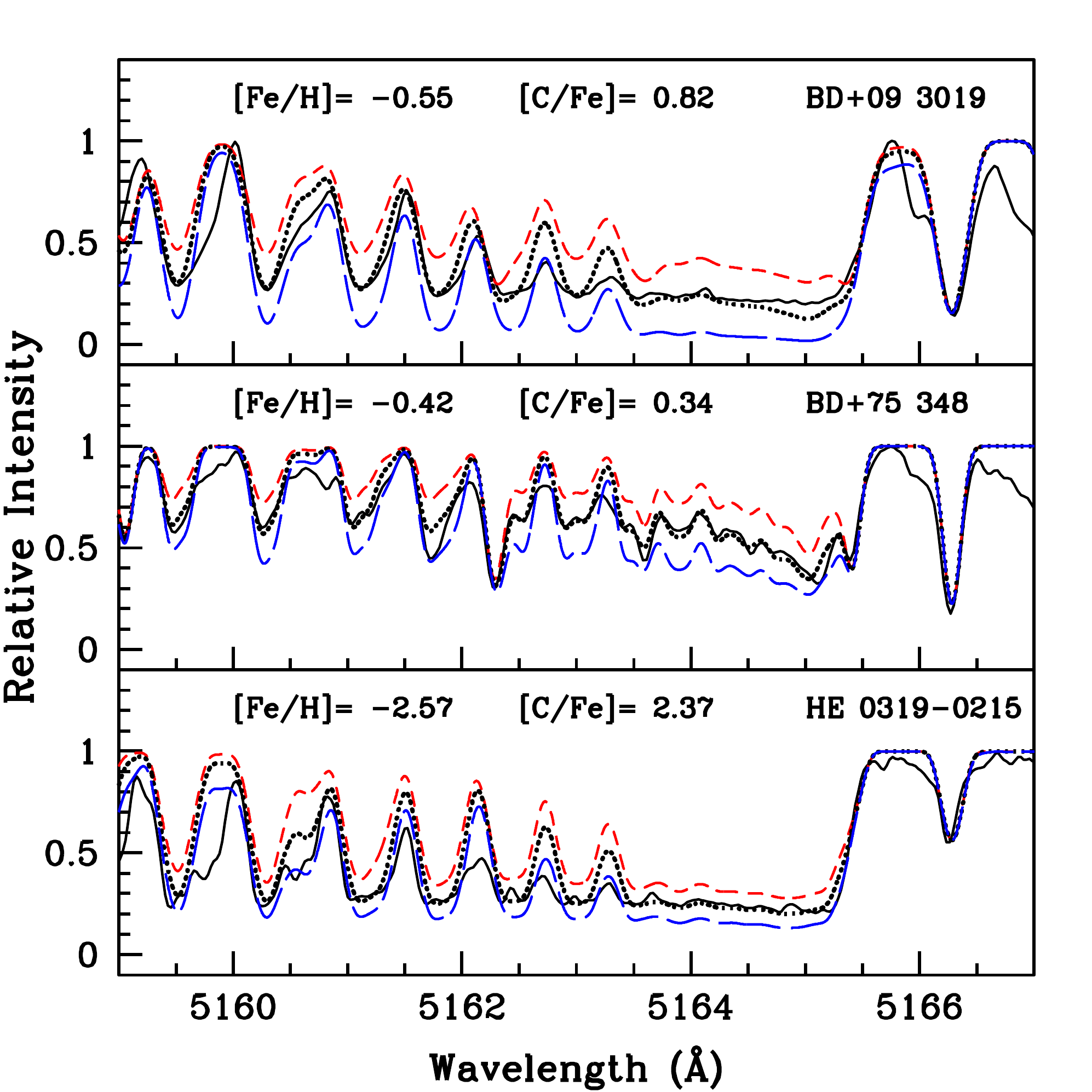}
        }%
        \subfigure[]{%
            \label{fig:C2_second}
            \includegraphics[height=8.0cm,width=8.5cm]{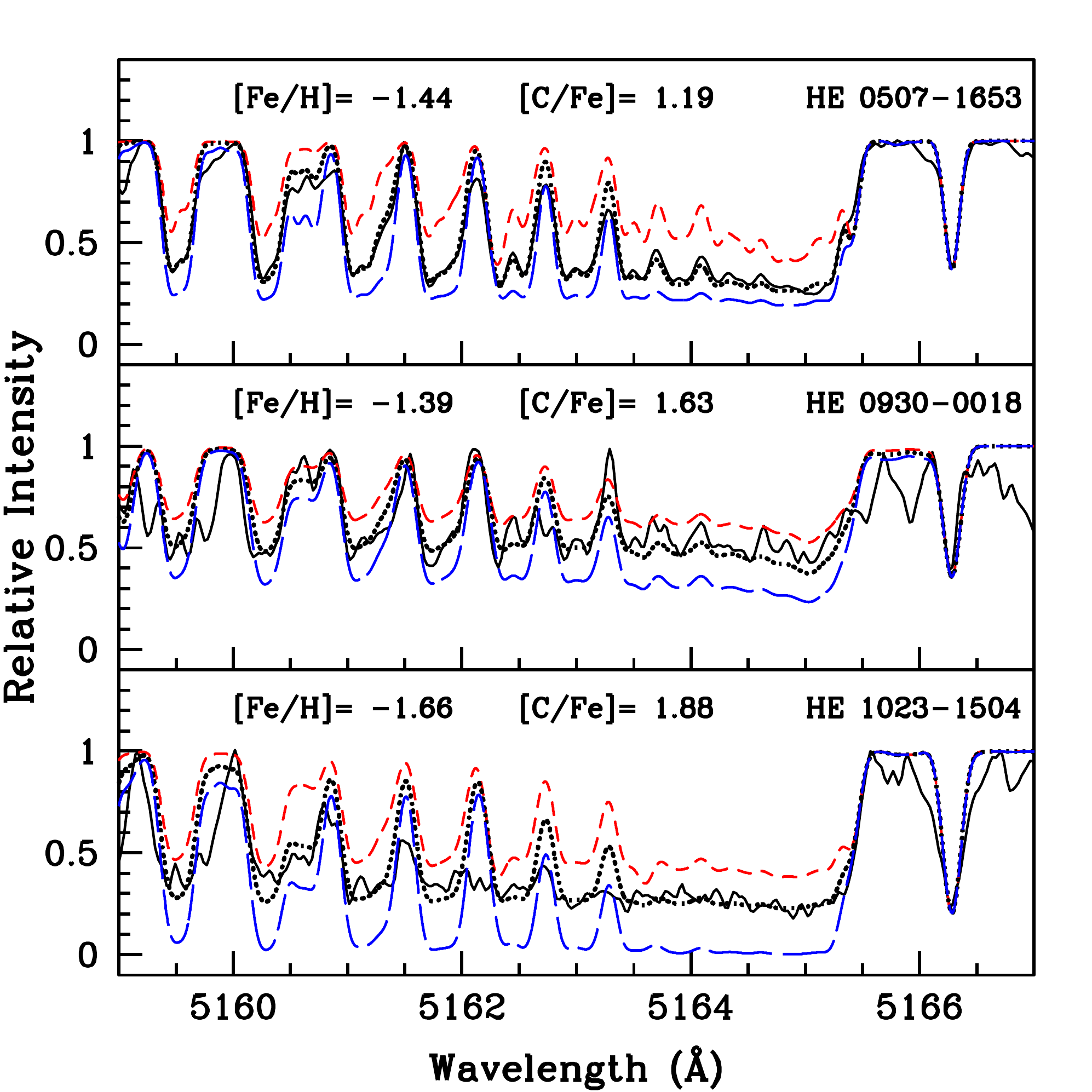}

        }\\ 

    \caption{ The spectral synthesis plot of C$_{2}$ band around 5165 {\rm \AA}. The dotted lines indicate the synthesized spectra and the solid lines indicate the observed spectra. Two alternative synthetic spectra are shown corresponding to $\Delta$[C/Fe] = +0.3 (long-dashed line) and $\Delta$[C/Fe] = $-$0.3 (short-dashed line) }%
   \label{fig:C2}
       \end{center}

\end{figure*}

The abundance of carbon is derived using three molecular bands of carbon, namely CH band around 4310 {\rm \AA} and C$_{2}$ bands near 5165 and 5635 {\rm \AA}. Figure~\ref{fig:C2} shows the spectral synthesis of C$_{2}$ molecular band around 5165 {\rm \AA}. Abundance of N is estimated from the CN molecular band around 4215 {\rm \AA}, adopting the carbon abundance derived using the C$_{2}$ and CH molecular bands. Abundance of oxygen could be determined only for HD~238020. We have used the [O~I] 6300 {\rm \AA} forbidden line for estimating the abundance of oxygen.

\begin{figure}
        \centering
        \includegraphics[height=8cm,width=9.2cm]{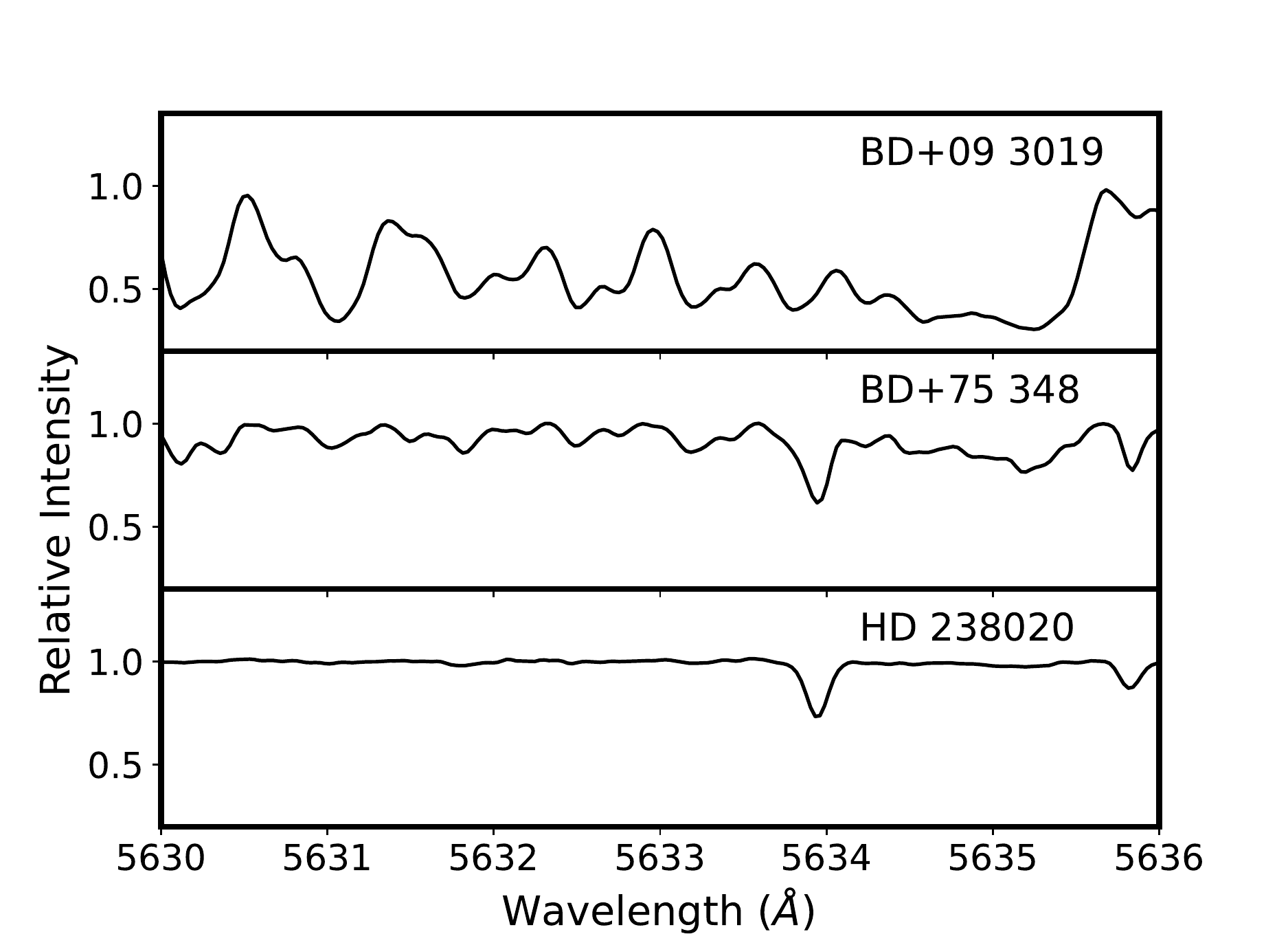}
        \caption{ Sample spectra of the C$_{2}$ band at 5635 {\rm \AA} of the three Ba stars. This band is not present in the star HD~238020.}
\label{fig:C2_comp}
\end{figure}

\begin{figure*}
     \begin{center}
\centering
        \subfigure[]{%
            \label{fig:CN_app}
            \includegraphics[height=7.5cm,width=9.2cm]{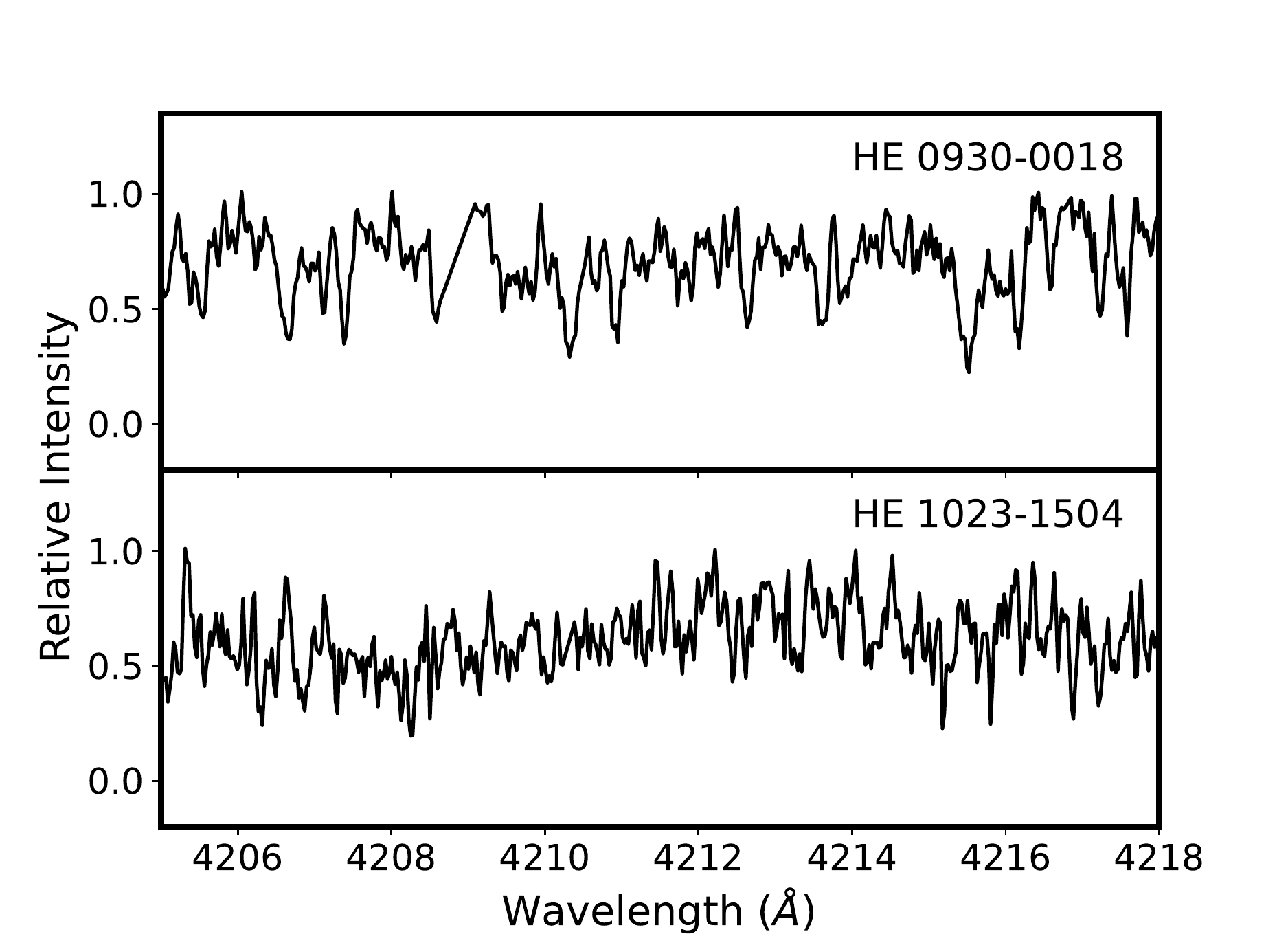}
        }%
        \subfigure[]{%
            \label{fig:CH_app}
            \includegraphics[height=7.5cm,width=9.2cm]{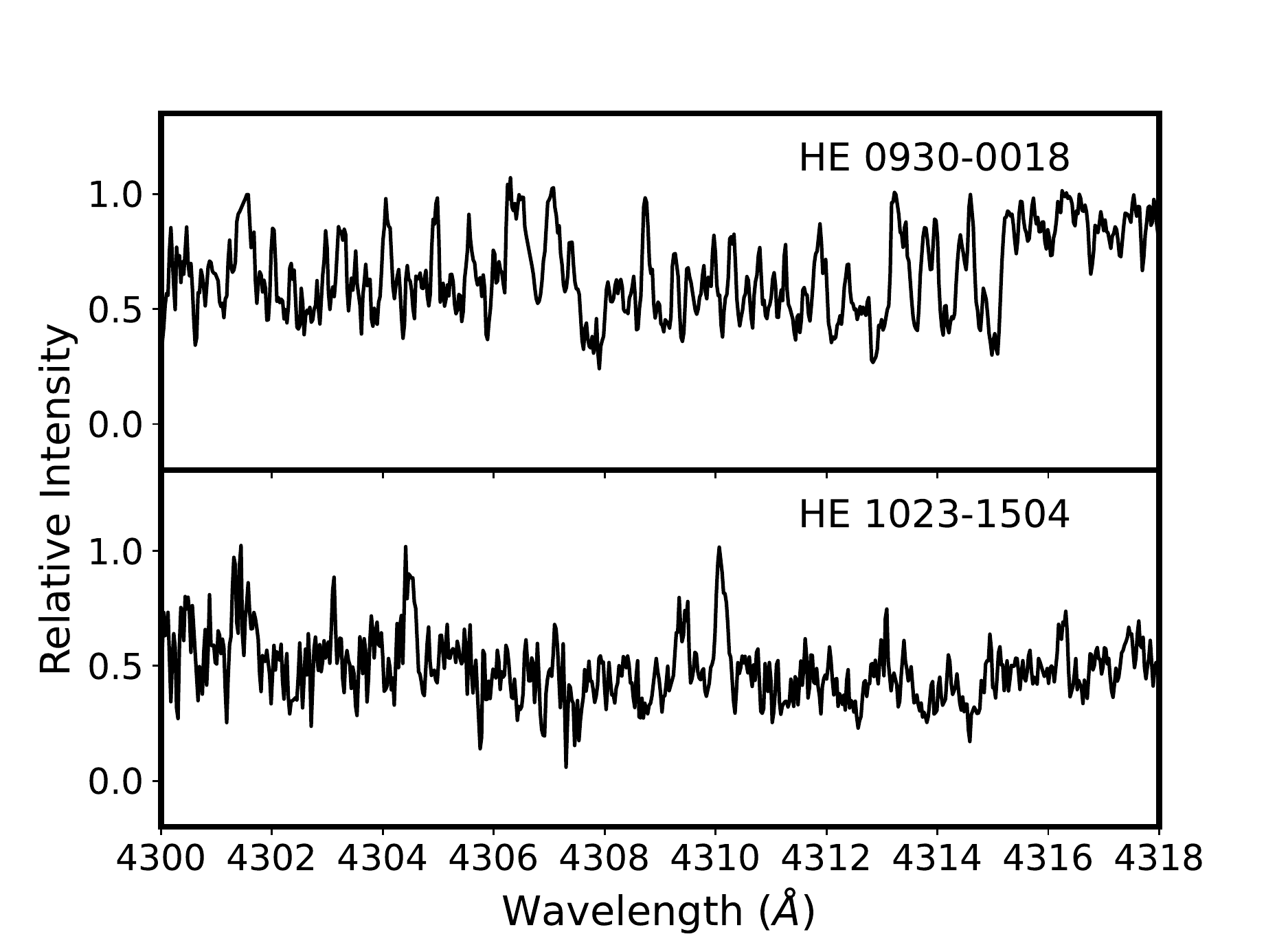}

        }\\ 

    \caption{ Sample spectra of HE~0930--0018 and HE~1023--1504 in the wavelength regions 4205 -- 4218 {\rm \AA} and 4300 -- 4318 {\rm \AA}. }%
   \label{fig:sample_spectrum_CH_CN}
       \end{center}

\end{figure*}

For the star HD~238020, the C$_{2}$ molecular band at 5635 {\rm \AA} and the CN molecular band are found to be too weak to be used for abundance estimation. In  Figure~\ref{fig:C2_comp} we have shown a comparison of the C$_{2}$ band at 5635 {\rm \AA} for the stars BD+75~348, BD+09~3019 and HD~238020 to demonstrate the absence of the band in HD~238020.  We could not use the CH molecular band in HE~0507--1653 and CN molecular band in BD+09~3019 for abundance estimation as the bands are found to be saturated. For HE~0930--0018 and HE~1023--1504, the CH and the CN molecular bands could not be used for abundance determination due to the low S/N (signal to noise ratio) in the particular region of the spectra (Figure~\ref{fig:sample_spectrum_CH_CN}). BD+09~3019, HE~0319--0215, HE~0507--1653, HE~0930--0018, and HE~1023--1504 are found to be carbon enhanced with [C/Fe] $>$  0.70. While BD+75~348 is mildly enhanced in carbon, abundance of carbon in HD~238020 is found to be solar. Nitrogen is enhanced with [N/Fe] $>$ 0.70 in all the three stars (BD+75~348, HE~0319--0215, HE~0507--1653), for which abundance of nitrogen could be estimated. The wavelengths, lower excitation potentials, and log~$gf$ values of different molecular transitions for the C$_{2}$ bands at 5165 {\rm \AA} \& 5635 {\rm \AA} and CN bands are adopted from \citet{Brooke_et_al_2013}, \citet{Ram_et_al_2014}, and \citet{Sneden_et_al_2014}. The molecular linelist for CH band at 4310 {\rm \AA}  used for our calculation using {\it{linemake}}\footnote{linemake contains laboratory atomic data (transition probabilities, hyperfine and isotopic substructures) published by the Wisconsin Atomic Physics and the Old Dominion Molecular Physics groups. These lists and accompanying line list assembly software have been developed by C. Sneden and are curated by V. Placco at \url{https://github.com/vmplacco/linemake}.} \citep{Linemake_2021} are from \citet{Masseron_et_al_2014}. The abundances of carbon derived from the different bands of C$_{2}$ and CH are found to be quite similar with the maximum difference of 0.16 dex for the star BD+09~3019. As shown in \citet{Goswami_et_al_1_2021}, the change in abundance of C derived from C$_{2}$ and CH bands varies in the $\pm$0.04 dex range with the change in temperature in the range $\pm$100~K. We observed negligible change in the abundance of C by varying log~$g$ and $\xi$ by $\pm$0.2 (cgs) and $\pm$0.2 kms$^{-1}$ respectively. While CH band shows a variation in the abundance of C by $\pm$0.04 dex with the metallicity variation in the range $\pm$0.2 dex, the C$_{2}$ bands show no variation.

\subsubsection{Na, Mg, Si, Ca, Sc, Ti, V}
Abundance of Na in the program stars ranges from [Na/Fe] = 0.05 to 0.95 dex, with the minimum [Na/Fe] for HD~238020 and the maximum for HE~0319--0215. Mg is moderately enhanced in all the program stars except for BD+09~3019, for which [Mg/Fe] = 1.26. We could estimate the abundance of Si only for HD~238020 using a single line Si~I 6237.32 {\rm \AA}, and found to be sub-solar with [Si/Fe] = --0.52.  This is somewhat surprising,  as this object is classified as a  mild Ba star with a probability of being a thin disk object (see section~\ref{sec:kinematics} on  Kinematic analysis), and, in the case of thin/thick disk stars $\alpha$-elements ([$\alpha$/Fe]), are known to lie between 0.2 dex to 0.3 dex for stars with --1.0~$<$~[Fe/H]~$<$~0 \citep{Edvardsson_et_al_1993}. Within the metallicity range  --1.0~$<$~[Fe/H]~$<$~0, in barium stars [Si/Fe] is found to be in the range 0.0 dex to 0.58 dex \citep{de_Castro_et_al_2016}.  Massive stars $>$ 10 M$_{\odot}$ produce $\alpha$-elements during hydrostatic oxygen burning and also during type~II supernovae explosions \citep{Woosley_&_Weaver_1986}; hence Si is also expected to show similar trend as seen in other $\alpha$-elements. Si is also known to show high scatter at all metallicities --4~$<$~[Fe/H]~$<$~0.  However, in metal-poor  Red Horizontal Branch stars (i.e., BPS~CS~22186--0005 ), [Si/Fe] is found to be as low as $-$0.25 dex \citep{Preston_et_al_2006}, 0.27 dex higher than that observed in HD~238020.  It will be worthwhile to investigate in future the anomalous low Si abundance observed in HD~238020 that now  remains unexplained. Ca is found to range from sub-solar to super-solar abundances with --0.14~$<$~[Ca/Fe]~$<$~0.83. Abundance of Sc is estimated using spectrum synthesis calculations of several Sc~II lines (Table~\ref{tab:Elem_linelist2}) whenever available. For five out of the seven objects, we have derived the abundance of Ti from the spectral lines of both the neutral (Ti~I) and ionized (Ti~II) species of Ti. The abundances of Ti derived from both the species are found to be very close for all the stars except for HE~0319--0215, with a difference of 0.14 dex. For BD+09~3019 \& HE~1023--1504 abundance of Ti could be derived only from Ti~I lines as Ti~II lines are found to be severely blended.  Abundance of V is estimated from spectrum synthesis calculation of V~I line at 5727.652 {\rm \AA} (Table~\ref{tab:Elem_linelist2}). We could only estimate an upper limit of V for the object HE~0319--0215. Abundance of V ranges from sub-solar to super-solar with --0.39~$<$~[V/Fe]~$<$~0.50.

\subsubsection{Cr, Mn, Co, Ni, Cu, Zn}
For HD~238020, we could estimate the abundance of Cr from both neutral (Cr~I) and ionized species (Cr~II) of Cr. In other stars, we could not detect the lines due to Cr~II, and abundance could be derived only from Cr~I lines. In the program stars, abundance of Cr ranges from --0.51 dex to 0.11 dex. We have derived the abundance of Mn using spectrum synthesis calculation of several lines due to Mn~I (Table~\ref{tab:Elem_linelist2}). Mn is found to be sub-solar in the program stars with --1.33~$<$~[Mn/Fe]~$<$~--0.15. For barium stars in the metallicity range --1~$<$~[Fe/H]~$<$~0.0 , [Mn/Fe] is found to lie in the range --0.55 dex to +0.37 dex \citep{Yang_et_al_2016, shejeela_agb_2020}. In our sample, the spectrum synthesis of  Mn~I 4754.042 {\rm \AA} line in BD+09~3019, a strong barium star returned a value [Mn/Fe] = --1.33, making it the most Mn deficient barium star. A few extremely metal-poor stars are however known to exhibit such  low Mn abundance (i.e,  HE 1012--1540 (--1.00 dex),  HE~1356--0622 (--0.98 dex)  \citep{Cohen_et_al_2008}).  Lines due to Co, Cu \& Zn are not usable for abundance determination in HE~0319--0215. A few samples of these lines of this star are shown in Figure~\ref{fig:CoCuZn}. For HE~1023--1504, we could not estimate the abundance of Cu due to the unavailability of good lines. In the other program stars Co ranges from sub-solar to solar abundances with --0.21~$<$~[Co/Fe]~$<$~0.14, Ni ranges from sub-solar to super-solar abundances with --0.30~$<$~[Ni/Fe]~$<$~0.70, [Cu/Fe] ranges from --0.20 dex to 0.38 dex and [Zn/Fe] ranges from --0.16~$<$~[Zn/Fe]~$<$~0.13.

\subsubsection{Sr, Y, Zr}
We could derive the abundance of Sr for all the stars except HE~1023--1504. Sr II 4215.52 {\rm \AA} line appears in the region of the CN molecular band around 4215 {\rm \AA}. That is why, it is not always possible to use this particular line for the objects with enhanced C and N abundances. We have used this line to estimate the abundance of Sr for HD~238020. For the stars BD+75~348, HE~0319--0215, HE~0507--1653 and HE~0930--0018 Sr~I 4607.33 {\rm \AA} line (Table~\ref{tab:Elem_linelist2}) is used for spectrum synthesis calculations. The abundance of Sr ([Sr/Fe]) in the program stars ranges from 0.08 dex to 1.67 dex. Y is found to be enhanced ([Y/Fe] $>$ 0.70) in all the program stars except HD~238020 (with [Y/Fe] = --0.26). For BD+75~348, HD~238020 and HE~0507--1653 we could estimate the abundance of Zr from the lines due to both neutral (Zr~I) and ionized (Zr~II) species. The abundance of Zr derived from the lines of both the species is closely similar, with the largest difference of 0.25 dex for BD+75~348. Abundance of Zr ranges from 0.05 dex to 1.63 dex in the program stars.

\subsubsection{Ba, La, Ce, Pr, Nd, Sm, Eu, Dy, Hf}
Abundances of the neutron-capture elements are enhanced in all the program stars except HD~238020. Abundances of Eu and Hf could not be estimated for HD~238020 as no lines could be detected. Abundances of Ba through Dy in HD~238020 are found to be moderately enhanced with 0.11~$<$~[X/Fe]~$<$~0.48.  

Abundance of Ba is derived using spectral synthesis calculations of Ba~II 5853.668 {\rm \AA} amd 6141.713 {\rm \AA} lines (Table~\ref{tab:Elem_linelist2}) whenever available. Abundance of Ba could not be derived for BD+09~3019 as the Ba lines are found to be too strong and saturated. In the program stars, [Ba/Fe] ranges from 1.09 dex to 2.08 dex. Abundance of La is derived using spectrum synthesis calculation of La~II 4921.776 {\rm \AA} line (Table~\ref{tab:Elem_linelist2}). La is found to be enhanced in the program stars with 1.09~$<$~[La/Fe]~$<$~2.25. [Ce/Fe] ranges from 1.34 dex to 2.50 dex in the program stars. Abundances of Pr, Nd and Sm are enhanced ([X/Fe] $>$ 0.90) in the program stars. Abundance of Eu is derived using spectrum synthesis calculations of Eu~II 6437.640 {\rm \AA} \& 6645.064 {\rm \AA} (Table~\ref{tab:Elem_linelist2}). Abundance of Eu could not be derived for HE~0930--0018 as no Eu lines could be found in the spectrum (Figure~\ref{fig:Eu_comp}). Eu is found to be enhanced ([Eu/Fe] $>$ 0.70) in the rest of the program stars with 0.70~$<$~[Eu/Fe]~$<$~1.05. Abundance of Dy could be derived only for BD+75~348, BD+09~3019 and HD~238020 and abundance of Hf could only be derived for BD+75~348.

{\footnotesize
\begin{table*}
\centering
\caption{\bf{Elemental abundances in BD+75~348, BD+09~3019 and HD~238020}}
\label{tab:abundances1}
\scalebox{0.80}{
\begin{tabular}{l r c|l r r|l r r|l r r}
\hline
& \multicolumn{6}{c}{BD+75~348} & \multicolumn{2}{c}{BD+09~3019} & \multicolumn{3}{c}{HD~238020}\\
\hline
Element & Z  & solar $log{\epsilon}^a$ & $log{\epsilon}$ &[X/H]& [X/Fe] & $log{\epsilon}$ & [X/H] & [X/Fe] & $log{\epsilon}$ & [X/H] & [X/Fe]\\
	    &    &                         &    (dex)       &       &        &   (dex)         &       &        &   (dex)         &       &\\
\hline
C (C$_{2}$, 5165 {\rm \AA}) & 6  & 8.43 & 8.35 (syn)    &$-$0.08& 0.34   & 8.70 (syn)      & 0.27  & 0.82   & 7.75 (syn)      &$-$0.68& 0.00   \\
C (C$_{2}$, 5635 {\rm \AA}) & 6  & 8.43 & 8.30 (syn)    &$-$0.13& 0.29   & 8.71 (syn)      & 0.28  & 0.83   & -               & -     & -      \\ 
C (CH, 4310 {\rm \AA})      & 6  & 8.43 & 8.30 (syn)    &$-$0.13& 0.29   & 8.55 (syn)      & 0.12  & 0.67   & 7.68 (syn)      &$-$0.75&$-$0.07 \\
N (CN, 4215 {\rm \AA})      & 7  & 7.83 & 8.20 (syn)    & 0.37  & 0.79   & (saturated)     & -     & -      & -               & -     & -      \\
O  {\sc i}  &  8 & 8.69 & -                  & -        & -     & -                  & -       & -    & 8.09 (syn)        &$-$0.60 & 0.08       \\ 
Na {\sc i}  & 11 & 6.24 & 6.28$\pm$0.12 (4)  & 0.04     & 0.46  & 5.95$\pm$0.20 (2)  & $-$0.29 & 0.26 & 5.61$\pm$0.11 (4) &$-$0.63 & 0.05   \\
Mg {\sc i}  & 12 & 7.60 & 7.38$\pm$0.17 (4)  & $-$0.22  & 0.20  & 8.31 (1)           & 0.71    & 1.26 & 7.29$\pm$0.08 (4) &$-$0.31 & 0.37   \\
Si {\sc i}  & 14 & 7.51 & -                  & -        & -     & -                  & -       &  -   & 6.31 (1)          &$-$1.20 &$-$0.52 \\
Ca {\sc i}  & 20 & 6.34 & 5.97$\pm$0.21 (16) & $-$0.37  & 0.05  & 5.65$\pm$0.00 (2)  & $-$0.69 &$-$0.14&5.83$\pm$0.19 (21)&$-$0.51 & 0.17   \\
Sc {\sc ii}$^{*}$ & 21 & 3.15 & 2.71$\pm$0.09 (4)  & $-$0.44  &$-$0.02& 3.15$\pm$0.00 (2)  & 0.00    & 0.55 & 2.47$\pm$0.13 (6) &$-$0.68 & 0.00 \\
Ti {\sc i}  & 22 & 4.95 & 4.65$\pm$0.20 (13) & $-$0.30  & 0.12  & 4.78$\pm$0.19 (2)  & $-$0.17 & 0.38 & 4.45$\pm$0.11 (14)&$-$0.50 & 0.18   \\
Ti {\sc ii} & 22 & 4.95 & 4.65$\pm$0.19 (10) & $-$0.30  & 0.12  & -                  &  -      &  -   & 4.40$\pm$0.20 (22)&$-$0.55 & 0.13   \\
V {\sc i}$^{*}$ & 23 & 3.93 & 3.12 (1)           & $-$0.81  &$-$0.39& 3.88 (1)           & $-$0.05 & 0.50 & 3.23 (1)          &$-$0.70 &$-$0.02 \\
Cr {\sc i}  & 24 & 5.64 & 5.23$\pm$0.15 (7)  & $-$0.41  & 0.01  & 4.85$\pm$0.18 (3)  & --0.79  &  --0.24 & 4.98$\pm$0.12 (13)&$-$0.66 & 0.02   \\
Cr {\sc ii} & 24 & 5.64 &  -                 & -        &  -    & -                  &  -      &  -   & 5.01 (1)          &$-$0.63 & 0.05   \\
Mn {\sc i}$^{*}$& 25 &5.43 & 4.58$\pm$0.02 (3)  & $-$0.85  &$-$0.43& 3.55 (1)        &--1.88   &--1.33 &4.60$\pm$0.03 (3) &$-$0.83 &$-$0.15   \\
Fe {\sc i}  & 26 & 7.50 & 7.09$\pm$0.20 (85) & $-$0.41  & -     & 6.95$\pm$0.18 (36) & $-$0.55 & -    & 6.83$\pm$0.15 (136)&$-$0.67 & -      \\
Fe {\sc ii} & 26 & 7.50 & 7.08$\pm$0.18 (12) & $-$0.42  & -     & 6.96$\pm$0.11 (3)  & $-$0.54 & -    & 6.82$\pm$0.16 (17)&$-$0.68 & -      \\
Co {\sc i}  & 27 & 4.99 & 4.49$\pm$0.10 (2)  & $-$0.50  &$-$0.08&  4.23 (1)          & --0.76  & --0.21& 4.30$\pm$0.20 (4) &$-$0.69 &$-$0.01 \\
Ni {\sc i}  & 28 & 6.22 & 5.77$\pm$0.24 (12) & $-$0.45  &$-$0.03&  5.65 (1)          & --0.57  & --0.02  & 5.60$\pm$0.21 (15)&$-$0.62 & 0.06   \\
Cu {\sc i}  & 29 & 4.19 & 3.66 (1)           & $-$0.53  &$-$0.11&  3.62 (1)          & --0.57  & --0.02  & 3.89 (1)          &$-$0.30 & 0.38   \\
Zn {\sc i}  & 30 & 4.56 & 4.12 (1)           & $-$0.44  &$-$0.02&  3.87 (1)          & --0.69  & --0.14  & 3.75$\pm$0.04 (2) &$-$0.81 &$-$0.13 \\
Sr {\sc i}  & 38 & 2.87 & 3.40 (syn)         & 0.53     & 0.95  & 3.99 (1)           & 1.12    & 1.67 & -                 & -      & -      \\
Sr {\sc ii} & 38 & 2.87 & -                  & -        & -     & -                  &  -      & -    & 2.27 (1)          &$-$0.60 & 0.08   \\
Y {\sc ii}  & 39 & 2.21 & 3.06$\pm$0.16 (10) & 0.85     & 1.27  & 3.52$\pm$0.20 (2)  & 1.31    & 1.86 & 1.27$\pm$0.12 (8) &$-$0.94 &$-$0.26 \\
Zr {\sc i}  & 40 & 2.58 & 3.16$\pm$0.08 (3)  & 0.58     & 1.00  & 2.98$\pm$0.14 (3)  & 0.40    & 0.95 & 2.01$\pm$0.06 (2) &$-$0.57 & 0.11   \\
Zr {\sc ii} & 40 & 2.58 & 3.41$\pm$0.09 (3)  & 0.83     & 1.25  & -                  &  -      & -    & 1.95$\pm$0.04 (2) &$-$0.63 & 0.05   \\
Ba {\sc ii}$^{*}$&56 & 2.18 & 3.38$\pm$0.13 (2)  & 1.20 & 1.62  & -                  &  -      & -    & 1.75 (1)          &$-$0.43 & 0.25   \\
La {\sc ii}$^{*}$& 57 & 1.10 & 2.60 (1)      & 1.50     & 1.92  & 2.80 (1)         & 1.70    & 2.25 & 0.90 (1)          &$-$0.20 & 0.48  \\
Ce {\sc ii} & 58 & 1.58 & 2.97$\pm$0.20 (13) & 1.39     & 1.81  & 3.07$\pm$0.16 (3)  & 1.49    & 2.04 & 1.09$\pm$0.11 (7) &$-$0.49 & 0.19   \\
Pr {\sc ii} & 59 & 0.72 & 2.17$\pm$0.22 (5)  & 1.45     & 1.87  & 2.65$\pm$0.15 (3)  & 1.93    & 2.48 & 0.50$\pm$0.20 (2) &$-$0.22 & 0.46   \\
Nd {\sc ii} & 60 & 1.42 & 2.63$\pm$0.18 (11) & 1.21     & 1.63  & 2.81$\pm$0.06 (2)  & 1.39    & 1.94 & 1.00$\pm$0.25 (12)&$-$0.42 & 0.26   \\
Sm {\sc ii} & 62 & 0.96 & 2.08$\pm$0.07 (3)  & 1.12     & 1.54  & 2.78$\pm$0.16 (3)  & 1.82    & 2.37 & 0.70$\pm$0.07 (4) &$-$0.26 & 0.42   \\
Eu {\sc ii}$^{*}$&63 & 0.52 & 0.80 (1)       & 0.28     & 0.70  & 1.02 (1)           & 0.50    & 1.05 & -                 &   -    & -      \\
Dy {\sc ii} & 66 & 1.10 & 1.49 (1)           & 0.39     & 0.81  & 2.43 (1)           & 1.33    & 1.88 & 0.53 (1)          &$-$0.57 & 0.11   \\
Hf {\sc ii} & 72 & 0.85 & 2.33 (1)           & 1.48     & 1.90  & -                  &  -      & -    & -                 &   -    & -      \\

\hline 
\end{tabular}}
 $^{*}$ abundance is derived using spectrum synthesis calculations. 
 $^{a}$ \citet{asplund2009}, Columns 4, 7 \& 10 present the abundances ($log{\epsilon}$) of the different elements along with the standard deviations (when more than one line is used to derive the abundance). The number inside the parenthesis in columns 4, 7 \& 10 shows the number of lines used for the abundance determination.\\
\end{table*}
}

{\footnotesize
\begin{table*}
\centering
\caption{\bf{Elemental abundances in HE~0319--0215, HE~0507--1653, HE~0930--0018 and HE~1023--1504}}
\label{tab:abundances2}
\scalebox{0.60}{
\begin{tabular}{l r c|l r r|l r r|l r r|l r r}
\hline
& \multicolumn{6}{c}{HE~0319--0215} & \multicolumn{2}{c}{HE~0507--1653} & \multicolumn{3}{c}{HE~0930--0018} & \multicolumn{3}{c}{HE~1023--1504}\\
\hline
Element & Z  & solar $log{\epsilon}^a$ & $log{\epsilon}$ &[X/H]& [X/Fe] & $log{\epsilon}$ & [X/H] & [X/Fe] & $log{\epsilon}$ & [X/H] & [X/Fe] & $log{\epsilon}$ & [X/H] & [X/Fe]\\
	    &    &                         &    (dex)       &       &        &   (dex)         &       &        &   (dex)         &       & &   (dex)         &       &\\
\hline
C (C$_{2}$, 5165 {\rm \AA}) & 6  & 8.43 & 8.23 (syn)    &$-$0.20& 2.37   & 8.18 (syn)      & $-$0.25 & 1.19   & 8.67 (syn)      & 0.24  & 1.63   
& 8.65 (syn)         & 0.22    &  1.88\\
C (C$_{2}$, 5635 {\rm \AA}) & 6  & 8.43 & 8.29 (syn)    &$-$0.14& 2.43   & 8.12 (syn)      & $-$0.31 & 1.13   & 8.70 (syn)      & 0.27  & 1.66   
& 8.72 (syn)         & 0.29    &  1.95\\ 
C (CH, 4310 {\rm \AA})      & 6  & 8.43 & 8.30 (syn)    &$-$0.13& 2.44   & (saturated)     & -       & -      & Low S/N         & -     & -      
& Low S/N            & -       & -    \\
N (CN, 4215 {\rm \AA})      & 7  & 7.83 & 7.33 (syn)    &$-$0.50& 2.07   & 7.80            & $-$0.03 & 1.41   & Low S/N         & -     & -      
& Low S/N            & -       & -    \\
Na {\sc i}  & 11 & 6.24 & 4.62$\pm$0.12 (2)  & $-$1.62  & 0.95  & 5.19$\pm$0.02 (3)  & $-$1.05 & 0.39 & 5.41$\pm$0.03 (2) &$-$0.83 & 0.56         & 5.06 (1)           & $-$1.18 & 0.48 \\
Mg {\sc i}  & 12 & 7.60 & 5.51 (1)           & $-$2.09  & 0.48  & 6.68$\pm$0.11 (3)  & $-$0.92 & 0.52 & 6.69 (1)          &$-$0.91 & 0.48         & 6.30$\pm$0.05 (syn, 2) & --1.30    &  0.36 \\
Ca {\sc i}  & 20 & 6.34 & 4.49$\pm$0.21 (4)  & $-$1.85  & 0.72  & 5.30$\pm$0.19 (8)  & $-$1.04 & 0.40 &5.34$\pm$0.11 (6) &$-$1.00 & 0.39          & 5.51$\pm$0.14 (3)  & $-$0.83 & 0.83 \\
Sc {\sc ii}$^{*}$& 21 & 3.15 &  0.55 (1)      &   $-$2.60  &  $-$0.03  & 1.73$\pm$0.03 (2)  & $-$1.42 & 0.02 &  1.58$\pm$0.08 (2)&  $-$1.57  &  $-$0.18         & 1.55 (1) &   $-$1.60  & 0.06    \\
Ti {\sc i}  & 22 & 4.95 & 3.16 (1)           & $-$1.79  & 0.78  & 3.88$\pm$0.11 (3)  & $-$1.07 & 0.37 & 3.46$\pm$0.21 (7) &$-$1.49 &$-$0.10       & 3.95$\pm$0.15 (3)  &$-$1.00  & 0.66 \\
Ti {\sc ii} & 22 & 4.95 & 3.30$\pm$0.12 (3)  & $-$1.65  & 0.92  & 3.85$\pm$0.23 (8)  & $-$1.10 & 0.34 & 3.45$\pm$0.20 (4) &$-$1.50 &$-$0.11       & -                  & -       & -    \\
V {\sc i}$^{*}$& 23 & 3.93 &  $<$ 1.60 (1) &  $<$ $-$2.33 &   $<$ 0.24 &  2.80 (1) &   $-$1.13 &   0.31  & 2.66 (1)          &$-$1.27 & 0.12         &    2.53 (1) &  $-$1.40  &    0.26   \\
Cr {\sc i}  & 24 & 5.64 &   3.18$\pm$0.19 (3)  &  $-$2.46 &    0.11  & 4.21$\pm$0.15 (6)  & $-$1.43 & 0.01 & 3.74$\pm$0.16 (3) &$-$1.90 &$-$0.51       & 3.98$\pm$0.11 (2)  &$-$1.66  & 0.00 \\
Mn {\sc i}$^{*}$& 25 & 5.43 &  2.15 (1)       &   $-$3.28 &   $-$0.71 & 3.75 (1)           & $-$1.68 &$-$0.24& 3.15 (1)         &$-$2.28 &$-$0.89       & 3.28 (1) & $-$2.15 & $-$0.49 \\
Fe {\sc i}  & 26 & 7.50 & 4.92$\pm$0.10 (30) & $-$2.58  & -     & 6.04$\pm$0.11 (36) & $-$1.46 & -    & 6.11$\pm$0.11 (19)&$-$1.39 & -            & 5.84$\pm$0.14 (20) &$-$1.66  & -    \\
Fe {\sc ii} & 26 & 7.50 & 4.94$\pm$0.05 (5)  & $-$2.56  & -     & 6.08$\pm$0.08 (4)  & $-$1.42 & -    & 6.11$\pm$0.09 (3) &$-$1.39 & -            & 5.84$\pm$0.07 (2)  &$-$1.66  & -    \\
Co {\sc i}  & 27 & 4.99 & -                  & -        & -     & 3.52 (1)           & $-$1.47 &$-$0.03& 3.49$\pm$0.12 (2) &$-$1.50 &$-$0.11      & 3.47 (1) & $-$1.52 & 0.14 \\
Ni {\sc i}  & 28 & 6.22 & 3.35$\pm$0.05 (syn, 2)  & $-$2.87 &  $-$0.30 & 4.76$\pm$0.18 (4)  & $-$1.46 &$-$0.02& 5.57$\pm$0.16 (3) &$-$0.65 & 0.74        &  5.26$\pm$0.18 (2) &   $-$0.96 &  0.70 \\
Cu {\sc i}  & 29 & 4.19 & -                  & -        & -     & 2.55 (1)           & $-$1.64 &$-$0.20& 2.85 (1) & $-$1.34  & 0.05   & -                  & -       & -    \\
Zn {\sc i}  & 30 & 4.56 & -                  & -        & -     & 3.21 (1)           & $-$1.35 & 0.09 & 3.30 (1)          &$-$1.26 & 0.13         & 2.74 (1) & $-$1.82 & $-$0.16 \\
Sr {\sc i}  & 38 & 2.87 & 1.97 (syn)         &  $-$0.90  &  1.67  & 2.97 (syn)           & 0.10    & 1.54 & 2.37 (syn) & $-$0.50 & 0.89    & -   & -  & -       \\
Y {\sc ii}  & 39 & 2.21 & 0.79$\pm$0.20 (2)  & $-$1.42  & 1.15  & 1.97$\pm$0.11 (6)  & $-$0.24 & 1.20 & 1.52$\pm$0.18 (3) &$-$0.69 & 0.70         & 1.67$\pm$0.21 (2)  &$-$0.54  & 1.12 \\
Zr {\sc i}  & 40 & 2.58 & -                  & -        & -     & 2.71$\pm$0.09 (2)  & 0.13    & 1.57 & 1.57 (1)          &$-$1.01 & 0.38         & -                  & -       & -    \\
Zr {\sc ii} & 40 & 2.58 & 0.63 (1)           & $-$1.95  & 0.62  & 2.77$\pm$0.21 (2)  & 0.19    & 1.63 & -                 & -      & -            & 2.23 (1)           &$-$0.35  & 1.31 \\
Ba {\sc ii}$^{*}$& 56 & 2.18 & 1.49 $\pm$0.12 (2)&$-$0.69&1.88 &2.80 $\pm$0.10 (2)& 0.62 & 2.06 &1.88 $\pm$0.32 (2)&$-$0.30 & 1.09     & 2.60 (1)           & 0.42    & 2.08 \\
La {\sc ii}$^{*}$& 57 & 1.10 & 0.25 (1)      &   --0.85  &  1.72 & 1.60 (1)         & 0.50    & 1.94 & 0.80 (1)        &$-$0.30 & 1.09         & 1.30 (1) & 0.20   & 1.86 \\
Ce {\sc ii} & 58 & 1.58 & 0.93$\pm$0.02 (2)  & $-$0.65  & 1.92  & 2.17$\pm$0.09 (9)  & 0.59    & 2.03 & 1.53$\pm$0.13 (3) &$-$0.05 & 1.34         & 2.42$\pm$0.19 (3)  & 0.84    & 2.50 \\
Pr {\sc ii} & 59 & 0.72 & 0.03 (1)           & $-$0.69  & 1.88  & 1.25 (1)           & 0.53    & 1.97 & 0.41$\pm$0.09 (2) &$-$0.31 & 1.08         & 1.60$\pm$0.11 (3) & 0.88 & 2.54 \\
Nd {\sc ii} & 60 & 1.42 & 0.74$\pm$0.22 (6)  & $-$0.68  & 1.89  & 2.01$\pm$0.15 (11) & 0.59    & 2.03 & 0.96$\pm$0.14 (8) &$-$0.46 & 0.93         & 1.91$\pm$0.13 (5)  & 0.49    & 2.15 \\
Sm {\sc ii} & 62 & 0.96 & 0.19$\pm$0.13 (4)  & $-$0.77  & 1.80  & 1.60$\pm$0.15 (4)  & 0.64    & 2.08 & 0.55$\pm$0.20 (2) &$-$0.41 & 0.98         & 1.24$\pm$0.14 (2)  & 0.28    & 1.94 \\
Eu {\sc ii}$^{*}$& 63 & 0.52 & $-$1.20 (1)      & $-$1.72  & 0.85  &0.11 $\pm$0.01 (2)&$-$0.41& 1.03& -                 &   -    & -            & $-$0.25 (1)        &$-$0.77  & 0.89 \\

\hline 
\end{tabular}}
 $^{*}$ abundance is derived using spectrum synthesis calculations. 
 $^{a}$ \citet{asplund2009}, Columns 4, 7, 10 \& 13 present the abundances ($log{\epsilon}$) of the different elements along with the standard deviations (when more than one line is used to derive the abundance). The number inside the parenthesis in columns 4, 7, 10 \& 13 shows the number of lines used for the abundance determination.\\
\end{table*}
}

\subsection{Kinematic analysis of the program stars}
\label{sec:kinematics}

\begin{figure}
        \centering
        \includegraphics[height=9.2cm,width=9.2cm]{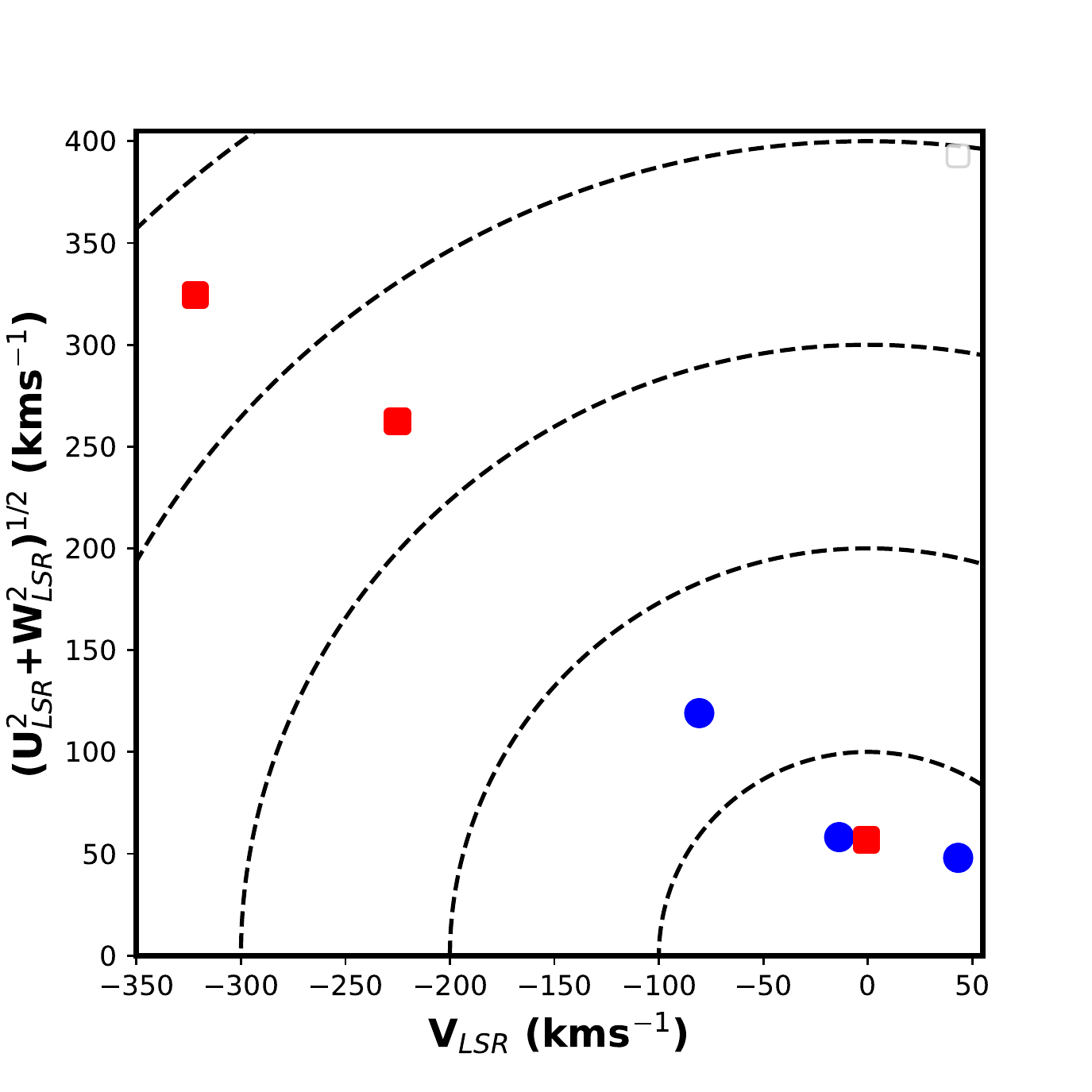}
        \caption{ Toomre diagram. Red filled squares represent the CEMP-{\it{s}} stars and blue filled circles represent Ba stars. The dashed curves connect the loci with same V$_{LSR}$.  }
\label{fig:toomre}
\end{figure}

We have carried out the kinematic analysis of the program stars, following the procedure described in \citet{Purandardas_et_al_2019} and \citet{Goswami_et_al_1_2021}, to know the Galactic population to which the program stars belong. The values of parallax ($\pi$) and proper motions ($\mu_{\alpha}$, $\mu_{\delta}$) are taken from the {\it{Gaia}} database and radial velocities of our estimates are considered for calculating the components of space velocity (U$_{LSR}$, V$_{LSR}$, W$_{LSR}$) with respect to Local Standard of Rest (LSR). The total space velocity (V$_{spa}$) is given by V$_{spa}$ = $\sqrt{U_{LSR}^{2} + V_{LSR}^{2} + W_{LSR}^{2}}$. We have calculated the probabilities of the program stars being in the halo, thin or thick disc using the procedures given by \citet{reddy2006}, \citet{bensby2003, bensby2004}, and \citet{mishenina2004}. For the probability determination, it is assumed that Gaussian distribution functions for U$_{LSR}$, V$_{LSR}$ and W$_{LSR}$, with given mean values and dispersions represent thin disc, thick disc and halo populations \citep{reddy2006}. A detailed discussion of the procedure can be found in our previous works \citep{Purandardas_et_al_2019, Goswami_et_al_1_2021}. The values of U$_{LSR}$, V$_{LSR}$, W$_{LSR}$, V$_{spa}$ and the probability estimates for the objects being members of the thin disc (P$_{thin}$), thick disc (P$_{thick}$) and halo (P$_{halo}$) population are presented in Table~\ref{tab:kinematicresults}. BD+75~348, HD~238020 and HE~0930--0018 show the probability of being thin disc objects, BD+09~3019 is a thick disc object and HE~0319--0215 \& HE~0507--1653 are halo objects. A toomre diagram is shown in Figure~\ref{fig:toomre}.

\begin{table*}
\centering
\caption{\bf {Spatial velocity and probability estimates}}
\begin{tabular}{lccccccc}
\hline
Star Name       & $U_{LSR} (km/s) $  & $V_{LSR} (km/s)$     &$ W_{LSR} (km/s)$    & $V_{spa} (km/s)$     & $P_{thin}$ & $P_{thick}$& $ P_{halo}$\\
\hline
BD+75~348       &--44.75 $\pm 1.17$  &    43.21 $\pm 0.93$   &   17.28 $\pm 1.07$   &  64.56 $\pm 0.10$  & 0.97        & 0.03       &  0.00\\
BD+09~3019      & 102.34 $\pm  4.47$ & $-$80.69 $\pm 4.10$   &$-$60.85 $\pm 3.04$   & 143.83 $\pm 0.41$  & 0.00        & 0.97       &  0.03\\
HD~238020       &--43.06 $\pm 1.41$  & $-$13.73 $\pm 1.02$   &$-$39.08 $\pm 2.60$   &  59.75 $\pm 2.95$  & 0.90        & 0.10       &  0.00\\
HE~0319$-$0215  & 305.94 $\pm 22.00$ & --321.86 $\pm 65.35$  &  107.59 $\pm 16.06$  & 456.91 $\pm 27.24$ & 0.00        & 0.00       &  1.00\\
HE~0507$-$1653  &--247.94 $\pm 1.03$ & --225.22 $\pm 2.55$   &$-$85.75 $\pm 3.89$   & 345.76 $\pm 3.39$  & 0.00        & 0.00       &  1.00\\
HE~0930$-$0018  &--53.49 $\pm 0.42$  &   --0.63 $\pm 0.47$   &   18.80 $\pm 0.40$   &  56.70 $\pm 0.27$  & 0.98        & 0.02       &  0.00\\
\hline
\label{tab:kinematicresults}	
\end{tabular}
\end{table*}

\section{Discussion}
\label{sec:discussion}
\subsection{Classification schemes of the mild/strong Ba and CEMP-s stars}
\label{sec:classification1}
\subsubsection{Ba stars}

After the first identification of Ba stars by \citet{Bidelman_Keenan_1951}, several authors \citep{ Pilachowski_1977, Sneden_et_al_1981, Lu_1991, Jorissen_et_al_1998, Yang_et_al_2016, de_Castro_et_al_2016, Escorza_et_al_2017, Jorissen_et_al_2019} have put forward classification criteria to distinguish the mild and strong Ba stars. \citet{Pilachowski_1977} reported abundance analysis of seven mild Ba stars and demonstrated that they show [s/Fe] $\geq$ 0.50, where [s/Fe] implies the average abundance of the available s-process elements. \citet{Sneden_et_al_1981} pointed out that limiting values of [s/Fe] for mild Ba stars and classical (strong) Ba stars are 0.21 and 0.73 respectively. Using the Ba index given by \citet{Warner_1965}, \citet{Lu_1991} classified the Ba stars with Ba indices 0.3 -- 1.5 and 2 -- 5 as mild and strong Ba stars, respectively. \citet{Jorissen_et_al_1998} classified the stars with Ba $\leq$ 2 as mild and Ba = 3 -- 5 as strong Ba stars. \citet{Escorza_et_al_2017} considered stars with Ba indices 1 and 2 as mild Ba stars and 3 -- 5 as strong Ba stars. However, avoiding the use of Ba indices, \citet{Yang_et_al_2016} used [Ba/Fe] as an indicator to distinguish the two groups. They considered stars showing [Ba/Fe] $>$ 0.60 as strong Ba stars. Their sample stars show 0.17 $<$ [Ba/Fe] $<$ 0.54 for mild Ba stars. \citet{de_Castro_et_al_2016} used [s/Fe] $\geq$ 0.25 as a distinguishing value between normal giants and Ba stars. \citet{Jorissen_et_al_2019} recommended that [La/Fe] and [Ce/Fe] for strong Ba stars are greater than equal to unity, and mild Ba stars always show [Ce/Fe] $\geq$ 0.2.

In this study, we use a similar criteria as that of \citet{Yang_et_al_2016}. But, as abundance of Ba is  not reported for many Ba stars in the literature due to the saturation of strong Ba lines in the spectra of Ba stars, we use [hs/Fe] to differentiate the mild and strong Ba stars. Here, hs (heavy-s) represents Ba, La, Ce and Nd. We have not used Pr and Sm in the calculations of hs, as r-process contributes more to these two elements than that of s-process. In order to reduce the systematic error in calculating [hs/Fe], we have  considered only those stars for which abundances of at least  two or more heavy-s process elements are reported in the literature. We consider stars with [hs/Fe] $>$ 0.60 as strong Ba stars and stars with 0.17 $<$ [hs/Fe] $<$ 0.60 as mild Ba stars. We have discarded the possibility of using [s/Fe] as a classifier, because [s/Fe] contains not only Pr and Sm, but also the ls elements Sr, Y, and Zr, and the  s-process AGB models cannot reproduce the ls elements satisfactorily. These elements are believed to  have contributions from several different sources.

\vskip 1.0 cm
\subsubsection{CEMP-s stars}

CEMP stars are metal-poor stars ([Fe/H] $<$ --1.0) with [C/Fe] $>$ 0.7. Several classification schemes have been proposed in the literature \citep{beers2005discovery, Jonsell_et_al_2006, Masseron_et_al_2010, abate2016cemp-rs, Frebel_review_2018, Hansen_et_al_2019, Goswami_et_al_1_2021} to identify the CEMP-s stars from the different subclasses of CEMP stars. However, it is difficult to distinguish the CEMP-s and CEMP-r/s stars (see \citet{Goswami_et_al_1_2021} for more details). \citet{beers2005discovery}, for the first time used [Ba/Fe], [Eu/Fe] and [Ba/Eu] to differentiate between the CEMP stars. Following it, \citet{Jonsell_et_al_2006}, \citet{Masseron_et_al_2010} and \citet{abate2016cemp-rs} used the same abundance ratios with different limiting values. \citet{Hansen_et_al_2019} used [Sr/Ba] to distinguish the subclasses. In this paper, we have followed the classification scheme discussed in \citet{Goswami_et_al_1_2021}-\\

\begin{itemize}
      \item CEMP: [C/Fe]$\geq$0.7
      \item CEMP-r/s: [Ba/Fe]$\geq$1.0, [Eu/Fe]$\geq$1.0
      \begin{itemize}
            \item [i)] 0.0$\leq$[Ba/Eu]$\leq$1.0 and/or 0.0$\leq$[La/Eu]$\leq$0.7
      \end{itemize}
      \item CEMP-s: [Ba/Fe]$\geq$1.0 
      \begin{itemize}
            \item [i.)] [Eu/Fe]$<$1.0, [Ba/Eu]$>$0.0 and/or  [La/Eu]$>$0.5
            \item [ii.)] [Eu/Fe]$\geq$1.0, [Ba/Eu]$>$1.0 and/or  [La/Eu]$>$0.7
      \end{itemize}
\end{itemize}

\subsection{Classification of the program stars}
\label{sec:classification2}

\subsubsection{BD+75~348, BD+09~3019 \& HD~238020}
From the visual inspection of the strong lines Sr~II 4077 {\rm \AA} and Ba~II 4554 {\rm \AA}, we classify these three objects as Ba stars. The [Fe/H] of BD+75~348, BD+09~3019 \& HD~238020 lies between --0.68 and --0.42. 
BD+75~348, with [C/Fe] $\sim$ 0.31, is not a carbon enhanced star, but it shows enhancement in neutron-capture elements. BD+09~3019, with [C/Fe] $\sim$ 0.77, shows enhancement in carbon, and the neutron-capture elements are also strongly enhanced in the star. The high value of [hs/Fe] ($>$ 0.60) put BD+75~348 and BD+09~3019 in the category of strong Ba stars. HD~238020, on the other hand, shows only mild enhancement in neutron-capture elements with solar abundance of carbon. The C/O of HD~238020 is greater than unity. With [hs/Fe] = 0.30, we classify this object as a mild Ba star.

\subsubsection{HE~0319--0215, HE~0507--1653, HE~0930--0018 \& HE~1023--1504}
\begin{figure*}
     \begin{center}
\centering
        \subfigure{%
            \label{fig:baeueu}
            \includegraphics[height=8.0cm,width=8.5cm]{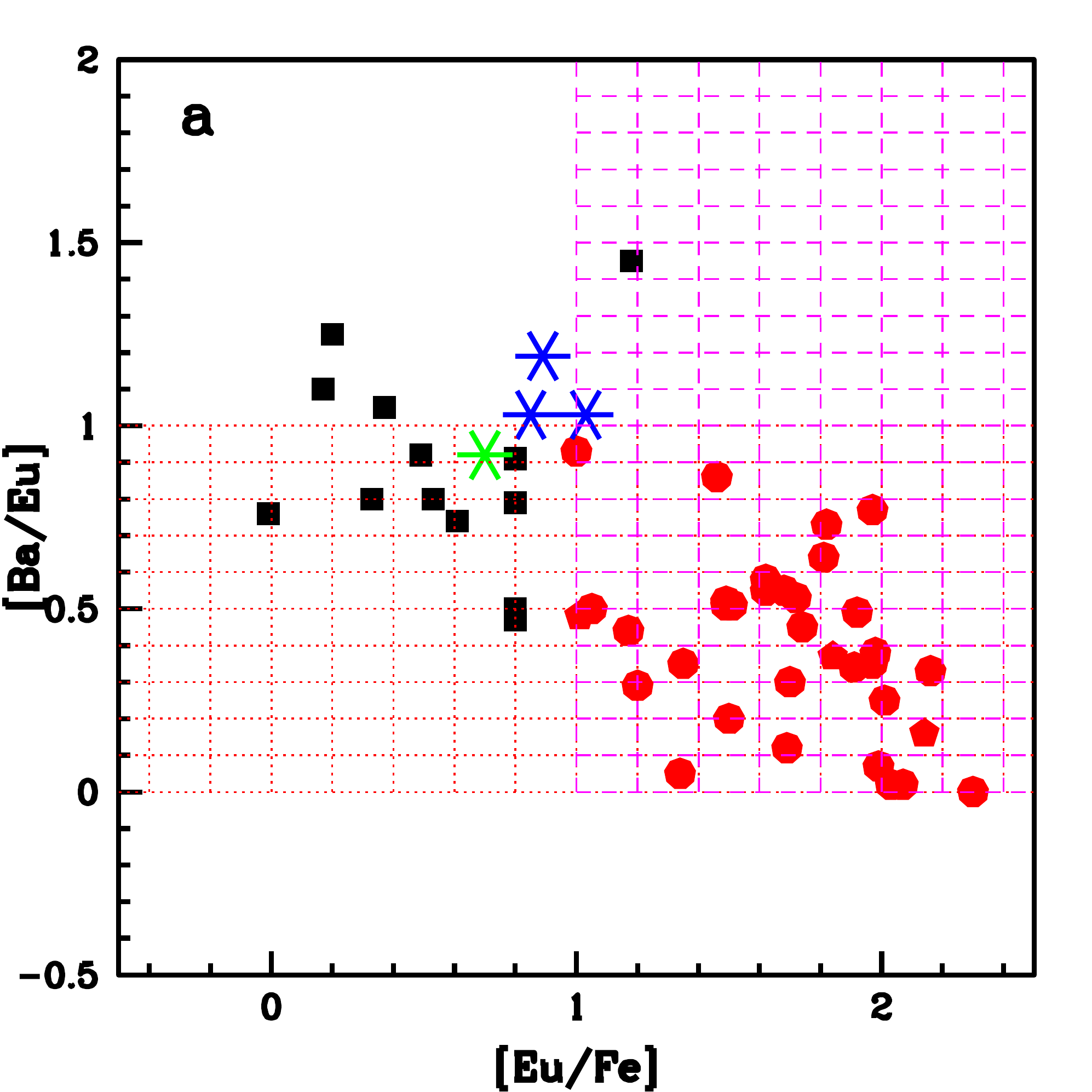}
        }%
        \subfigure{%
            \label{fig:laeubaeu}
            \includegraphics[height=8.0cm,width=8.5cm]{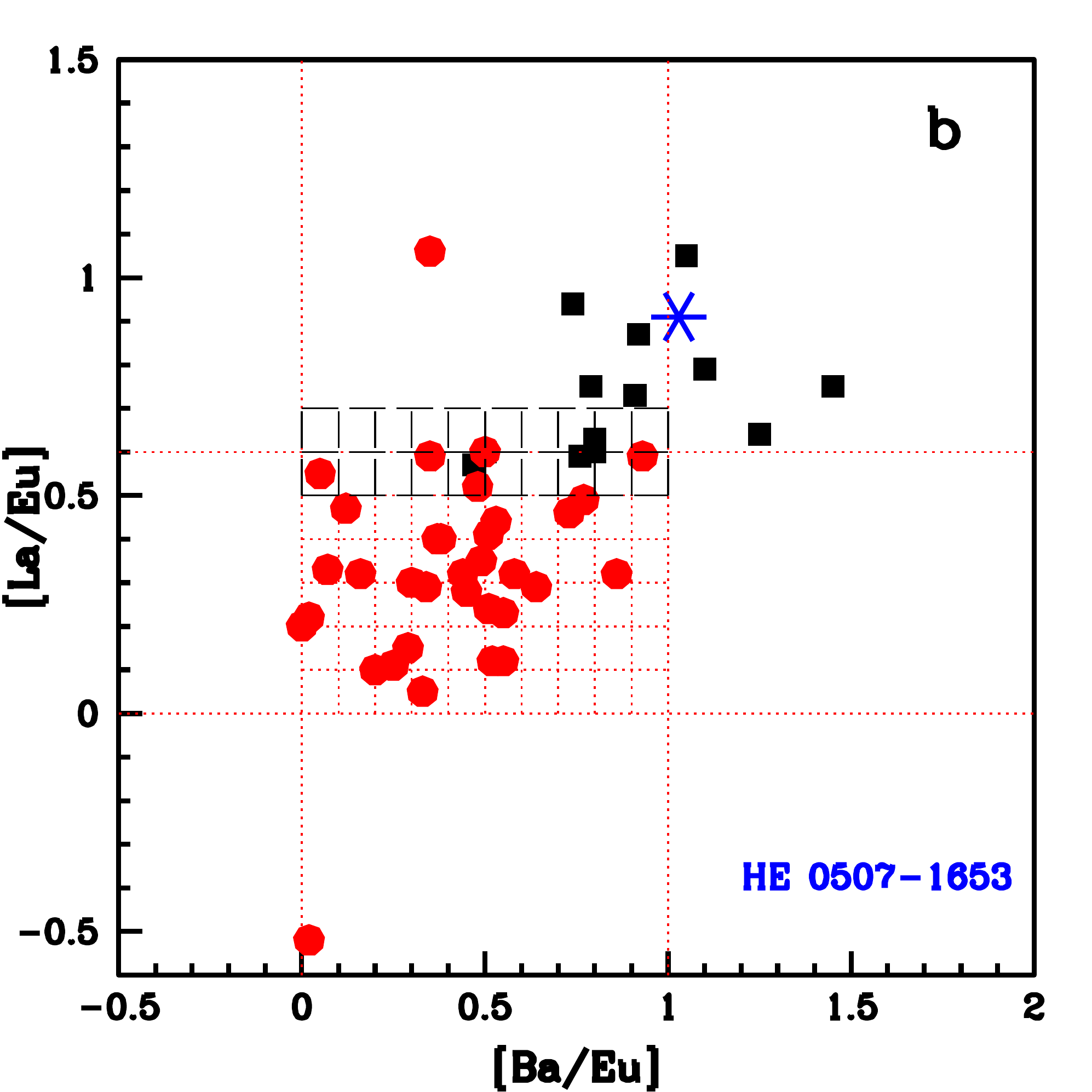}

        }\\ 

    \caption{ Filled red circles and filled black squares respectively represent literature CEMP-r/s and CEMP-s stars compiled by \citet{Goswami_et_al_1_2021}, the blue and the green stars represent the CEMP-s and Ba stars of this study. panel a: The grid formed by the dotted red lines represents the region of CEMP-r/s stars put forward by \citet{beers2005discovery}. The grid formed by the dashed magenta lines represents the region of CEMP-r/s stars put forward by \citet{abate2016cemp-rs}. panel b: Grid formed by the dotted red lines bound by 0.0 $<$ [La/Eu] $<$ 0.6 and 0.0 $<$ [Ba/Eu] $<$ 1.0 indicates the region defined for CEMP-r/s stars by \citet{Goswami_et_al_1_2021}. The grid formed by the black dashed lines bound by 0.5 $<$ [La/Eu] $<$ 0.7 represents the region where [Eu/Fe] $>$ 1.0 classifies the stars as CEMP-r/s and [Eu/Fe] $<$ 1.0 classifies the stars as CEMP-s.  }%
   \label{fig:classification}
       \end{center}

\end{figure*}

HE~0319--0215, HE~0507--1653, HE~0930--0018 \& HE~1023--1504 are metal-poor (MP: [Fe/H]~$<$~--1.0) and very metal-poor (VMP: [Fe/H]~$<$~--2.0) stars with --1.39~$<$[Fe/H]~$<$~--2.57. All these stars are carbon enhanced ([C/Fe]~$>$~1.0) and enriched in neutron-capture elements. All four stars fall in the category of CEMP-s stars following the classification criteria of \citet{Goswami_et_al_1_2021}. Although [Eu/Fe] is greater than unity in HE~0507--1653 and the classification scheme of \citet{abate2016cemp-rs} (Figure~\ref{fig:baeueu}) puts it in the category of CEMP-r/s stars based on [Ba/Eu]~$>$~0.0. Figure~\ref{fig:laeubaeu} shows that this star is a CEMP-s star from the classification scheme of \citet{Goswami_et_al_1_2021}. We tried to fit i-process models \citep{Hampel_et_al_2016} calculated for neutron-densities, n $\sim$ 10$^{9}$ -- 10$^{15}$ cm$^{-3}$ to the observed abundance pattern of HE~0507--1653. We found that the model with n = 10$^{9}$ cm$^{-3}$ gives the minimum $\chi^{2}$ value. So, this star cannot be a CEMP-r/s star. However, we will discuss in Section~\ref{sec:primary_mass} that s-process yields produced by an M = 2.0 M$_{\odot}$ AGB can satisfactorily reproduce the observed abundance pattern of the star, placing it in the CEMP-s category. This verifies the upper limit on [Ba/Eu] ($>$ 1.0) for CEMP-r/s stars put forward by our previous work \citep{Goswami_et_al_1_2021}. Classifications of the program stars along with some important abundance ratios, such as [ls/Fe], [hs/Fe], [hs/ls], [Eu/Fe], [Ba/Eu] \& [La/Eu] are presented in Table~\ref{tab:abundanceratios}.

{\footnotesize
\begin{table*}
\centering
\caption{\bf{Observed abundance ratios and classifications of the program stars}}
	\label{tab:abundanceratios}
\scalebox{0.95}{
\begin{tabular}{lcccccccc}
\hline
Star Name       & [Fe/H]  & [ls/Fe]  & [hs/Fe]  & [hs/ls] & [Eu/Fe] & [Ba/Eu] &  [La/Eu] &  Classification \\             
\hline
BD+75~348       & --0.42  & 1.16    &  1.75    &  0.59   &  0.70   &  0.92   &  1.22   &  Strong Ba star \\
BD+09~3019      & --0.55  & 1.49    &  2.08    &  0.59   &  1.05   &   -     &  1.20   &  Strong Ba star \\
HD~238020       & --0.68  &--0.04   &  0.30    &  0.34   &    -    &   -     &   -     &   Mild Ba star  \\
HE~0319--0215   & --2.57  & 1.15    &  1.85    &  0.70   &  0.85   &  1.03   &  0.87   &   CEMP-s star   \\
HE~0507--1653   & --1.44  & 1.46    &  2.02    &  0.56   &  1.03   &  1.03   &  0.91   &   CEMP-s star   \\
HE~0930--0018   & --1.39  & 0.66    &  1.11    &  0.45   &    -    &   -     &   -     &   CEMP-s star   \\
HE~1023--1504   & --1.66  & 1.22    &  2.15    &  0.93   &  0.89   &  1.19   &  0.97   &   CEMP-s star   \\
\hline

\end{tabular}}

\end{table*}
}

\subsection{Masses of the program stars}
\label{sec:secondary_mass} 
The masses of the program stars are estimated from the position of the stars on the HR diagram (log(L/L$_{\odot}$) versus log(T$_{eff}$) plot). In order to estimate the luminosities (L) of the program stars using Equation~\ref{eqn:luminosity} the visual magnitudes (V) of the stars are taken from SIMBAD, the parallaxes ($\pi$) are taken from {\it{Gaia}}\footnote{\url{https://gea.esac.esa.int/archive/}}, the interstellar extinction (A$_{v}$) values are calculated from the formula given in \citet{Chen_et_al_1998}, the bolometric corrections (BC) are calculated using the empirical calibrations of \citet{alonso1999effective}. 

\begin{equation}
\label{eqn:luminosity}
log(L/L_{\odot}) = 0.4(M_{bol\odot} - V - 5 - 5log(\pi) + A_{v} - BC)
\end{equation}

\begin{figure*}
     \begin{center}
\centering
        \subfigure[]{%
            \label{fig:hr_ba_first}
            \includegraphics[height=7.0cm,width=8.5cm]{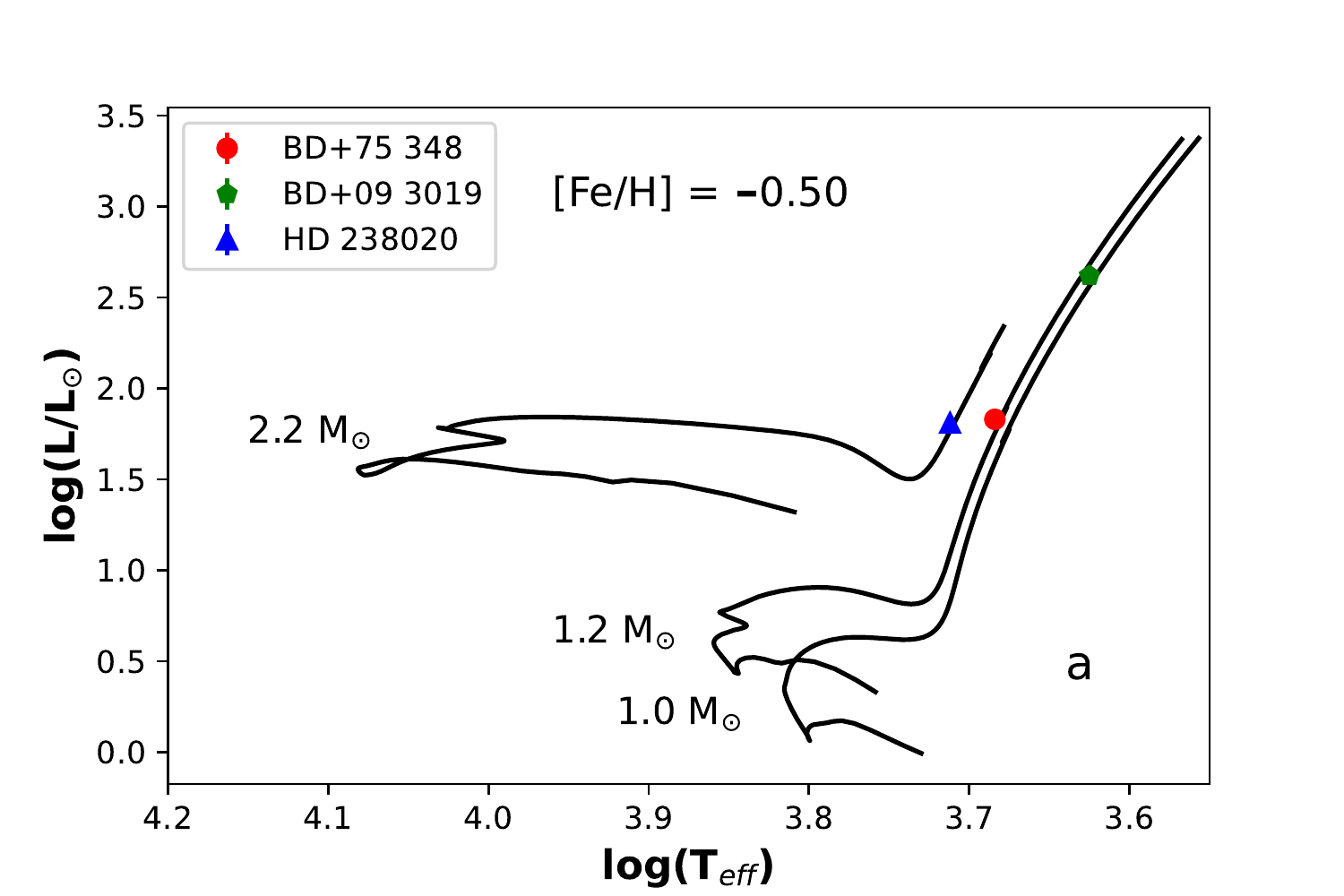}
        }%
        \subfigure[]{%
            \label{fig:hr_cemp_second}
            \includegraphics[height=7.0cm,width=8.5cm]{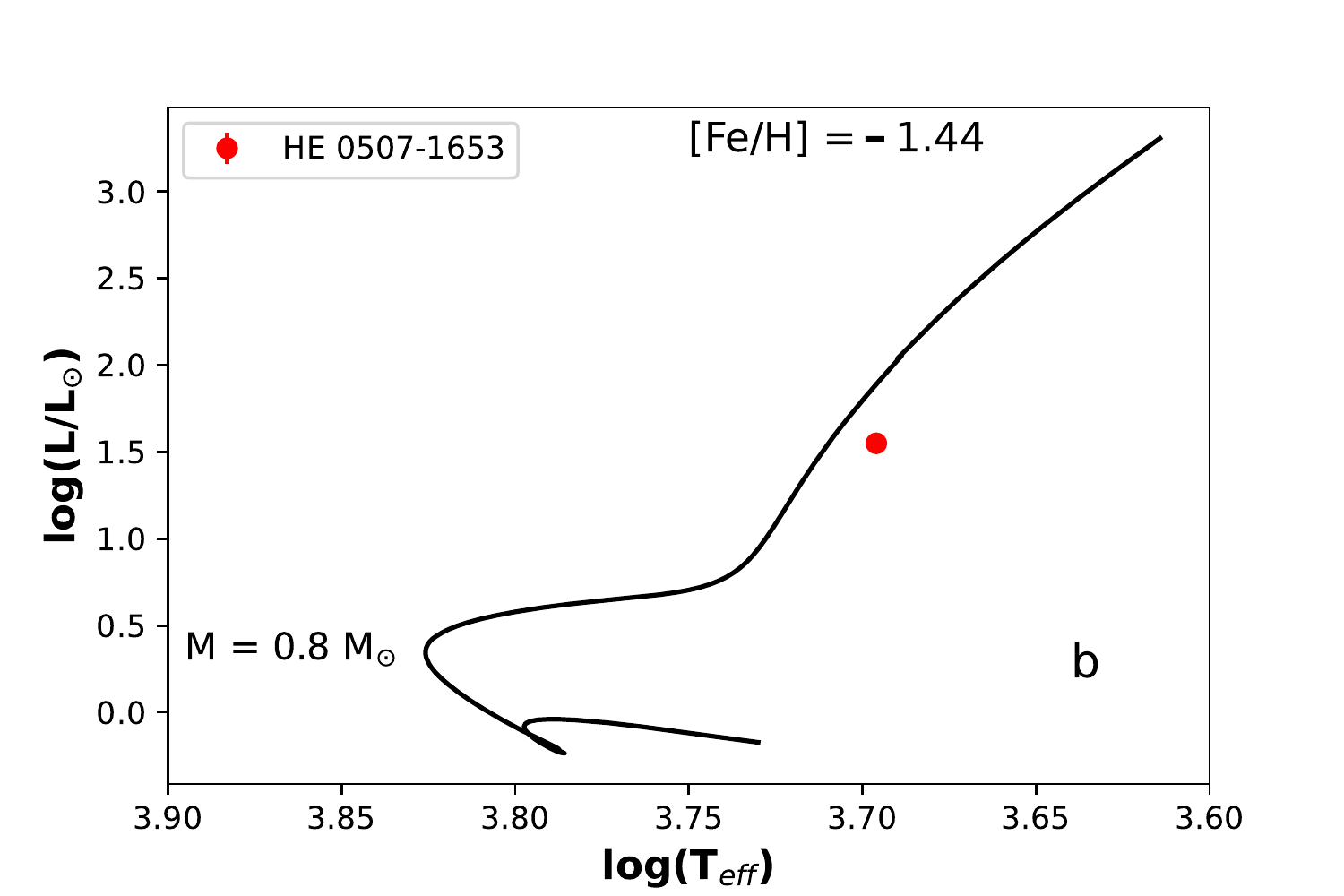}

        }\\ 

    \caption{ Hertzsprung-Russel diagram. Panel (a) shows evolutionary tracks for different masses at [Fe/H] $\sim$ --0.50. Red filled circle, green filled pentagon and blue filled triangle represent the positions of the Ba stars of our study. In panel (b), an evolutionary track for M = 0.8 M$_{\odot}$ at [Fe/H] = --1.44 is shown. The red filled circle represents a CEMP-s star HE~0507--1653. }%
   \label{fig:mass_ba}
       \end{center}

\end{figure*}

We have used the updated BASTI-IAC evolutionary tracks\footnote{\url{http://basti-iac.oa-abruzzo.inaf.it/}} \citep{Hidalgo_et_al_2018} generated, including overshooting and diffusion, for the corresponding metallicities of the program stars. For log(T$_{eff}$), we have used our spectroscopic T$_{eff}$ estimates. Figure~\ref{fig:hr_ba_first} shows the appropriate evolutionary tracks corresponding to the Ba stars of our study. Figure~\ref{fig:hr_cemp_second} shows the evolutionary track corresponding to HE~0507--1653, a CEMP-s star of our study. We can see that the star falls towards the right side of the evolutionary track. This is because the evolutionary tracks highly depend on the opacity in the stellar atmospheres and the BASTI-IAC evolutionary tracks are not calculated considering high carbon abundances. So, these evolutionary tracks are not suitable for carbon-enhanced stars. As the abundances of C or O or other $\alpha$-elements increase, formation of  complex molecules in the stellar atmosphere becomes more prominent. Due to this, it may be possible that the opacity increases and the radiation pressure pushes the outer layers to make the star larger. Hence, temperature decreases and the star moves towards right side of the HR diagram. The use of the evolutionary tracks customized to the observed abundances of the stars is out of the scope of this paper. The masses of the program stars including log(L/L$_{\odot}$) estimates, are presented in Table~\ref{tab:mass_estimation}. Due to the precise parallax values provided by {\it{Gaia}}, the uncertainties in the estimation of Luminosities are so small that we cannot even see the error bars in Figure~\ref{fig:mass_ba}. However, the evolutionary tracks are available at mass intervals of 0.1 M$_{\odot}$ and 0.5 M$_{\odot}$ for the mass ranges 0.8 -- 3 M$_{\odot}$ and 3 -- 6 M$_{\odot}$ respectively. So, we consider an uncertainty in mass of $\pm$0.05 M$_{\odot}$ and $\pm$0.25 M$_{\odot}$ for the mass ranges 0.8 -- 3 M$_{\odot}$ and 3 -- 6 M$_{\odot}$ respectively.

\begin{table*}
\centering
\caption{\bf {Masses of the program stars}}
\begin{tabular}{lcccccc}
\hline
Star Name       & Parallax          &  A$_{v}$  &    BC      &  log(L/L$_{\odot}$) & log$T_{eff}$ & Mass (M$_{\odot}$)\\
                & (mas)             &           &            &                     &              &                   \\
\hline
BD+75~348       & 1.6672$\pm$0.0359 &  0.147608 & --0.320606 & 1.83$\pm$0.02      &  3.684       & 1.20    \\
BD+09~3019      & 0.4105$\pm$0.0183 &  0.000000 & --0.690934 & 2.62$\pm$0.04      &  3.625       & 1.00    \\
HD~238020       & 2.4767$\pm$0.0205 &  0.000000 & --0.225678 & 1.81$\pm$0.01      &  3.712       & 2.20    \\
HE~0319--0215   & 0.0842$\pm$0.0158 &  0.034626 & --0.443833 & 2.80$\pm$0.15      &  3.667       & 0.80    \\
\hline
\label{tab:mass_estimation}	
\end{tabular}
\end{table*}

\subsection{Initial masses of the binary companion (primary)}
\label{sec:primary_mass}
\begin{figure*}
     \begin{center}
\centering
        \subfigure[]{%
            \includegraphics[height=8.0cm,width=8.5cm]{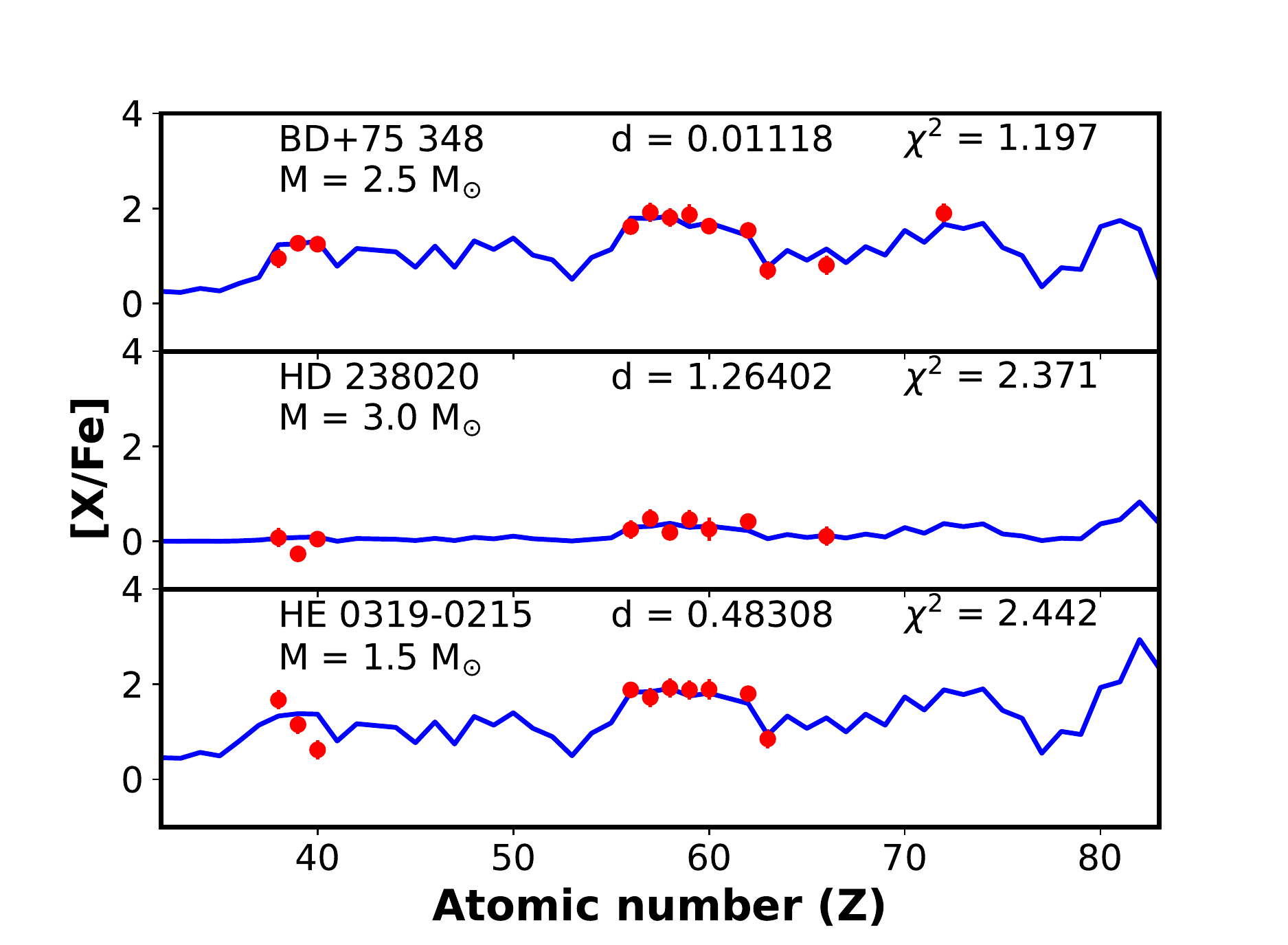}
        }%
        \subfigure[]{%
            \includegraphics[height=8.0cm,width=8.5cm]{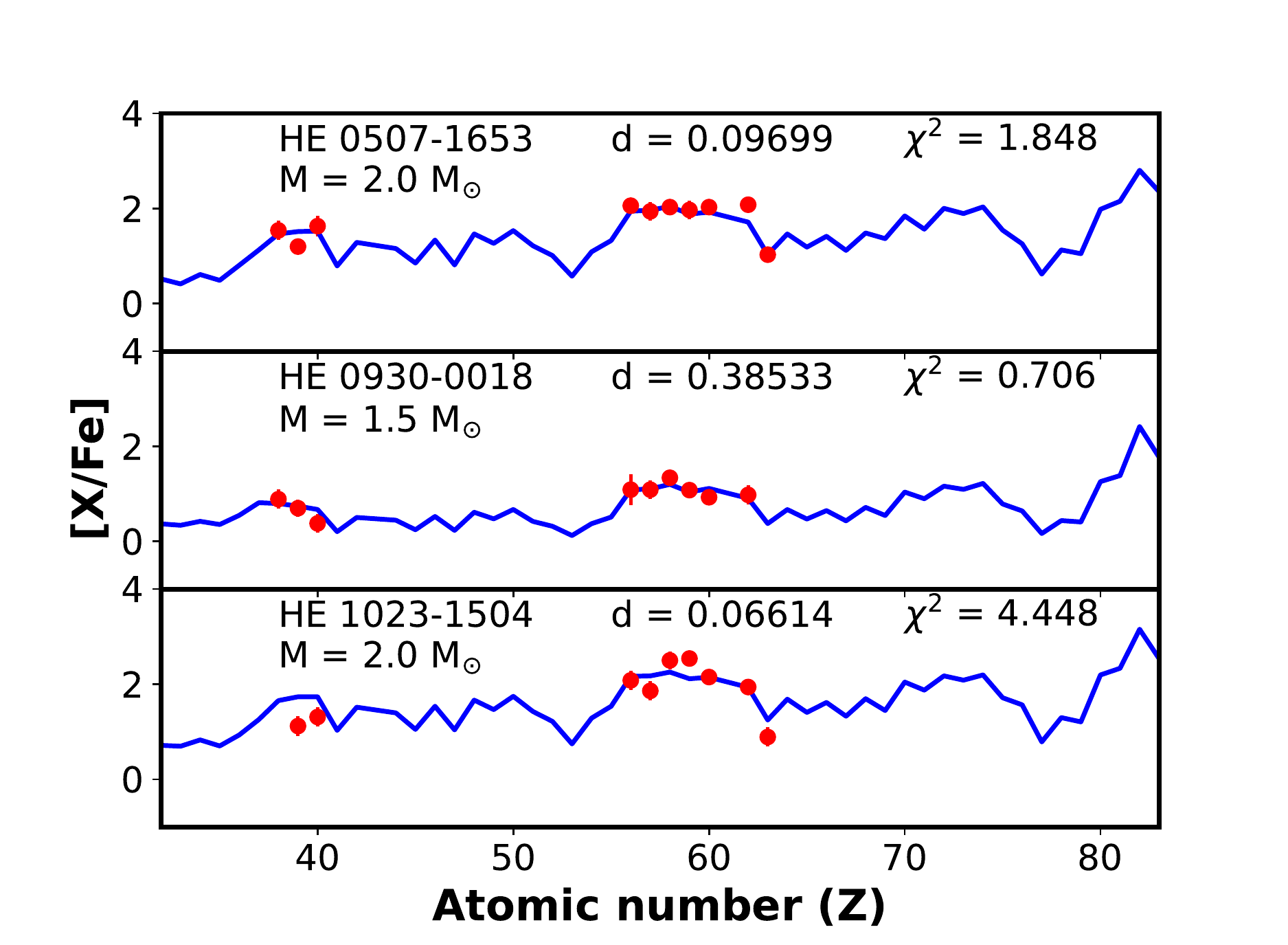}

        }\\ 

    \caption{ Best fit parametric-models of the program stars. The points with error bars indicate the observed abundances. }%
   \label{fig:model_fit}
       \end{center}

\end{figure*}

We have performed a parametric-model-based analysis, using AGB yields of the FRUITY\footnote{\url{http://fruity.oa-teramo.inaf.it/}} models \citep{Straniero_et_al_2006, Cristallo_et_al_2009, Cristallo_et_al_2011, Cristallo_et_al_2015} to estimate the initial masses of the companions of the program stars. We have followed the procedure discussed in \citet{shejeela_agb_2020}. We have calculated the s-process yields, using FRUITY models, at different masses (1.3 -- 6.0 M$_{\odot}$) corresponding to the metallicities of the program stars. Model yields are calculated considering the conditions of standard $^{13}$C pockets \citep{Cristallo_et_al_2015} and no initial rotational velocities of the AGB stars. The observed abundances of the neutron-capture elements of a program star are then fitted with the following parametric model function as discussed in \citet{Husti_et_al_2009}--

\begin{equation}
\label{eqn:parametric_model_function}
\left[\dfrac{X}{Fe}\right] = log\left(10^{[X/Fe]^{ini}} \times (1-10^{-d}) + 10^{[X/Fe]^{AGB}-d}\right)
\end{equation}

where, $\left[\dfrac{X}{Fe}\right]$ is the abundance of the element X in the program star, $[X/Fe]^{ini}$ is the initial abundance (before mass-transfer) of the element X in the program star, $[X/Fe]^{AGB}$ is the AGB yield of the element X, $d$ is the dilution factor and a free parameter here.  We find the model which gives the best fit to the observational data by varying $d$ for each set of AGB model yields with different masses. In order to find the best fit, we calculated $\chi^{2}$ for each AGB model. The corresponding mass of the AGB model for which we get the minimum value of $\chi^{2}$ is the required mass of the companion AGB of the program star. The best-fitting models with the AGB masses, corresponding dilution factors and minimum $\chi^{2}$ values of the program stars are shown in Figure~\ref{fig:model_fit}.

\subsection{Mass distribution of primary and secondary stars}
\label{sec:mass_distribution}
The mass distribution of Ba stars has been previously studied by several groups \citep{Han_et_al_1995, Mennessier_et_al_1997, Jorissen_et_al_1998, Escorza_et_al_2017, Jorissen_et_al_2019}. Using the same procedures discussed in Section~\ref{sec:secondary_mass} and Section~\ref{sec:primary_mass}, we have evaluated the mass distributions of Ba stars (secondary stars) and the primaries (companion AGBs) of both Ba and CEMP-s stars. For this analysis, we have selected a sample of 228 Ba stars and 36 CEMP-s stars from several sources in the literature. The atmospheric parameters and elemental abundances of the Ba stars are collected from \citet{Allen_Barbuy_2006}, \citet{de_Castro_et_al_2016}, \citet{Yang_et_al_2016}, \citet{Karinkuzhi_et_al_2018} and \citet{shejeela_agb_2020}. We have calculated the [hs/Fe] values for the sample of 228 Ba stars and removed the ones having [hs/Fe] $<$ 0.17. We believe that this lower limit (0.17) of [hs/Fe] will ensure the selection of true Ba stars from the sample. The sample of Ba stars ranges from [Fe/H] = --1.11 to +0.23. In Table~\ref{tab:Ba_stars}, we have presented the list of Ba stars used for this study. Column 6 of Table~\ref{tab:Ba_stars} gives the [hs/Fe] that we have calculated and column 13 shows the classification. The atmospheric parameters and elemental abundances of CEMP-s stars are taken from our previously compiled CEMP-s stars' list in \citet{Goswami_et_al_1_2021} (Table~\ref{tab:CEMP-s_stars}). We could not estimate the masses of the sample of CEMP-s stars due to the inadequacy of the evolutionary tracks, as discussed in Section~\ref{sec:secondary_mass}. Note that the method of deriving the masses of Ba and CEMP-s stars using evolutionary tracks in HR diagram has a serious drawback. The evolutionary tracks that we use do not take care of the mass-transfer event taking place when the star was in main-sequence stage.

{\footnotesize
\begin{table*}
\centering
\caption{\bf{Masses of the Ba stars and initial masses of their AGB companions.}}
\label{tab:Ba_stars}
\scalebox{0.86}{
\begin{tabular}{lccccccccccccc}
\hline

Star Name & T$_{eff}$ & logg & [Fe/H] & [ls/Fe] & [hs/Fe] & [hs/ls] & M$_{Ba}$ & M$^{ini}_{AGB}$ & $\chi^{2}$ & dil & N & Type & Ref \\
          &  (K)      &(cgs) &        &         &         &        &(M$_{\odot}$)&(M$_{\odot}$)&            &   &   &      &     \\
\hline
BD-01302   & 4200 & 1.10 &-0.64 &-0.14 & 0.17 & 0.310 &  -   & 1.5 & 0.67 & 1.85697   & 5 & m & 1  \\
BD-18821   & 5000 & 2.30 &-0.27 & 0.70 & 1.27 & 0.570 & 2.30 & 3.0 & 1.92 & 0.18030  & 5 & s & 1  \\
BD-083194  & 4900 & 3.00 &-0.10 & 0.95 & 1.52 & 0.570 & 2.00 &  -  &  -   &    -      & 5 & s & 1  \\
\hline
\end{tabular}}
References: 1. \citet{de_Castro_et_al_2016}.\\          
\textbf{Note:} This table is available in its entirety  online only.
A portion is shown here for guidance regarding its form and content.\\

\end{table*}}

\begin{figure*}
     \begin{center}
\centering
        \subfigure[]{%
            \label{fig:secondary_first}
            \includegraphics[height=7.5cm,width=9.0cm]{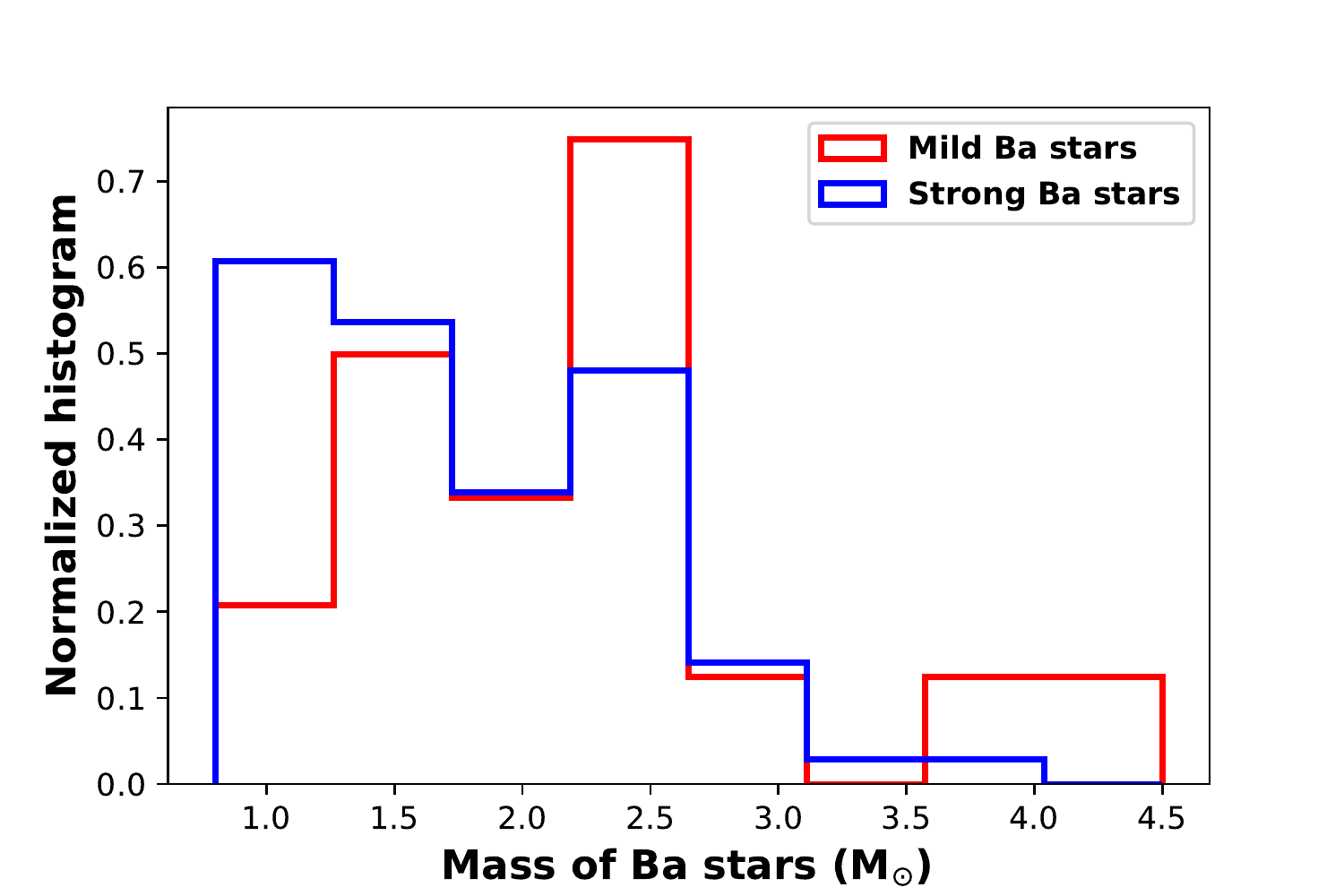}
        }%
        \subfigure[]{%
            \label{fig:secondary_second}
            \includegraphics[height=7.5cm,width=9.0cm]{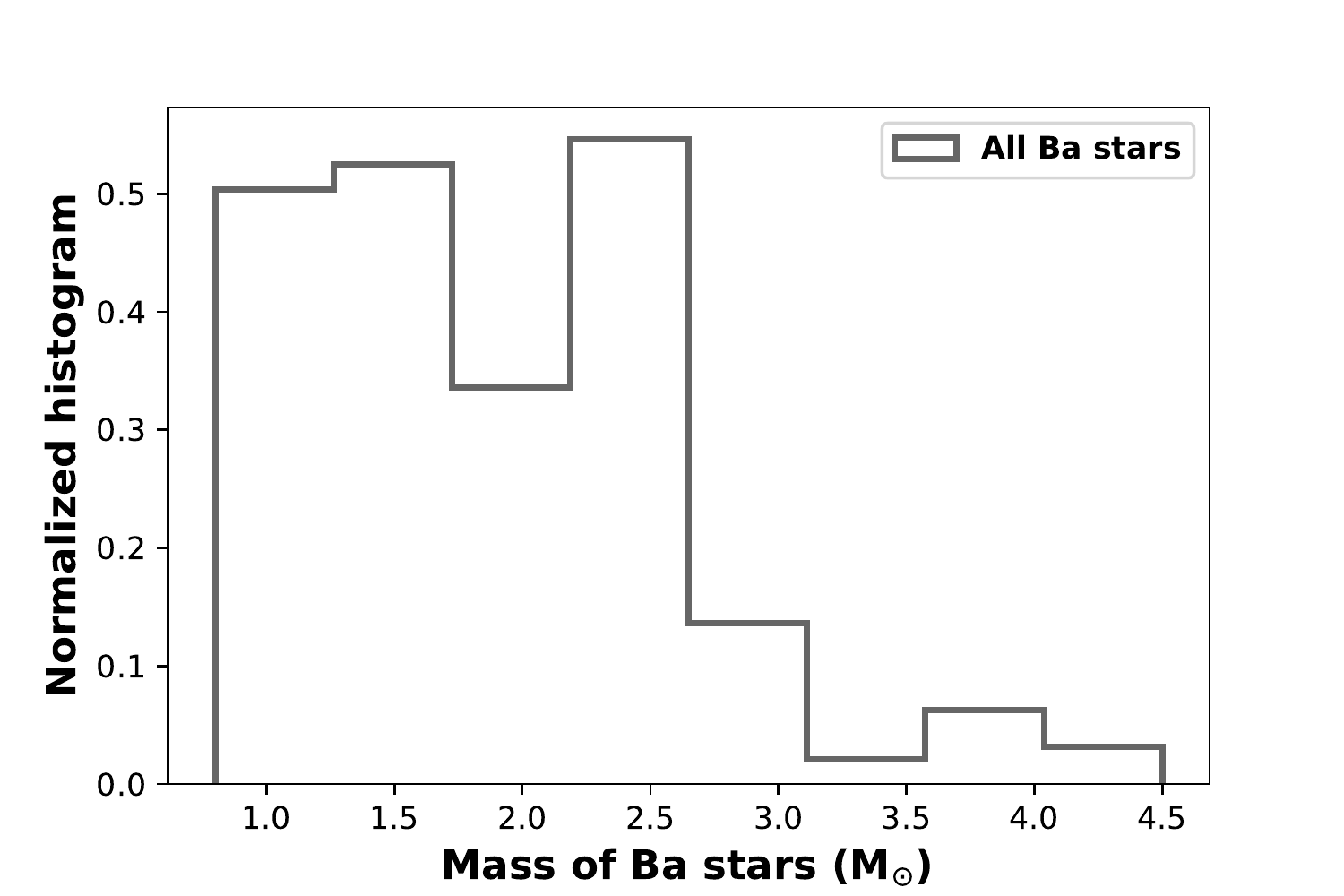}
        }\\ 
        \subfigure[]{%
            \label{fig:primary_third}
            \includegraphics[height=7.5cm,width=9.0cm]{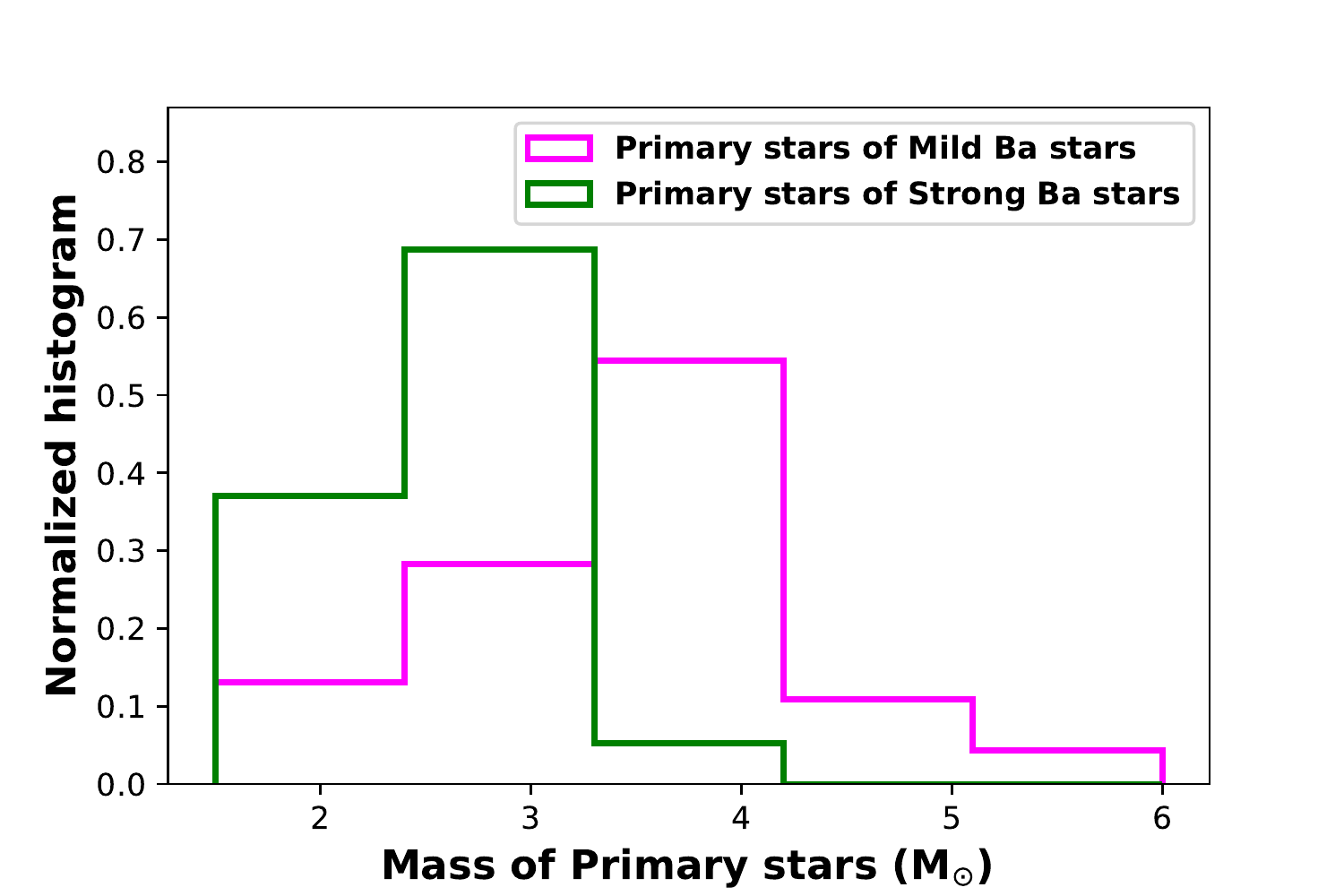}
        }%
        \subfigure[]{%
            \label{fig:pri_sec_fourth}
            \includegraphics[height=7.5cm,width=9.0cm]{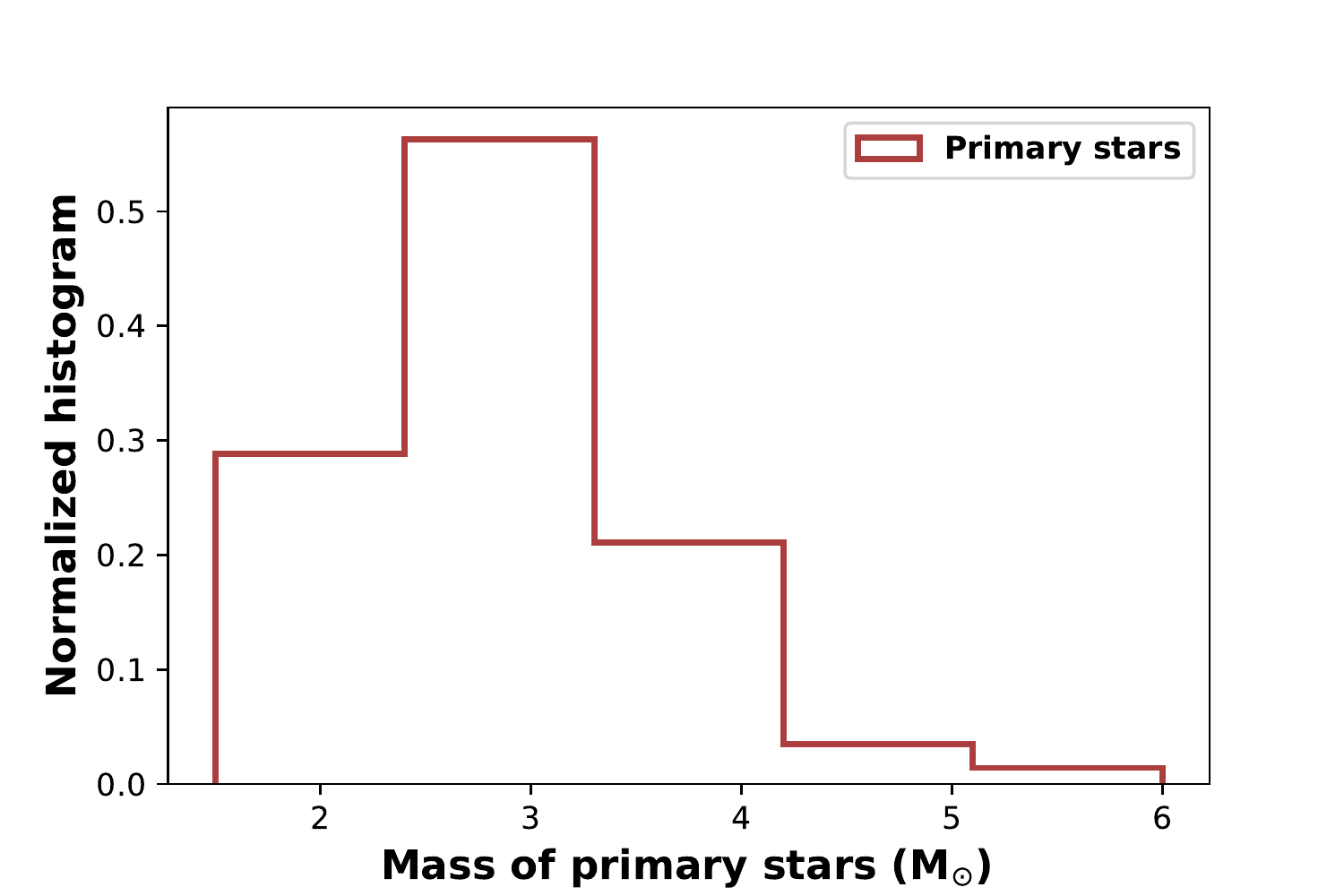}

        }\\ 

    \caption{Mass distributions of Ba stars and their progenitor AGBs (primary stars). Panel (a) shows the mass distributions of mild and strong Ba stars. Panel (b) shows the mass distribution of both mild and strong Ba stars as a whole. Panel (c) shows the mass distributions of the primary stars of mild and strong Ba stars. Panel (d) shows the mass distribution of the primary stars of both mild and strong Ba stars as a whole.   }%
   \label{fig:ba_primary_secondary}
       \end{center}

\end{figure*}

\begin{figure}
        \centering
        \includegraphics[height=7cm,width=9cm]{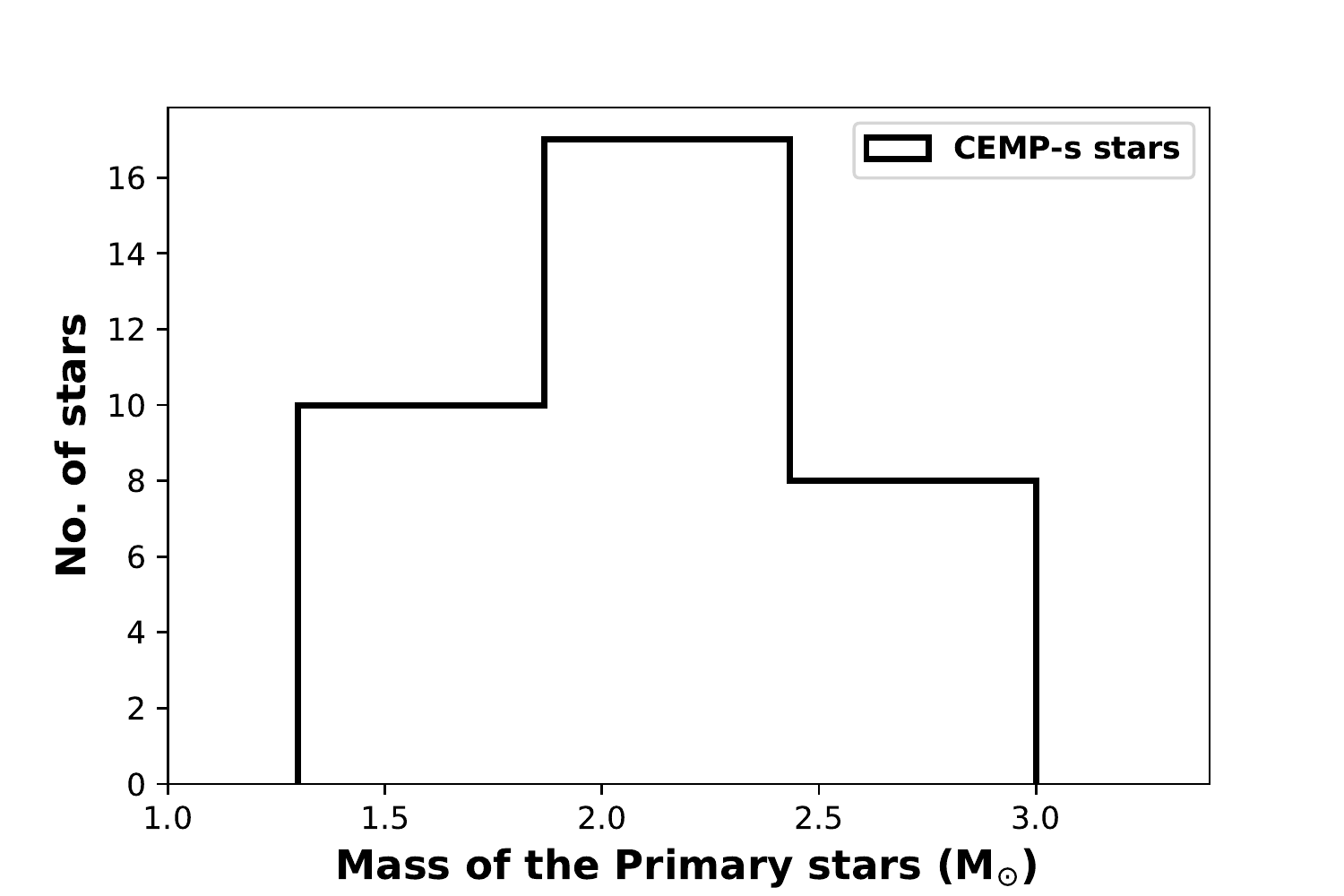}
        \caption{ Mass distribution of the primary companions of the CEMP-s stars. }
\label{fig:cemp_primary}
\end{figure}

\subsubsection{Mass distribution of Ba stars}

\citet{Han_et_al_1995} from theoretical analysis showed that the average mass of strong Ba stars is expected to be about 1.8 M$_{\odot}$. They found a different peak (1.7 M$_{\odot}$) when  all the Ba stars, including mild and strong Ba stars were considered. \citet{Mennessier_et_al_1997} advocated that mild Ba stars are clump giants with 2.5 -- 4.5 M$_{\odot}$ mass range and these stars are a population of high and low mass objects dominated by high mass objects with a small tail of less massive objects. They found that strong Ba stars are giants in the mass range 1.0 -- 3.0 M$_{\odot}$. \citet{Jorissen_et_al_1998} also reported different mass values for mild and strong Ba stars. Supposing the mass of the companion WD as 0.60 $\pm$ 0.04 M$_{\odot}$, they estimated the masses of mild and strong Ba stars to be 1.90 $\pm$ 0.20 and 1.50 $\pm$ 0.20 respectively. Recently, \citet{Escorza_et_al_2017} derived the mass distribution of Ba stars using a large sample. In order to find the masses of the Ba stars they generated the evolutionary tracks using {\it{STAREVOL}} at a metallicity [Fe/H] = --0.25. They don't see any difference in the mass distribution of mild and strong Ba stars, unlike the previous claims. They found that the mass distribution of Ba stars peak at 2.5 M$_{\odot}$ with a tail at higher masses upto 4.5 M$_{\odot}$. \citet{Jorissen_et_al_2019} also presented the mass distribution of the Ba stars they studied using similar method as \citet{Escorza_et_al_2017}, with the difference that they used the evolutionary tracks calculated at the metallicities of the individual stars. \citet{Jorissen_et_al_2019} found that the Ba stars range from 1.0 -- 3.0 M$_{\odot}$ with a tail up to 5 M$_{\odot}$. This tail is comprised of mild Ba stars, mostly of [Fe/H] $\geq$ --0.1. \citet{Jorissen_et_al_2019} mentioned that in order to get the accurate mass distribution, it is important to use the evolutionary tracks corresponding to the metallicities of the respective stars.

\begin{figure*}
     \begin{center}
\centering
        \subfigure[]{%
            \label{fig:hr_first}
            \includegraphics[height=7.0cm,width=8.5cm]{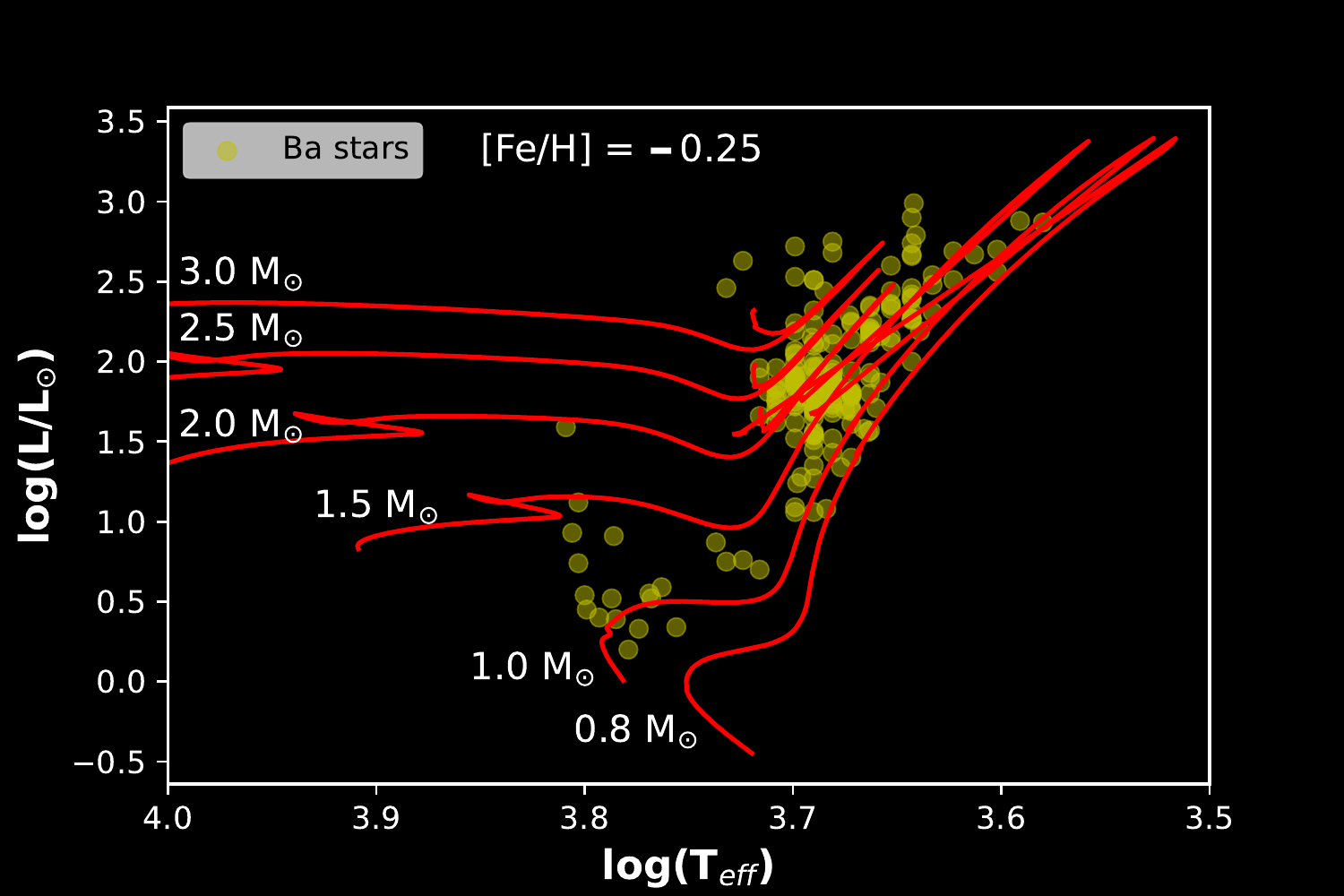}
        }%
        \subfigure[]{%
            \label{fig:hr_second}
            \includegraphics[height=7.0cm,width=8.5cm]{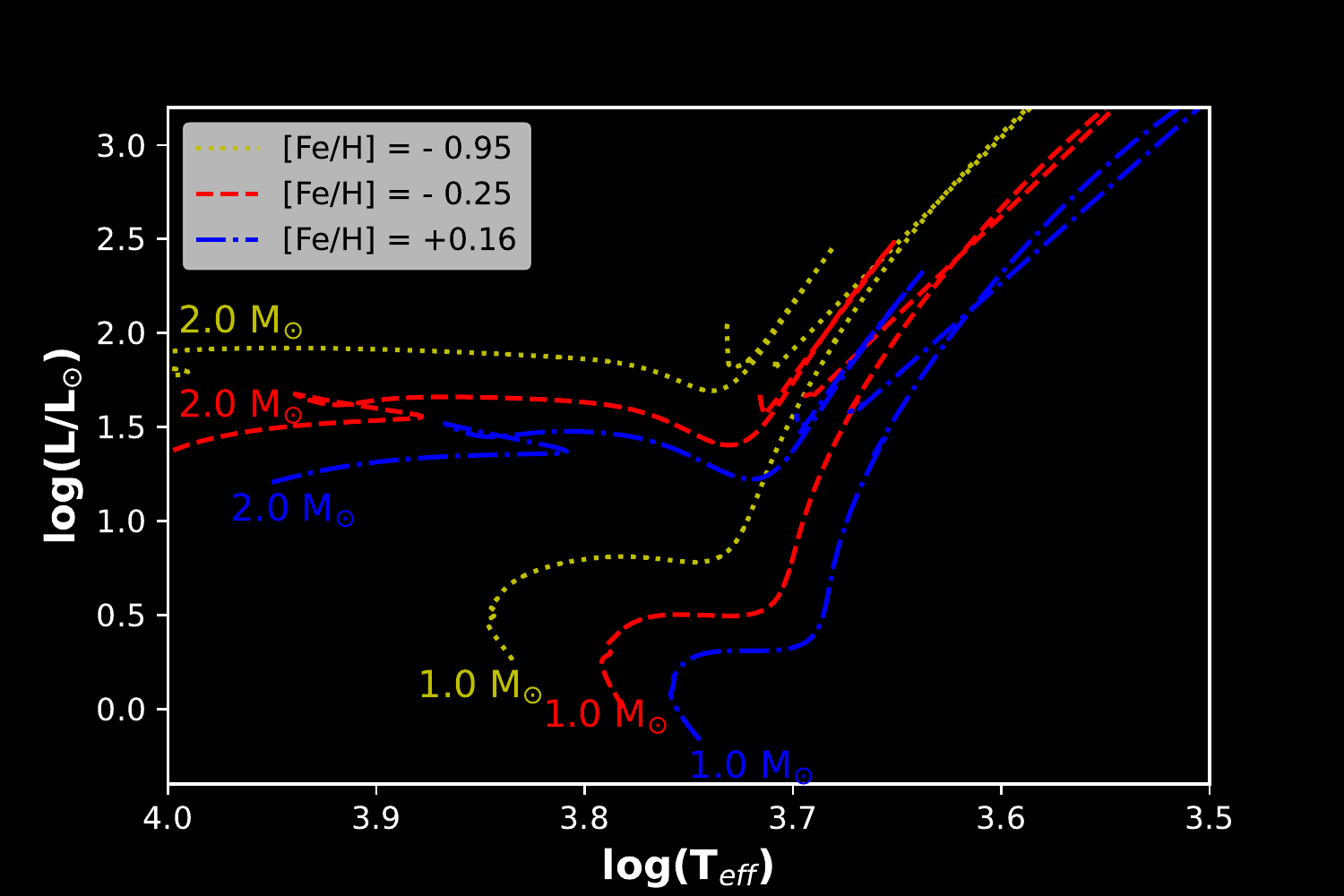}

        }\\ 

    \caption{ Hertzsprung-Russel diagram. Panel (a) shows evolutionary tracks for different masses at [Fe/H] = -0.25. Yellow points represent the positions of the Ba stars compiled from the literature. In panel (b), we have compared the evolutionary tracks for M = 1.0 M$_{\odot}$ \& 2.0 M$_{\odot}$ at three different metallicities [Fe/H] = --0.95, --0.25 and +0.16. }%
   \label{fig:hr_ba_comparison}
       \end{center}

\end{figure*}

Following the procedure described in Section~\ref{sec:secondary_mass} we have estimated the masses of the sample of Ba stars. We have used the BASTI-IAC evolutionary tracks at eleven different metallicities, [Fe/H] = +0.16, +0.05, --0.05, --0.15, --0.25, --0.35, --0.45, --0.55, --0.63, --0.72, --0.95 covering the metallicity range of the sample. We could estimate the masses of 205 Ba stars in total, out of which 52 are mild Ba stars and 153 are strong Ba stars. The [Fe/H] and T$_{eff}$ values of the stars are collected from the literature. The derived masses of the Ba stars are presented in Column 8 of Table~\ref{tab:Ba_stars}. Similar to that of \citet{Escorza_et_al_2017} and \citet{Jorissen_et_al_2019} we also found that mild and strong Ba stars occupy the same mass range with the tails of strong and mild Ba stars going to 4.0$\pm$0.25 M$\odot$ and 4.5$\pm$0.25 M$\odot$ respectively (Figure~\ref{fig:secondary_first}). Figure~\ref{fig:hr_first} shows the positions of the sample of Ba stars in the HR diagram with evolutionary tracks of different masses at [Fe/H]~=~--0.25. Figure~\ref{fig:secondary_second} shows the whole sample of Ba stars without distinguishing between mild and strong Ba stars. We found that the mass distribution of the Ba stars cannot be explained by a single Gaussian distribution. Rather, the Ba stars are distributed throughout the mass range with a disorderly manner. While \citet{Escorza_et_al_2017} found the masses of the Ba stars to peak at 2.5 M$_{\odot}$, in our case we found the average value of the distribution to be at 1.9 M$_\odot$. Note that \citet{Escorza_et_al_2017} observed a lack of Ba stars in the mass range 1.0 -- 2.0 M$_{\odot}$. But, neither our study nor the study by \citet{Jorissen_et_al_2019} confirms that observation. The difference seen in this study and the study by \citet{Escorza_et_al_2017} may come from two main sources. Firstly, the evolutionary tracks used in both the studies are different. While \citet{Escorza_et_al_2017} used STAREVOL code to generate the evolutionary tracks, we have used the BASTI-IAC evolutionary tracks (see Figure~9 of \citet{Escorza_et_al_2017} for a comparison of different sets of evolutionary tracks). Secondly, \citet{Escorza_et_al_2017} estimated the masses using the evolutionary tracks of a single [Fe/H] (=--0.25) and our estimation is based on eleven different metallicities close to the metallicities of the sample stars. For instance, Figure~\ref{fig:hr_second} shows how different the evolutionary track of a particular mass can be at different metallicities.

\vskip 1.0 cm
\subsubsection{Mass distribution of primary stars}

We have estimated the masses of the companion AGBs of the sample of Ba and CEMP-s stars following the same procedure described in Section~\ref{sec:primary_mass}. For this study, we assume that the enhancement of neutron-capture elements in all these stars is due to a mass-transfer event in a binary system where the slightly massive companion (primary star) evolves through the AGB phase, produces neutron-capture elements and transfers the elements to the secondary star, which alters the surface composition of the secondary star significantly. If the uncertainty in the abundance  for a particular element is not reported in the literature, we considered the uncertainty to be 0.2 dex. The number of neutron-capture elements for which abundance estimates are available in the literature range from 4 to 14 (column 12 of Table~\ref{tab:Ba_stars} and column 8 of Table~\ref{tab:CEMP-s_stars}).

{\it{Ba Stars:}} We could estimate the masses of the primary companions of 158 Ba stars. Out of these, 52 are mild Ba stars and 106 are strong Ba stars. We have divided the sample of Ba stars into seven groups based on the metallicities of the stars and used FRUITY models at z = 0.020, 0.014, 0.010, 0.008, 0.006, 0.003, 0.002. We have presented the initial masses of the primary stars (M$^{ini}_{AGB}$), $\chi^{2}$, dilution factor (d) and the number of elements used for the analysis (N) in columns 9, 10, 11 and 12 respectively of Table~\ref{tab:Ba_stars}. When we plot the histograms of primary companions' masses for mild and strong Ba stars separately, we can clearly see two peaks (Figure~\ref{fig:primary_third}). The mass range of primaries for mild Ba stars (1.5 $<$ M$_{\odot}$ $<$ 6.0) is greater than that (1.5 $<$ M$_{\odot}$ $<$ 4.0) of strong Ba stars. The masses of primaries of strong Ba stars peak at 2.5~M$_{\odot}$ with a standard deviation of 0.51~M$_{\odot}$ and that of mild Ba stars peak at 3.7~M$_{\odot}$ with a standard deviation of 1.03~M$_{\odot}$. If we consider the Ba star sample as a whole and not distinguish between mild and strong Ba stars, then the primaries of Ba stars peak at 2.9 M$_{\odot}$ with a standard deviation of 1.15 M$_{\odot}$ (Figure~\ref{fig:pri_sec_fourth}). We note that \citet{Stancliffe_2021}  studied the observed properties of barium star sample of \citet{de_Castro_et_al_2016} considering models involving mass transfer from an AGB companion.  The extent of dilution of accreted material  as the star evolves  have been  examined  and its impact on the surface abundances are  discussed. He could best fit 32 objects  from the sample using ejecta from 2.5 M$_{\odot}$ AGB stars and 36 objects using ejecta from 3 M$_{\odot}$ AGB stars. The accretion masses are found to be  broadly consistent with the  results obtained from hydrodynamical simulations of wind mass transfer in binary systems of \citet{Liu_et_al_2017}

{\it{CEMP-s Stars:}} The metallicities of the sample of 36 CEMP-s stars range from [Fe/H] = --3.00 to --1.29. We have divided the sample into five groups based on the metallicities of the stars. We have used FRUITY models at z= 0.001, 0.0003, 0.0001, 0.00005, 0.00002. We have presented the initial masses of the primary stars (M$^{ini}_{AGB}$), $\chi^{2}$, dilution factor (d) and the number of elements used for the analysis (N) in columns 5, 6, 7 and 8 respectively of Table~\ref{tab:CEMP-s_stars}. The primary mass distribution of CEMP-s stars peaks at 2.03 M$_{\odot}$ with a standard deviation of 0.49 M$_{\odot}$.

{\footnotesize
\begin{table*}
\centering
\caption{\bf{Initial masses of the AGB companions of CEMP-s stars.}}
\label{tab:CEMP-s_stars} 
\scalebox{1.0}{
\begin{tabular}{lccccccc}
\hline

\hline
\hline
Star Name     & T$_{eff}$ & log$g$ & [Fe/H] & M$^{ini}_{AGB}$ & $\chi^{2}$ & dil. & N \\
              &  (K)      & (cgs)  &        & (M$_{\odot}$)   &           &      &   \\
\hline
BD+04 2466    &  5100 & 1.80 & -1.92 & 3.0 & 3.61 & 0.26884 &  7  \\
BS 16077-0077 &  5900 & 3.19 & -2.05 & 2.0 & 1.89 & 1.14484 &  9  \\
CD-27 14351   &  4320 & 0.50 & -2.71 & 1.3 & 8.02 & 0.24628 &  8  \\
\hline
\end{tabular}}

\textbf{Note:} This table is available in its entirety online only.
A portion is shown here for guidance regarding its form and content.\\

\end{table*}}

\subsection{Formation scenarios of mild and strong Ba stars}
\label{sec:formation}
\citet{de_Castro_et_al_2016} considered that mild Ba stars are formed from the ISM mildly enhanced with s-process elements. However, studies \citep{Jorissen_et_al_1998, Jorissen_et_al_2019} have shown that both mild and strong Ba stars are formed in binary systems and AGB mass-transfer is responsible for the enhanced abundances of heavy elements. So, we can safely discard this formation scenario. From theoretical analysis, \citet{Han_et_al_1995} showed that while most mild Ba stars are formed from the wind accretion and wind exposure channels, the strong Ba stars are formed from the wind accretion, wind exposure and stable Roche lobe overflow (RLOF) channels. \citet{Yang_et_al_2016} discussed two possible formation scenarios for the formation of mild Ba stars- i) the mild enhancement could be explained by weaker neutron-exposure in the progenitor AGB and/or ii) less accretion efficiency due to longer orbital period which means larger distance between the binary companions. However, from long term radial velocity monitoring programs \citet{Jorissen_et_al_1998, Jorissen_et_al_2019} have shown that mild Ba stars are not limited to long period systems only. The orbital periods of strong Ba stars are generally shorter than that of mild Ba stars, but there is no tight correlation. In fact, there is a clear overlap in the orbital periods of these two subclasses. From Table~8 of \citet{Jorissen_et_al_2019}, we can see that the period ranges of mild and strong Ba stars are 80 -- 22065 days and 185.7 -- 8523 days respectively. So, a scenario having longer orbital period cannot explain the formation of a mild Ba star. However, we note that the ranges of both orbital periods and progenitor masses of mild Ba stars are larger than that of strong Ba stars.

\begin{figure*}
     \begin{center}
\centering
        \subfigure[]{%
            \label{fig:model_comparison}
            \includegraphics[height=7.5cm,width=9.0cm]{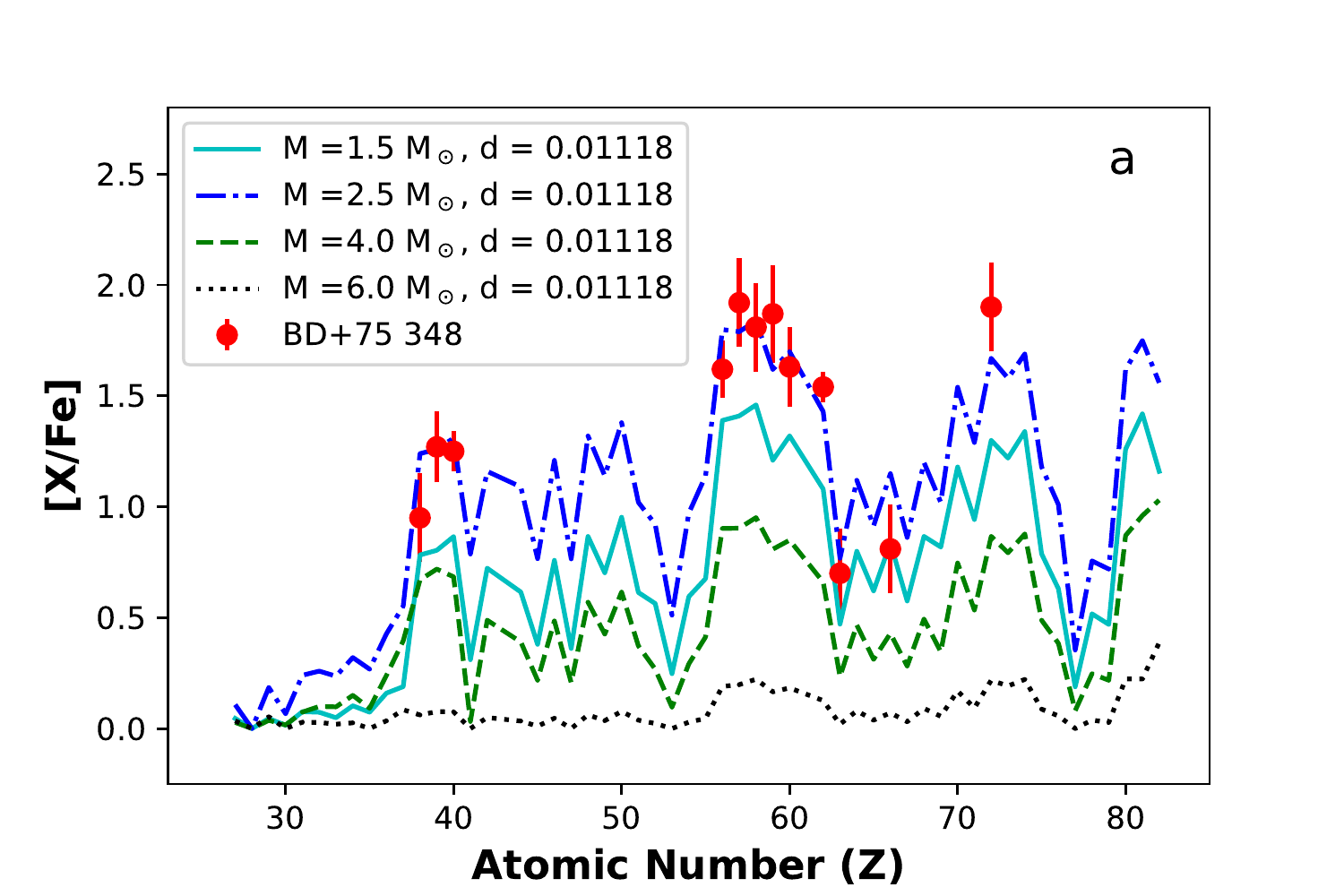}
        }%
        \subfigure[]{%
            \label{fig:model_comparison_metallicity}
            \includegraphics[height=7.5cm,width=9.0cm]{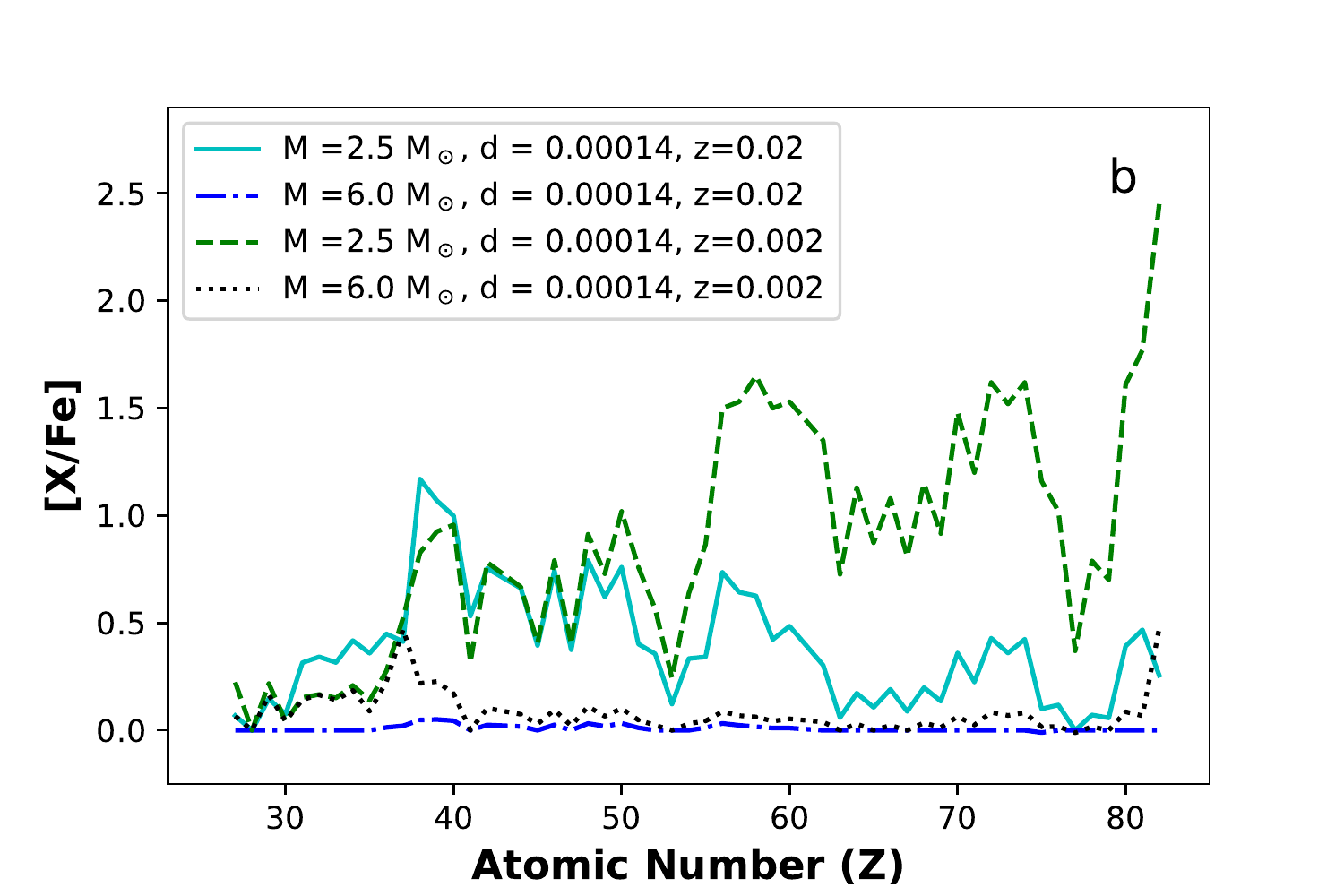}

        }\\ 

    \caption{ Panel a: Comparison of the s-process AGB model yields obtained at z = 0.006 for different masses using the same dilution factor (d). The points with error bars indicate the observed abundances of BD+75~348. Panel b: Comparison of the s-process AGB model yields obtained at two metallicities z = 0.02 and z = 0.002 for two masses M = 2.5 \& 6.0 M$_{\odot}$ using the same dilution factor (d) }%
   \label{fig:comparison_model}
       \end{center}

\end{figure*}

\begin{figure}
        \centering
        \includegraphics[height=7cm,width=9cm]{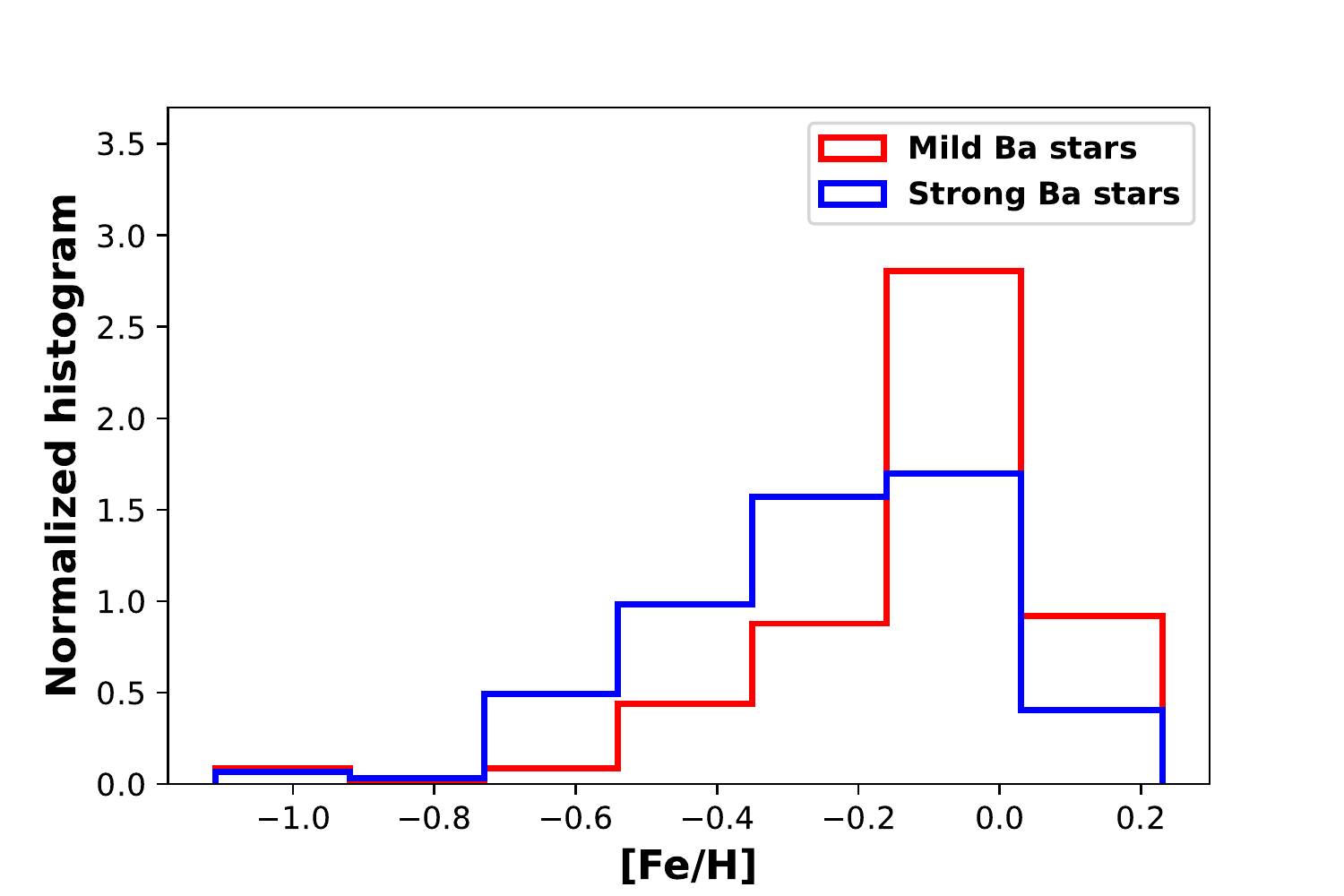}
        \caption{Metallicity distribution of the sample of mild and strong Ba stars.}
\label{fig:metallicity_ba_comparison}
\end{figure}

The difference in the peaks of primaries for mild and strong Ba stars (Figure~\ref{fig:primary_third}) indicates that companions of mild Ba stars are more massive than those of strong Ba stars. This is not surprising because, from Figure~\ref{fig:model_comparison}, it can be seen that massive (e.g. 4, 5, 6 M$_{\odot}$) AGBs yield less than the low-mass AGBs with same dilution factor. So, we can say that the dominant factor controlling the abundance peculiarities of mild Ba stars is the initial mass of the companion. Mass-transfer from 4.0 -- 6.0 M$_{\odot}$ AGB companions can describe the formation of mild Ba stars. The WD masses of $\sim$ 1.0 M$_{\odot}$ around Ba stars also point towards a massive (5 M$_{\odot}$) companion AGB \citep{Jorissen_et_al_2019}. However, there is an overlap from 1.3 -- 4.0 M$_{\odot}$ in the case of primary masses of mild and strong Ba stars. In this mass range, the metallicity of the system and dilution (hence the distance of the binary companions) play crucial roles for the formation of mild Ba stars. From Figure~\ref{fig:metallicity_ba_comparison}, we see that the mild Ba stars are distributed towards higher metallicities than that of strong Ba stars and Figure~\ref{fig:model_comparison_metallicity} shows the metallicity dependence of the AGB models. At higher metallicities, s-process efficiency decreases and hence AGB yields decrease.

\section{Conclusions}
\label{sec:conclusion}
 
We have conducted detailed spectroscopic analysis of seven stars based on high-quality, high-resolution spectra. Out of the seven program stars, we have classified two stars (BD+09~3019 and BD+75~348) as strong Ba stars, one star (HD~238020) as mild Ba star and the other four stars (HE~0319--0215, HE~0507--1653, HE~0930--0018 and HE~1023--1504) as CEMP-s stars. Note that HE~0507--1653 shows [Eu/Fe] $>$ 1.0 and according to the classification criteria of \citet{abate2016cemp-rs} this star belongs to the category of CEMP-r/s stars, but the classification criteria that we proposed in \citet{Goswami_et_al_1_2021} place this star in the CEMP-s subclass. While we could not get a reasonable fit for this object with i-process models with higher ($\geq$ 10$^{12-15}$ cm$^{-3}$) neutron-densities, s-process AGB model with initial mass 2.0 M$\odot$ satisfactorily reproduces the observed abundance pattern. Kinematic analysis shows that BD+75~348, HD~238020 and HE~0930--0018 belong to the thin disc population, BD+09~3019 is a thick disc object and HE~0319--0215 \& HE~0507--1653 show the probability of being halo objects.
 
We have derived the mass distribution of Ba stars and found that a single Gaussian cannot describe the mass distribution. The average mass of the distribution is found to be 1.9 M$_{\odot}$ with the tails of strong and mild Ba stars going up to 4.0 M$\odot$ and 4.5 M$\odot$ respectively. We confirm the previous claim by \citet{Escorza_et_al_2017} that mild and strong Ba stars occupy the same mass range, but we do not confirm their claim that there is a lack of Ba stars in the mass range 1.0 -- 2.0 M$_{\odot}$. Using a parametric-model-based analysis we have derived the mass distributions of the AGB progenitors of the Ba and CEMP-s stars. To the best of our knowledge, this is the first attempt to derive the initial mass distribtuions of the primary companions of these stars. We found that the mass distributions of the progenitor AGBs of the mild and strong Ba stars peak at different values. The peaks of the progenitor mass distributions of mild and strong Ba stars are at 3.7 M$_{\odot}$ and 2.5 M$_{\odot}$ respectively. We can, therefore, say that the initial mass of the companion AGB is the dominant factor controlling the heavy elements' enhancement in mild Ba stars. However, we cannot neglect the orbital periods and metallicities of the binary systems as clear overlap can be seen in the progenitor mass distributions and orbital periods of mild and strong Ba stars. The mass distribution of progenitor AGBs of CEMP-s stars peak at 2.03 M$_{\odot}$ with a standard deviation of 0.49 M$_{\odot}$.

\noindent

\begin{acknowledgements}

This work made use of the SIMBAD astronomical database, operated at CDS, Strasbourg, France, the NASA ADS, USA and data from the European Space Agency (ESA) mission Gaia (\url{https://www.cosmos.esa.int/gaia}), processed by the Gaia Data Processing and Analysis Consortium (DPAC, \url{https://www.cosmos.esa.int/web/gaia/dpac/consortium}). We thank the referee for useful comments and suggestions. 
We are thankful to Melanie Hampel for providing us with the  {\it{i}}-process yields in the form of number fractions.
Funding from DST SERB project No. EMR/2016/005283 is gratefully acknowledged. 
\end{acknowledgements}

\bibliography{sample}{}
\bibliographystyle{aasjournal}

\appendix

{\footnotesize
\begin{table*}
\centering
{\bf{Table 4. Equivalent widths (in m\r{A}) of Fe lines used for deriving atmospheric parameters.}}
\label{tab:Fe_linelist_appendix} 
\scalebox{0.73}{
\begin{tabular}{ccccccccccc}
\hline
Wavelength&Element&E$_{low}$ &   log gf  &  BD+75~348 & BD+09~3019 & HD~238020  &HE~0319--0215&HE~0507--1653&HE~0930--0018&HE~1023--1504  \\
(\r{A})   &       & (eV)     &           &            &            &            &             &             &             &               \\
\hline 
4187.04   & Fe I  & 2.45     & --0.548   & -          & -          & -          & -          &124.8 (6.12)& -          & -               \\
4202.03   &       & 1.48     & --0.708   & -          & -          & -          & -          &179.4 (6.02)& -          & -               \\
4203.57   &       & 1.01     & --3.869   & -          & -          & 73.1 (6.72)& -          & -          & -          & -               \\
4210.343  &       & 2.48     & --0.870   &174.2 (7.17)& -          &133.3 (6.95)& -          & -          & -          & -               \\
4213.65   &       & 2.85     & --1.290   & -          & -          & 97.4 (6.88)& -          & -          & -          & -               \\
4217.55   &       & 3.43     & --0.510   & -          & -          & 97.9 (6.75)& -          & -          & -          & -               \\
\hline
\end{tabular}}

The numbers in the parenthesis in columns 5 to 11 give the derived abundances from the respective line.\\

\end{table*}}

{\footnotesize
\begin{table*}
{\textit{-continued}}
\centering
\scalebox{0.73}{
\begin{tabular}{ccccccccccc}
\hline
Wavelength&Element&E$_{low}$ &   log gf  &  BD+75~348 & BD+09~3019 & HD~238020  &HE~0319--0215&HE~0507--1653&HE~0930--0018&HE~1023--1504  \\
(\r{A})   &       & (eV)     &           &            &            &            &             &             &             &               \\
\hline 
4994.13   &       & 0.92     & --3.080   &154.5 (6.97)& -          &122.6 (6.79)& 74.4 (4.83)& -          & -          & -               \\
5002.79   &       & 3.40     & --1.580   &105.2 (7.38)& -          & 65.4 (6.77)& -          & -          & -          & -               \\
5006.12   &       & 2.83     & --0.615   & -          & -          & -          & 84.2 (5.07)&112.5 (5.99)& -          & -               \\
5014.94   &       & 3.94     & --0.250   & -          & -          & 92.1 (6.67)& -          & -          & -          & -               \\
5049.82   &       & 2.28     & --1.420   &167.1 (7.15)& -          &132.4 (6.93)& -          & -          & -          & -               \\
5074.75   &       & 4.22     & --0.200   & -          & -          & 93.8 (6.95)& -          & -          & -          & -               \\
\hline
\end{tabular}}
\end{table*}}

{\footnotesize
\begin{table*}

{\textit{-continued}}
\centering
\scalebox{0.73}{
\begin{tabular}{ccccccccccc}
\hline
Wavelength&Element&E$_{low}$ &   log gf  &  BD+75~348 & BD+09~3019 & HD~238020  &HE~0319--0215&HE~0507--1653&HE~0930--0018&HE~1023--1504  \\
(\r{A})   &       & (eV)     &           &            &            &            &             &             &             &               \\
\hline 
5554.89   &       & 4.55     & --0.440   & -          & -          & 67.3 (6.89)& -          & -          & -          & -               \\
5557.98   &       & 4.47     & --1.280   & -          & -          & 38.8 (7.07)& -          & -          & -          & -               \\
5560.21   &       & 4.43     & --1.190   & 60.9 (7.15)& -          & 35.4 (6.87)& -          & -          & -          & -               \\
5562.71   &       & 4.43     & --0.580   & 77.1 (6.86)& -          & -          & -          & -          & -          & -               \\
5567.39   &       & 2.61     & --2.800   & -          & -          & 62.0 (6.97)& -          & -          & -          & -               \\
5576.09   &       & 3.43     & --1.000   & -          & -          & 97.8 (6.88)& -          & -          & -          & -               \\
\hline
\end{tabular}}

\end{table*}}

{\footnotesize
\begin{table*}
\centering
{\bf{Table 6. Equivalent widths (in m\r{A}) of lines used for calculation of elemental abundances.}}
\label{tab:Elem_linelist1_appendix} 
\scalebox{0.73}{
\begin{tabular}{ccccccccccc}
\hline
Wavelength&Element&E$_{low}$ &   log gf  &  BD+75~348 & BD+09~3019 & HD~238020  &HE~0319--0215&HE~0507--1653&HE~0930--0018&HE~1023--1504  \\
(\r{A})   &       & (eV)     &           &            &            &            &             &             &             &               \\
\hline 
5682.633  & Na I  & 2.10     & --0.700  &115.8 (6.25)&172.5 (6.08)& 64.4 (5.62)& 22.4 (4.71)& 45.2 (5.19)&112.1 (5.39)& 51.5 (5.06)\\
5688.205  &       & 2.10     & --0.450  &140.3 (6.42)&170.6 (5.81)& 87.1 (5.77)& 25.3 (4.53)& 58.7 (5.17)&136.8 (5.43)& -          \\
6154.23   &       & 2.10     & --1.560  & 59.1 (6.14)& -          & 16.1 (5.50)& -          & -          & -          & -          \\
6160.75   &       & 2.10     & --1.260  & 88.1 (6.30)& -          & 30.1 (5.55)& -          & 19.7 (5.21)& -          & -          \\
4571.10   & Mg I  & 0.00     & --5.691  &181.2 (7.59)& -          &131.6 (7.27)& 62.7 (5.51)&113.8 (6.62)& -          & -          \\
4702.99   &       & 4.35     & --0.666  &178.2 (7.37)& -          &161.8 (7.38)& -          & -          & -          & -          \\
\hline
\end{tabular}}

The numbers in the parenthesis in columns 5 to 11 give the derived abundances from the respective line.\\

\end{table*}}

{\footnotesize
\begin{table*}
{\textit{-continued}}
\centering
\scalebox{0.73}{
\begin{tabular}{ccccccccccc}
\hline
Wavelength&Element&E$_{low}$ &   log gf  &  BD+75~348 & BD+09~3019 & HD~238020  &HE~0319--0215&HE~0507--1653&HE~0930--0018&HE~1023--1504  \\
(\r{A})   &       & (eV)     &           &            &            &            &             &             &             &               \\
\hline 
5185.91   &       & 1.89     & --1.350  & 99.0 (4.50)& -          & 84.4 (4.15)& -          & 64.0 (3.51)& -          & -          \\
5226.54   &       & 1.57     & --1.300  & -          & -          &112.9 (4.43)& -          & -          & 76.0 (3.44)& -          \\
5336.77   &       & 1.58     & --1.700  &108.8 (4.68)& -          & 92.7 (4.33)& -          & 95.0 (4.17)& -          & -          \\
5381.02   &       & 1.57     & --2.080  & -          & -          & 79.4 (4.38)& -          & -          & -          & -          \\
5418.75   &       & 1.58     & --1.999  & 84.5 (4.41)& -          & 70.1 (4.09)& -          & -          & -          & -          \\
4545.95   & Cr I  & 0.94     & --1.370  & -          & -          & 92.5 (5.06)& -          & -          & -          & -          \\
\hline
\end{tabular}}

\end{table*}}

{\footnotesize
\begin{table*}
{\textit{-continued}}
\centering
\scalebox{0.73}{
\begin{tabular}{ccccccccccc}
\hline
Wavelength&Element&E$_{low}$ &   log gf  &  BD+75~348 & BD+09~3019 & HD~238020  &HE~0319--0215&HE~0507--1653&HE~0930--0018&HE~1023--1504  \\
(\r{A})   &       & (eV)     &           &            &            &            &             &             &             &               \\
\hline 
4483.89   &       & 0.86     &   0.010  & 86.8 (2.55)& -          & 25.6 (0.97)& -          & 77.4 (2.07)& -          & -          \\
4486.91   &       & 0.30     & --0.474  &115.4 (3.14)& -          & 34.0 (1.04)& -          & -          & -          & -          \\
4515.85   &       & 1.06     & --0.612  & 65.0 (2.76)& -          &  9.3 (1.24)& -          & -          & -          & -          \\
4560.96   &       & 0.68     & --0.546  &101.4 (3.28)& -          & -          & -          & -          & -          & -          \\
4562.36   &       & 0.48     &   0.081  &120.4 (2.88)& -          & 49.3 (1.04)& -          & 96.3 (2.06)& -          & -          \\
4628.16   &       & 0.52     &   0.008  & -          & -          & 42.5 (0.99)& -          & -          & -          & -          \\
\hline
\end{tabular}}

\end{table*}}

{\footnotesize
\begin{table*}
\centering
{\bf{Table 13. Masses of the Ba stars and initial masses of their AGB companions.}}
\label{tab:Ba_stars_appendix}
\scalebox{0.86}{
\begin{tabular}{lccccccccccccc}
\hline

Star Name & T$_{eff}$ & logg & [Fe/H] & [ls/Fe] & [hs/Fe] & [hs/ls] & M$_{Ba}$ & M$^{ini}_{AGB}$ & $\chi^{2}$ & dil & N & Type & Ref \\
          &  (K)      &(cgs) &        &         &         &        &(M$_{\odot}$)&(M$_{\odot}$)&            &   &   &      &     \\
\hline
BD-01302   & 4200 & 1.10 &-0.64 &-0.14 & 0.17 & 0.310 &  -   & 1.5 & 0.67 & 1.85697   & 5 & m & 2  \\
BD-18821   & 5000 & 2.30 &-0.27 & 0.70 & 1.27 & 0.570 & 2.30 & 3.0 & 1.92 & 0.18030  & 5 & s & 2  \\
BD-083194  & 4900 & 3.00 &-0.10 & 0.95 & 1.52 & 0.570 & 2.00 &  -  &  -   &    -      & 5 & s & 2  \\
BD-094337  & 4800 & 2.60 &-0.24 & 1.31 & 1.50 & 0.190 & 1.80 & 2.5 & 0.26 & 0.42156  & 5 & s & 2  \\
BD-142678  & 5200 & 3.10 & 0.01 & 0.94 & 0.94 & 0.010 & 3.00 &  -  &  -   &    -      & 5 & s & 2  \\
BD+092384  & 4900 & 2.50 &-0.98 & 0.57 & 0.88 & 0.310 & 0.80 & 3.0 & 0.90 & 0.46181  & 5 & s & 2  \\
\hline
\end{tabular}}
References: 1. \citet{Allen_Barbuy_2006}, 2. \citet{de_Castro_et_al_2016}. 3. \citet{Yang_et_al_2016}, 4. \citet{Karinkuzhi_et_al_2018}, 5. \citet{shejeela_agb_2020}, 6. This work \\

\end{table*}}

{\footnotesize
\begin{table*}
{\textit{-continued}}
\centering
\scalebox{0.86}{
\begin{tabular}{lccccccccccccc}
\hline
Star Name & T$_{eff}$ & logg & [Fe/H] & [ls/Fe] & [hs/Fe] & [hsls] & M$_{Ba}$ & M$^{ini}_{AGB}$ & $\chi^{2}$ & dil & N & Type & Ref \\
          &  (K)      &(cgs) &        &         &         &        &(M$_{\odot}$)&(M$_{\odot}$)&            &   &   &      &     \\
\hline
HD 32901   & 4400 & 1.60 &-0.44 & 0.29 & 0.92 & 0.630 & 0.80 & 3.0 & 0.52 & 0.63315  & 5 & s & 2  \\
HD 33709   & 5000 & 2.10 &-0.20 & 0.36 & 0.45 & 0.090 & 4.00 &  -  &  -   &    -      & 5 & m & 2  \\
HD 35993   & 5100 & 2.90 &-0.05 & 0.91 & 1.22 & 0.310 & 2.40 &  -  &  -   &    -      & 5 & s & 2  \\
HD 36650   & 4880 & 2.40 &-0.02 & 0.62 & 0.84 & 0.210 & 1.90 & 3.0 & 2.61 & 0.18718  & 5 & s & 5  \\
HD 38488   & 4400 & 2.00 & 0.05 & 0.83 & 0.98 & 0.140 & 2.30 &  -  &  -   &    -      & 5 & s & 2  \\
HD 39778   & 5000 & 2.50 &-0.12 & 0.78 & 0.98 & 0.200 & 2.40 &  -  &  -   &    -      & 5 & s & 2  \\
\hline
\end{tabular}}

\end{table*}}

{\footnotesize
\begin{table*}
\centering
{\textit{-continued}}
\scalebox{0.86}{
\begin{tabular}{lccccccccccccc}
\hline

Star Name & T$_{eff}$ & logg & [Fe/H] & [ls/Fe] & [hs/Fe] & [hsls] & M$_{Ba}$ & M$^{ini}_{AGB}$ & $\chi^{2}$ & dil & N & Type & Ref \\
          &  (K)      &(cgs) &        &         &         &        &(M$_{\odot}$)&(M$_{\odot}$)&            &   &   &      &     \\
\hline
HD 95193   & 5000 & 2.70 & 0.04 & 0.61 & 0.54 &-0.070 & 2.60 & 3.0 & 1.16 & 0.20525  & 7 & m & 2, 3  \\
HD 95345   & 4500 & 1.80 &-0.16 & 0.01 & 0.23 & 0.220 & 1.50 & 5.0 & 0.54 & 0.11124  & 5 & m & 2  \\
HD 100503  & 4000 & 2.00 &-0.72 & 1.18 & 1.67 & 0.490 &  -   & 2.0 & 6.43 & 0.00100 &12 & s & 4  \\
HD 102762  & 4400 & 1.70 &-0.17 & 0.85 & 1.51 & 0.650 & 2.30 &  -  &  -   &    -      & 5 & s & 2  \\
HD 107264  & 4500 & 1.50 &-0.19 & 0.74 & 1.23 & 0.480 & 2.60 &  -  &  -   &    -      & 5 & s & 2  \\
HD 107270  & 5400 & 2.70 & 0.05 & 0.69 & 0.32 &-0.370 & 4.00 &  -  &  -   &    -      & 5 & m & 2  \\
\hline
\end{tabular}}

\end{table*}}

{\footnotesize
\begin{table*}
\centering
{\textit{-continued}}
\scalebox{0.86}{
\begin{tabular}{lccccccccccccc}
\hline

Star Name & T$_{eff}$ & logg & [Fe/H] & [ls/Fe] & [hs/Fe] & [hsls] & M$_{Ba}$ & M$^{ini}_{AGB}$ & $\chi^{2}$ & dil & N & Type & Ref \\
          &  (K)      &(cgs) &        &         &         &        &(M$_{\odot}$)&(M$_{\odot}$)&            &   &   &      &     \\
\hline
HD 184001  & 5000 & 2.50 &-0.21 & 0.60 & 0.61 & 0.010 & 2.20 & 2.5 & 0.04 & 0.81789 & 5 & s & 2  \\
HD 187308  & 4900 & 2.50 &-0.08 & 0.65 & 0.70 & 0.060 & 2.10 & 3.0 & 1.93 & 0.08263 & 5 & s & 2  \\
HD 187762  & 4800 & 2.40 &-0.30 & 0.23 & 0.54 & 0.310 & 1.50 &  -  &  -   &    -     & 5 & m & 2  \\
HD 188985  & 6090 & 4.30 &-0.30 & 1.08 & 1.16 & 0.070 & 0.90 & 2.5 & 0.71 & 0.46841 &14 & s & 1  \\
HD 193530  & 4400 & 1.60 &-0.17 & 0.66 & 0.84 & 0.180 & 1.80 & 2.0 & 1.77 & 0.17296 & 5 & s & 2  \\
HD 196445  & 4400 & 1.40 &-0.19 & 0.99 & 1.42 & 0.430 & 1.90 &  -  &  -   &    -     & 5 & s & 2  \\
\hline
\hline
\end{tabular}}

\end{table*}}

{\footnotesize
\begin{table*}
\centering
{\bf{Table 14. Initial masses of the AGB companions of CEMP-s stars.}}
\label{tab:CEMP-s_stars_appendix} 
\scalebox{1.0}{
\begin{tabular}{lccccccc}
\hline

\hline
\hline
Star Name     & T$_{eff}$ & log$g$ & [Fe/H] & M$^{ini}_{AGB}$ & $\chi^{2}$ & dil. & N \\
              &  (K)      & (cgs)  &        & (M$_{\odot}$)   &           &      &   \\
\hline
BD+04 2466    &  5100 & 1.80 & -1.92 & 3.0 & 3.61 & 0.26884 &  7  \\
BS 16077-0077 &  5900 & 3.19 & -2.05 & 2.0 & 1.89 & 1.14484 &  9  \\
CD-27 14351   &  4320 & 0.50 & -2.71 & 1.3 & 8.02 & 0.24628 &  8  \\
CD-38 2151    &  4600 & 0.90 & -2.03 & 3.0 & 6.52 & 0.14247 &  9  \\
CS 22880-0074 &  5850 & 3.80 & -1.93 & 3.0 & 2.74 & 0.34538 & 11  \\
CS 22942-0019 &  5000 & 2.40 & -2.64 & 1.3 & 5.30 & 0.55686 & 10  \\
\hline
\hline
\end{tabular}}

\end{table*}}

\begin{figure}
        \centering
        \includegraphics[height=7cm,width=9cm]{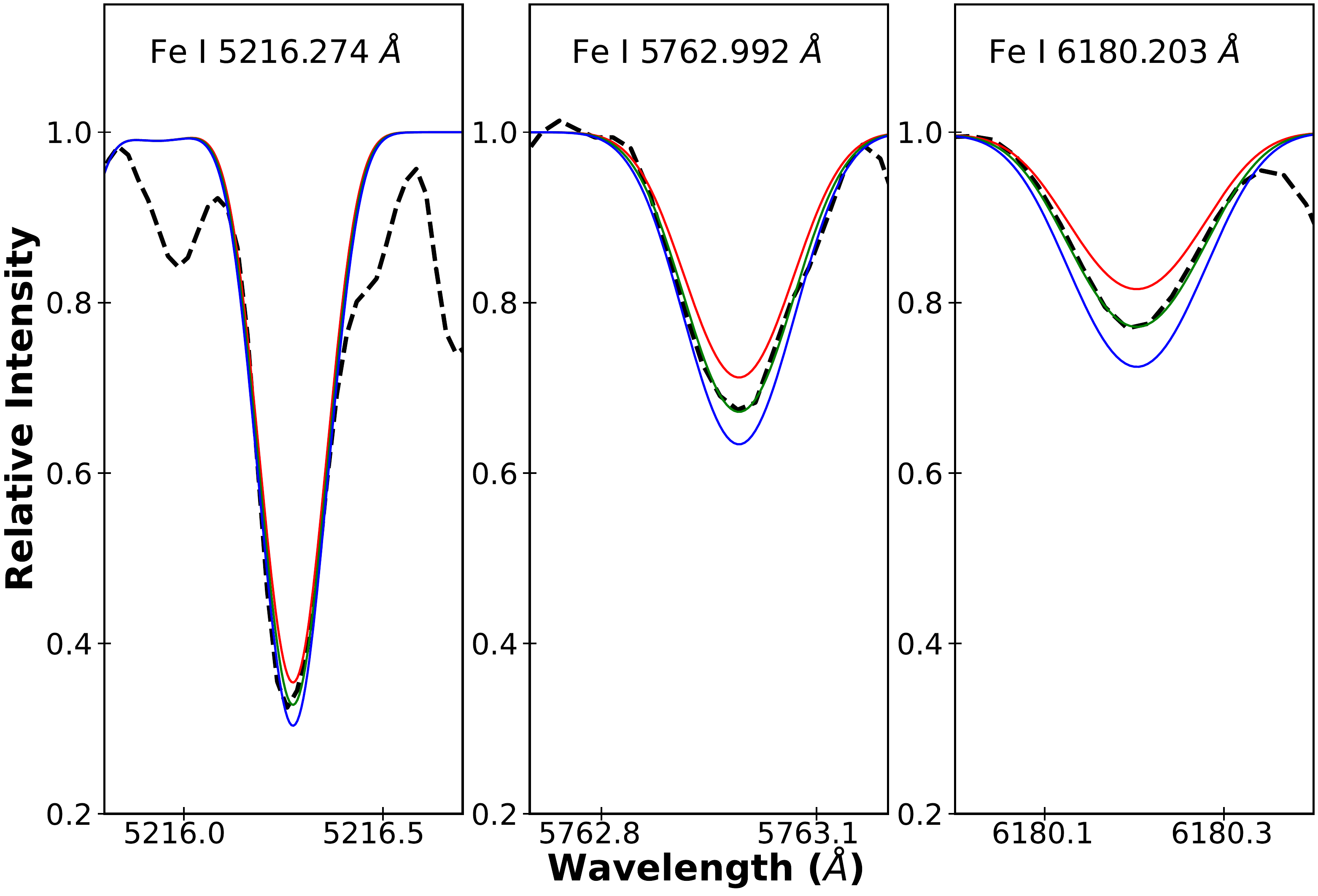}
        \caption{ Spectral synthesis plots for a few Fe lines of the spectrum of HE~1023--1504. }
\label{fig:fe_HE1023-1504}
\end{figure}

\begin{figure}
        \centering
        \includegraphics[height=8cm,width=9.2cm]{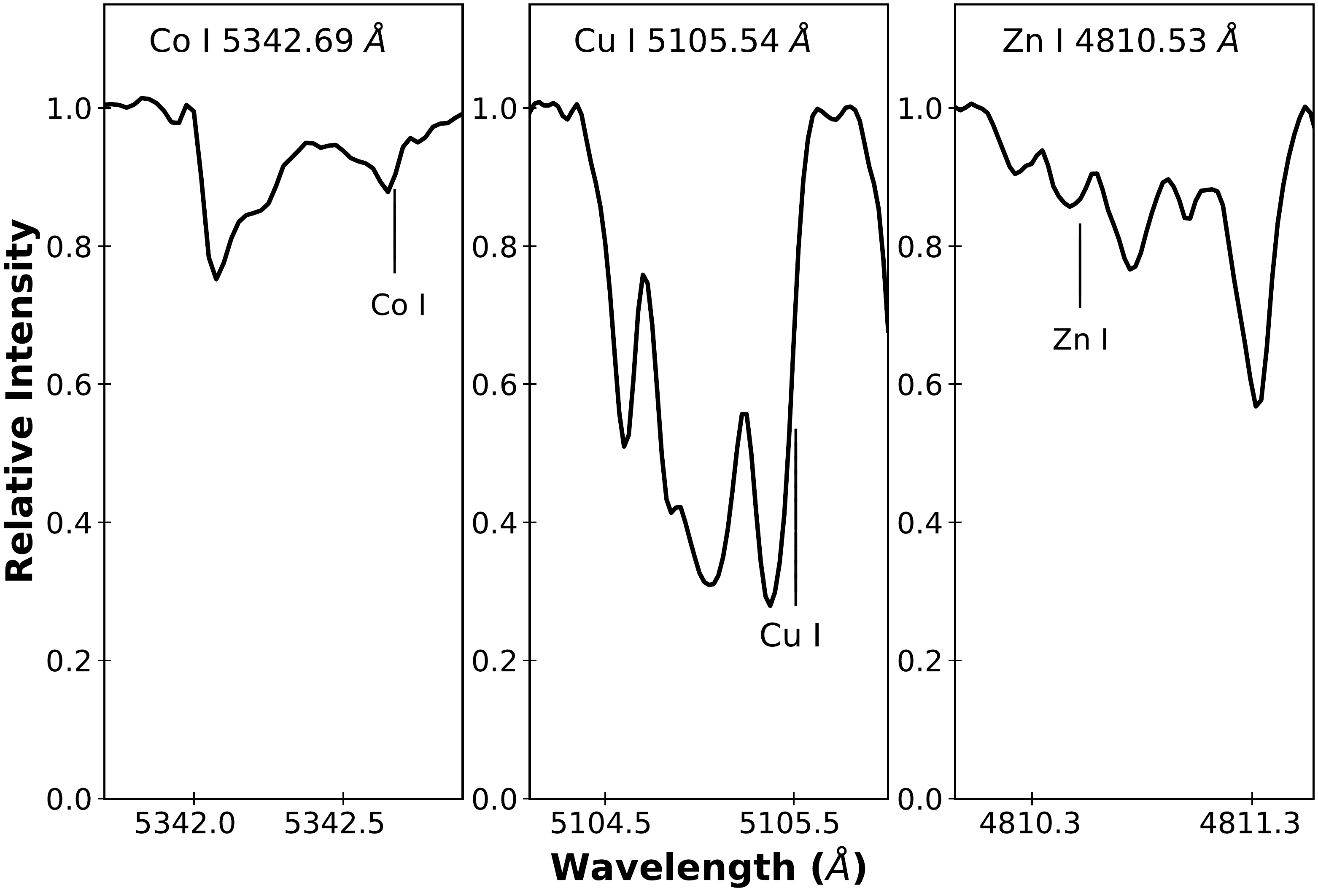}
        \caption{ Sample spectra of the Co, Cu and Zn lines of HE~0319--0215.}
\label{fig:CoCuZn}
\end{figure}

\begin{figure}
        \centering
        \includegraphics[height=7cm,width=9cm]{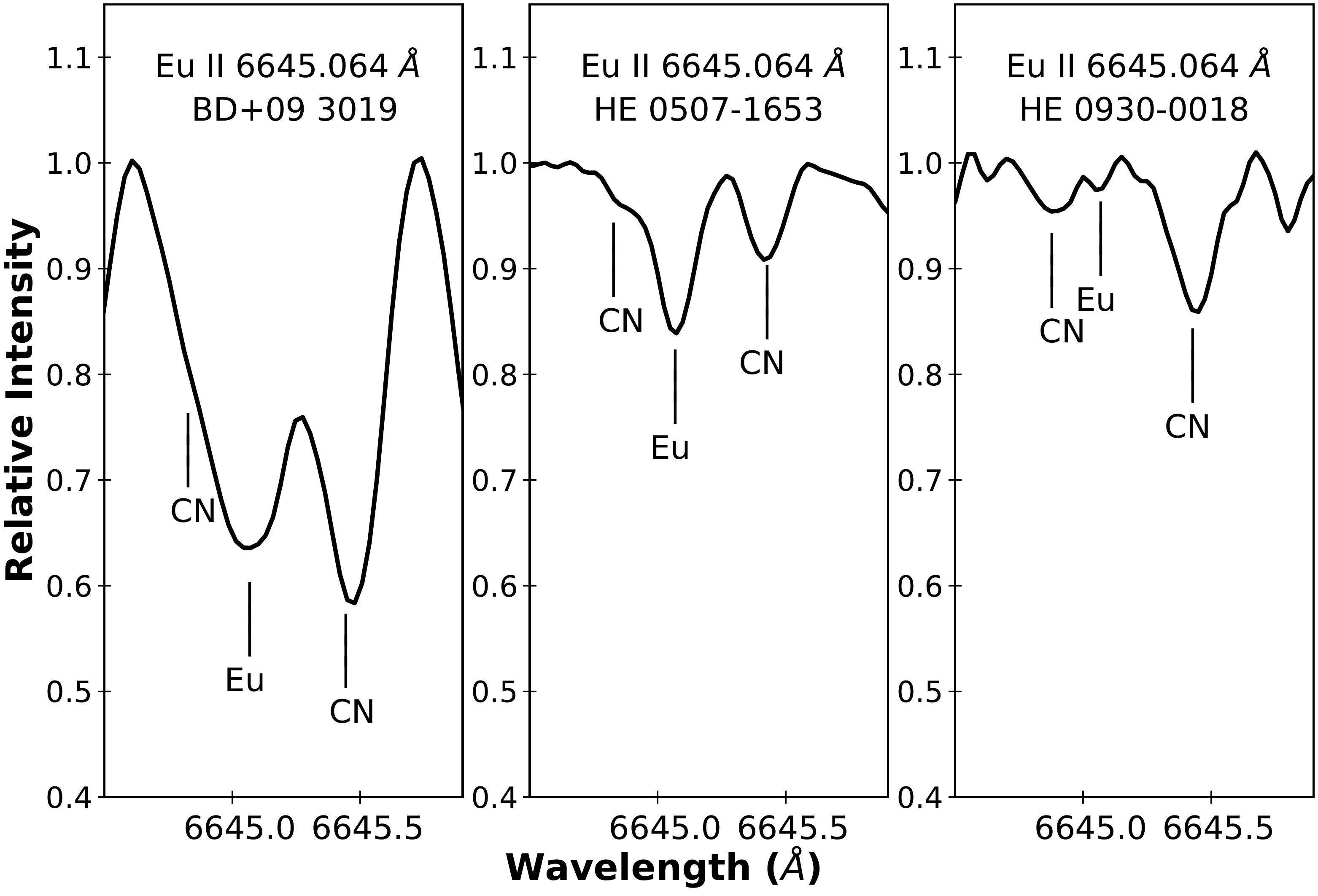}
        \caption{ Sample spectra of the Eu II line at 6645.064 {\rm \AA} in the stars BD+09~3019, HE~0507--1653 and HE~0930--0018. This line is very weak (not usable) in the star HE~0930--0018.}
\label{fig:Eu_comp}
\end{figure}

\end{document}